\documentclass[11pt]{article}
\usepackage{authblk}
\usepackage{caption}
\usepackage[caption=false]{subfig}

\usepackage{amsmath,amsfonts,amsthm,amssymb,graphics,graphicx,mathtools,epstopdf,etoolbox,hyperref,indentfirst,doi,enumitem,mleftright,xcolor,booktabs,footnote,relsize}

\definecolor{quantumviolet}{rgb}{0.32, 0.14, 0.49}
\hypersetup{
colorlinks = true,
citecolor= quantumviolet,
urlcolor= quantumviolet,
linkcolor = quantumviolet
}

\usepackage[
backend=bibtex,
style=alphabetic,
giveninits=true, 
maxbibnames=99,
minalphanames=3,
date=year
]{biblatex}
\renewbibmacro{in:}{}

\AtEveryBibitem{\clearfield{issue}} 

\newbibmacro{string+doiurl}[1]{%
	\iffieldundef{doi}{%
		\iffieldundef{url}{
			#1%
		}{%
			\href{\thefield{url}}{#1}%
		}%
	}{%
		\href{http://doi.org/\thefield{doi}}{#1}%
	}%
}
\DeclareFieldFormat{title}{\usebibmacro{string+doiurl}{\mkbibemph{#1}}}
\DeclareFieldFormat[article,thesis,incollection,inproceedings,manual]{title}%
{\usebibmacro{string+doiurl}
	{#1} 
}
\usepackage{algorithm}
\usepackage[noend]{algpseudocode}
\floatname{algorithm}{Protocol} 
\algrenewcommand\alglinenumber[1]{\normalsize #1.} 

\newcounter{algsubstate}

\newenvironment{algsubstates}
  {\setcounter{algsubstate}{0}%
   \renewcommand{\State}{%
     \refstepcounter{algsubstate}%
     \Statex {\normalsize\arabic{ALG@line}.\arabic{algsubstate}.}\kern5pt}
     }
  {}

\makeatletter
\newenvironment{breakablealgorithm}
  {
    ~\\[\parskip]
     \refstepcounter{algorithm}
     \hrule height.8pt depth0pt \kern0pt
     \renewcommand{\caption}[2][\relax]{
       {\raggedright\textbf{\fname@algorithm~\thealgorithm} ##2\par}%
       \ifx\relax##1\relax 
         \addcontentsline{loa}{algorithm}{\protect\numberline{\thealgorithm}##2}%
       \else 
         \addcontentsline{loa}{algorithm}{\protect\numberline{\thealgorithm}##1}%
       \fi
       \kern5pt\hrule\kern2pt
     }
  }{
     \kern2pt\hrule\relax
    ~\\[\parskip]
  }
\makeatother

\newcommand{\ceil}[1]{\left\lceil #1 \right\rceil}
\newcommand{\floor}[1]{\left\lfloor #1 \right\rfloor}
\newcommand{\ket}[1]{\left| #1 \right>}
\newcommand{\bra}[1]{\left< #1 \right|}
\newcommand{\ketbra}[2]{\ket{#1} \! \bra{#2}}
\newcommand{\pure}[1]{\ketbra{#1}{#1}}
\newcommand{\tr}[2][]{\operatorname{Tr}_{#1}\!\left[#2\right]} 
\newcommand{\binh}{h_2} 

\newcommand{\acos}{\cos^{-1}}

\newcommand{\Bt}{B'}
\newcommand{\Btstr}{{\str{B}'}}
\newcommand{\Ct}{C'}
\newcommand{\Ctstr}{{\str{C}'}}
\newcommand{\ctstr}{{\str{c}'}}
\newcommand{\calG}{\mathcal{G}}

\newcommand{\calZ}{\mathcal{Z}}
\newcommand{\cdfBin}[3]{B_{#1,#2}(#3)} 
\newcommand{\cmax}{\mathrm{EC}_\mathrm{max}}
\newcommand{\constr}{\nu} 
\newcommand{\cont}{\eps_\mathrm{con}}
\newcommand{\cperp}{\beta}
\newcommand{\defvar}{\coloneqq} 
\newcommand{\dop}[1]{\operatorname{S}_{#1}} 
\newcommand{\dtol}{\delta_\mathrm{tol}}
\newcommand{\dsou}{\delta_\mathrm{IID}}
\newcommand{\eps}{\varepsilon}
\newcommand{\expvaltr}[1]{\tr{#1 \rho_{AB}}} 
\newcommand{\Fj}{} 
\newcommand{\fmax}{f_\mathrm{max}}
\newcommand{\fmin}{f_\mathrm{min}}
\newcommand{\freq}{\operatorname{freq}}
\newcommand{\fZS}[3]{C_{#1,#2}(#3)}
\newcommand{\g}{g} 
\newcommand{\hash}{\operatorname{\mathtt{hash}}}
\newcommand{\Hmin}{H_\mathrm{min}}
\newcommand{\Hmax}{H_\mathrm{max}}
\newcommand{\id}{\mathbb{I}} 
\newcommand{\idmap}{\mathcal{I}} 
\newcommand{\idk}{\mathbb{U}} 
\newcommand{\inX}{\{0,1\}}
\newcommand{\inYg}{\{0,1\}}
\newcommand{\inYt}{\{2,3\}}
\newcommand{\keyw}{\tau} 
\newcommand{\len}{\operatorname{len}}
\newcommand{\lin}{r} 
\newcommand{\lkey}{\ell_\mathrm{key}}
\newcommand{\map}{\mathcal{M}}
\newcommand{\mPA}{\mathcal{M}_\mathrm{PA}}
\newcommand{\Max}{\operatorname{Max}}
\newcommand{\Min}{\operatorname{Min}}
\newcommand{\no}[1]{#1^c} 
\newcommand{\norm}[1]{\left\lVert#1\right\rVert} 
\newcommand{\optdom}{\mathcal{D}}
\newcommand{\Og}{\Omega_\mathrm{g}} 
\newcommand{\Oh}{\Omega_\mathrm{h}} 
\newcommand{\OPE}{\Omega_\mathrm{PE}} 
\newcommand{\OpPE}{\Omega'_\mathrm{PE}} 
\newcommand{\OpPEnot}{\Omega'^c_\mathrm{PE}} 
\newcommand{\p}{p} 
\newcommand{\pBell}{\mu} 
\newcommand{\pct}{\%} 
\newcommand{\pd}{P} 
\newcommand{\perr}{p_\mathrm{err}}
\newcommand{\pr}[2][]{\Pr_{#1}\!\left[#2\right]}
\newcommand{\pvm}{P} 
\newcommand{\pvmB}{Q} 
\newcommand{\q}{q} 
\newcommand{\rightsemicirc}{\put(1.5,2.5){\oval(4,4)[r]}\phantom{\circ}}
\newcommand{\rot}[1]{R_{#1}} 
\newcommand{\semicset}{S_{\rightsemicirc}}
\newcommand{\str}[1]{\mathbf{#1}} 
\newcommand{\smf}[1]{\vartheta_{#1}} 
\newcommand{\suchthat}{\text{ s.t.}} 
\newcommand{\thA}{{\theta_A}}
\newcommand{\thB}{{\theta_B}}
\newcommand{\upto}[1]{\left[#1\right]} 
\newcommand{\Var}{\operatorname{Var}}

\newcommand{\wexp}{w_\mathrm{exp}}
\newcommand{\wtrue}{w_\mathrm{true}}
\newcommand{\Yt}{Y'}
\newcommand{\Ytstr}{{\str{Y}'}}

\newcommand{\ecom}{\eps^\mathrm{com}}
\newcommand{\esound}{\eps^\mathrm{sou}}
\newcommand{\ecorr}{\eps^\mathrm{cor}_\mathrm{QKD}}
\newcommand{\esecr}{\eps^\mathrm{sec}_\mathrm{QKD}}
\newcommand{\eh}{\eps_\mathrm{h}}
\newcommand{\ecEC}{\eps^\mathrm{com}_\mathrm{EC}}
\newcommand{\ecPE}{\eps^\mathrm{com}_\mathrm{PE}}
\newcommand{\eEA}{\eps_\mathrm{EA}}
\newcommand{\ePA}{\eps_\mathrm{PA}}
\newcommand{\es}{\eps_s}
\newcommand{\ez}{\tilde{\eps}_s} 

\newtheorem{theorem}{Theorem}

\newtheorem{proposition}{Proposition}
\theoremstyle{definition} 
\newtheorem{definition}{Definition}

\begin{document}

\title{\textbf{\textcolor{quantumviolet}{Improved DIQKD protocols with finite-size analysis}}}

\renewcommand\Affilfont{\itshape\small} 

\author[1]{Ernest Y.-Z.\ Tan}
\author[2,3]{Pavel Sekatski}
\author[4]{Jean-Daniel Bancal}
\author[5]{Ren\'{e} Schwonnek}
\author[1]{Renato Renner}
\author[4]{Nicolas Sangouard}
\author[6,7]{Charles C.-W.\ Lim}
\affil[1]{Institute for Theoretical Physics, ETH Z\"{u}rich, Switzerland}
\affil[2]{Department of Physics, University of Basel, Klingelbergstrasse 82, 4056 Basel, Switzerland}
\affil[3]{Department of Applied Physics, University of Geneva, Chemin de Pinchat 22, 1211 Geneva, Switzerland}
\affil[4]{Universit\'e Paris-Saclay, CEA, CNRS, Institut de physique th\'eorique, 91191, Gif-sur-Yvette, France}
\affil[5]{Naturwissenschaftlich-Technische Fakult\"{a}t, Universit\"{a}t Siegen, Germany}
\affil[6]{Department of Electrical \& Computer  Engineering, National University of Singapore, Singapore}
\affil[7]{Centre for Quantum Technologies, National University of Singapore, Singapore}

\date{}

\maketitle

\begin{abstract}
The security of finite-length keys is essential for the implementation of device-independent quantum key distribution (DIQKD). Presently, there are several finite-size DIQKD security proofs, but they are mostly focused on standard DIQKD protocols and do not directly apply to the recent improved DIQKD protocols based on noisy preprocessing, random key measurements, and modified CHSH inequalities. Here, we provide a general finite-size security proof that can simultaneously encompass these approaches, using tighter finite-size bounds than previous analyses. In doing so, we develop a method to compute tight lower bounds on the asymptotic keyrate for any such DIQKD protocol with binary inputs and outputs. With this, we show that positive asymptotic keyrates are achievable up to depolarizing noise values of $9.33\pct$, exceeding all previously known noise thresholds. We also develop a modification to random-key-measurement protocols, using a pre-shared seed followed by a ``seed recovery'' step, which yields substantially higher net key generation rates by essentially removing the sifting factor. Some of our results may also improve the keyrates of device-independent randomness expansion.
\end{abstract}

\section{Introduction}

Device-independent quantum key distribution (DIQKD) is a cryptographic concept based on the observation that if some quantum devices violate a Bell inequality, then it is possible to distill a secret key from the devices' outputs, even when the devices are not fully characterized~\cite{BHK05,PAB+09,arx_Sca13}. This is a stronger form of security than that offered by standard QKD protocols, which assume that the devices are performing measurements within specified tolerances~\cite{SBC+09}. 
In recent years, experimental and theoretical developments have brought the possibility of a physical DIQKD demonstration closer to fruition. In particular, a series of experiments have achieved Bell inequality violations while closing the fair-sampling loophole, using NV-centre~\cite{HBD+15}, photonic~\cite{GMR+13,CMA+13,SMC+15,GVW+15,LZL+18,SLT+18}, and cold-atom~\cite{RBG+17} implementations.
On the theoretical front, several protocol modifications have been explored that improve the asymptotic keyrates and noise tolerance of DIQKD. 
We focus on those studied in~\cite{HST+20} (\emph{noisy preprocessing}, in which a small amount of trusted noise is added to the device outputs),~\cite{SGP+21} (\emph{random key measurements}, in which more than one measurement basis is used to generate the key, preventing an adversary from optimally attacking both bases simultaneously), and~\cite{WAP21,SBV+21} (\emph{modified CHSH inequalities}, which certify more entropy than the standard CHSH inequality).

Furthermore, earlier device-independent security proofs~\cite{PAB+09} required the assumption that the device behaviour across multiple rounds is independent and identically distributed (IID), sometimes referred to as the assumption of \emph{collective attacks}. With the development of a result known as the \emph{entropy accumulation theorem} (EAT)~\cite{DFR20,ARV19,DF19}, this assumption can now\footnote{To be precise, other proof techniques~\cite{NPS14,VV14,JMS20} are available that do not require the IID assumption, but the asymptotic keyrates given by those techniques are lower than that of the EAT.} be removed for some DIQKD protocols --- security proofs based on the EAT are valid against general attacks (sometimes referred to as \emph{coherent attacks}), as long as the device behaviour can be modelled in a sequential manner. The EAT also provides explicit bounds on the finite-size behaviour, hence concretely addressing the question of what sample size is required for a secure demonstration of DIQKD. (Another recent result that can be applied to obtain such bounds against general attacks is the \emph{quantum probability estimation} technique~\cite{ZKB18}, but in this work we focus on the EAT.)

However, thus far there has been no comprehensive analysis which simultaneously encompasses the various approaches mentioned above. In this work, we present a finite-size security proof that can be applied to protocols which incorporate all of these proposed improvements. In addition, we show that such protocols can achieve substantially higher depolarizing noise tolerance than all previous proven results. We now highlight the main contributions of our work.

\subsection{Summary of key results}

We consider a protocol based on one-way error correction (Protocol~\ref{prot:DIQKD}) that combines noisy preprocessing and random key measurements, and we perform a finite-size analysis against general attacks (Theorem~\ref{th:DIQKD}) as well as collective attacks (Theorem~\ref{th:collective}). Our approach is similar to that used in~\cite{ARV19}; however, we slightly modify the analysis in order to relax the theoretical requirements for the error-correction step, and our bounds are tighter since we use an updated version~\cite{DF19,LLR+21} of the EAT.
Roughly speaking, the proofs rely on lower bounds on the asymptotic keyrates, and hence in Sec.~\ref{sec:1rndbnd} we also describe an algorithm to compute such bounds, which can be applied independently of our finite-size analysis. This algorithm improves over previous results in~\cite{HST+20,SGP+21,WAP21,SBV+21} by having all of the following properties simultaneously: it applies to arbitrary 2-input 2-output protocols, it accounts for noisy preprocessing and random key measurements, and it provably converges to a tight bound\footnote{In fact, similar to~\cite{WAP21}, we find numerical evidence that the bound in~\cite{SGP+21} is not entirely tight, and also that a conjecture proposed there regarding Eve's optimal attack may not be true after all; see Sec.~\ref{sec:algorithmsummary}. } (for protocols of this form).

For simplicity, the protocol we present only uses the CHSH inequality rather than the modified CHSH inequalities of~\cite{WAP21,SBV+21}, because our results suggest (see Sec.~\ref{sec:bestbnd}) that within the 2-input 2-output scenario, there may not be much prospect for improvement by using the latter, at least for the depolarizing-noise model. However, the finite-size analysis we perform can (like~\cite{BRC20}) be applied to protocols based on arbitrary Bell inequalities --- in Sec.~\ref{sec:nonCHSH}, we briefly explain the relevant adjustments in that case. Similarly, while we only performed explicit computations of the asymptotic keyrates for CHSH-based protocols, our approach as described in Sec.~\ref{sec:1rndbnd} can be applied to all 2-input 2-output Bell inequalities.

With our formulas for the finite-size keyrates, we computed the keyrates for several scenarios. In particular, we studied the keyrates that could be achieved if the honest devices had performance described by the estimated parameters in~\cite{MvDR+19} for the NV-centre~\cite{HBD+15} and cold-atom~\cite{RBG+17} loophole-free Bell tests (detailed keyrate plots are shown in Fig.~\ref{fig:experiments} of Sec.~\ref{sec:finiteresults}). For photonic experiments, it is currently somewhat unclear whether the protocol here provides improvements over the results in~\cite{HST+20,WAP21,SBV+21} --- there are some complications which we explain in Sec.~\ref{sec:1rndresults}. We also plot some finite-size keyrates for honest devices subject to a simple depolarizing-noise model (see Fig.~\ref{fig:depol} of Sec.~\ref{sec:finiteresults}).

From those computations, we found that our security proof would require the \cite{HBD+15} and \cite{RBG+17} experiments to run for approximately $n\sim10^8$ and $10^{10}$ rounds respectively in order to certify a positive finite-size keyrate against general attacks. While this is a marked improvement over the basic \cite{PAB+09}~protocol (which yields zero asymptotic keyrate for those experiments), these requirements still appear to be quite far outside the reach of those implementations --- for reference, the \cite{HBD+15} experiment had sample size $n=245$ (over a 220-hour period), while the \cite{RBG+17} experiment collected a data set of size $n=10,000$ (over 2 measurement runs in a 10-day period) and also a larger data set of size $n=55,568$ (over multiple measurement runs in a 7-month period).
Our keyrate plots (Fig.~\ref{fig:experiments}) also showed that changing the various security parameters (discussed in detail in Sec.~\ref{sec:secdef}) by several orders of magnitude only results in fairly small changes to the keyrate, so it appears unlikely that the keyrates could be substantially improved by relaxing these security requirements.

To improve on these results, in Sec.~\ref{sec:mods} we describe two modifications of Protocol~\ref{prot:DIQKD}. Firstly, Protocol~\ref{prot:preshared} is a modification based on a pre-shared key, which achieves a \emph{net} key generation rate approximately double that of Protocol~\ref{prot:DIQKD}, by overcoming a crucial disadvantage of random-key-measurement protocols (namely, the sifting factor). Secondly, Protocol~\ref{prot:optcoll} is a modification that is optimized for the collective-attacks assumption --- by changing the protocol itself for this scenario rather than just the security proof, we were able to further improve the keyrate. 
However, we find that even with these modifications and relaxed security requirements, it appears that the required number of rounds for a positive finite-size keyrate is still impractically large, at about $n\sim10^6$~to~$10^7$ for the \cite{RBG+17} parameters (see Fig.~\ref{fig:protIID} in Sec.~\ref{sec:mods}). 

As another immediate consequence of our work, we computed a lower bound on the asymptotic noise tolerance of our protocols against depolarizing noise, by using our algorithm for evaluating the asymptotic keyrates. Our results (shown in Fig.~\ref{fig:keyrates} of Sec.~\ref{sec:1rndresults}) certify that positive asymptotic keyrates are possible for depolarizing-noise values of up to $9.33\pct$ at least. In comparison, the previous best thresholds were $8.34\pct$~\cite{WAP21} (using noisy preprocessing and modified CHSH inequalities) or $8.2\pct$~\cite{SGP+21} (using random key measurements). Our improvement over these results is hence of similar magnitude to their improvements over the basic~\cite{PAB+09} protocol, which has a threshold of $7.15\pct$. 
Furthermore, our bounds are in fact close to the highest possible bounds allowed by simple convexity arguments --- we prove in Sec.~\ref{sec:bestbnd} that all protocols of this general form cannot achieve a depolarizing-noise threshold beyond $9.57\pct$, so the threshold we obtained is not far from the absolute highest possible value in this setting.

Finally, we remark that several results from this work can also be used for DI randomness expansion (DIRE)~\cite{arx_colbeckthesis,PAM+10,BRC20,LLR+21}. Specifically, the algorithm we use for computing the asymptotic DIQKD keyrates can also be used to obtain tighter bounds on the asymptotic keyrates for DIRE, after which a finite-size analysis could be performed using the EAT~\cite{BRC20,LLR+21}. Moreover, the ``key recovery'' process in the modified protocol (Protocol~\ref{prot:preshared}) can be applied in the context of DIRE, hence allowing one to improve the keyrates by using the random-key-measurement approach of~\cite{SGP+21}. (This idea was also independently proposed in a separate work~\cite{arx_BRC21} at a similar time.) We discuss this in detail in Sec.~\ref{sec:preshared}.

\subsection{Paper structure}

In Sec.~\ref{sec:prelim}, we state the definitions and notation we use in this work. In Sec.~\ref{sec:prot}, we describe the main protocol we consider, and state the main theorem which bounds the finite-size keyrates (Theorem~\ref{th:DIQKD}), followed by presenting some plots of the resulting values. We give the security proof for this finite-size bound in Sec.~\ref{sec:finite}. In Sec.~\ref{sec:1rndbnd} we present the algorithm to compute the asymptotic keyrates (i.e.~the leading-order terms in Theorem~\ref{th:DIQKD}), and describe the resulting depolarizing-noise thresholds as well as an upper bound on these thresholds. Finally, in Sec.~\ref{sec:mods} we discuss several variations, such as a modified random-key-measurement protocol which bypasses the sifting factor, and security proofs against collective attacks instead of general attacks.

\section{Preliminaries}
\label{sec:prelim}

We define some basic notation in Table~\ref{tab:notation}, and state some further definitions below.
We take all systems to be finite-dimensional, but we will not impose any bounds on the system dimensions unless otherwise specified.
\begin{table}[h!]
\caption{List of notation}
\def\arraystretch{1.5} 
\setlength\tabcolsep{.28cm}
\centering
\begin{tabular}{c l}
\toprule
\textit{Symbol} & \textit{Definition} \\
\toprule
$\log$ & Base-$2$ logarithm \\
\hline
$H(\cdot)$ & Base-$2$ von Neumann entropy \\
\hline
$D(\cdot \Vert \cdot)$ & Base-$2$ quantum relative entropy \\
\hline
$\norm{\cdot}_p$ & Schatten $p$-norm \\
\hline
$\floor{\cdot}$ (resp.~$\ceil{\cdot}$) & Floor (resp.~ceiling) function \\
\hline
$X\geq Y$ (resp.~$X>Y$) & $X-Y$ is positive semidefinite (resp.~positive definite)\\
\hline
$\dop{=}(A)$ (resp.~$\dop{\leq}(A)$) & Set of normalized (resp.~subnormalized) states on register $A$ \\
\hline
$\idk_A$ & Maximally mixed state on register $A$
\\
\hline
$\upto{n}$ & Indices from $1$ to $n$, i.e.~$\{1,2,\dots,n\}$ \\
\hline
$A_{\upto{n}}$ & Registers $A_1 \dots A_n$ \\
\hline
$\no{\Omega}$ & Complement (i.e.~negation) of an event $\Omega$ \\
\hline
$\freq$ & See Definition~\ref{def:freq} \\
\hline
$\cdfBin{n}{p}{k}$ & See Definition~\ref{def:binom} \\
\hline
$\rho_{\land\Omega}$ and $\rho_{|\Omega}$ & See Definition~\ref{def:cond} \\
\toprule
\end{tabular}
\def\arraystretch{1}
\label{tab:notation}
\end{table}

\newpage

\begin{definition}\label{def:freq}
(Frequency distributions) For a string $\str{z}\in\mathcal{Z}^n$ on some alphabet $\mathcal{Z}$, $\freq_{\str{z}}$ denotes the following probability distribution on $\mathcal{Z}$:
\begin{align}
\freq_{\str{z}}(z) \defvar \frac{1}{n} \sum_{j=1}^{n} \delta_{z,z_j}.
\end{align}
\end{definition}

\begin{definition}\label{def:binom}
(Binomial distribution) Let $X\sim\operatorname{Binom}(n,p)$ denote a random variable $X$ following a binomial distribution with parameters $(n,p)$, i.e.~$X$ is the sum of $n$ IID Bernoulli random variables $X_j$ with $\pr{X_j=1}=p$.
We denote the corresponding cumulative distribution function as
\begin{align}
\cdfBin{n}{p}{k} \defvar \pr[X\sim\operatorname{Binom}(n,p)]{X \leq k}.
\end{align}
\end{definition}

\begin{definition}\label{def:cond}
(Conditioning on classical events) For a classical-quantum state $\rho \in \dop{\leq}(CQ)$ in the form
$\rho_{CQ} = \sum_c \pure{c} \otimes \omega_c$ 
for some $\omega_c \in \dop{\leq}(Q)$,
and an event $\Omega$ defined on the register $C$, we define the following ``conditional states'':
\begin{align}
\rho_{\land\Omega} \defvar \sum_{c\in\Omega} \pure{c} \otimes \omega_c, \qquad\qquad \rho_{|\Omega} \defvar \frac{\tr{\rho}}{\tr{\rho_{\land\Omega}}} \rho_{\land\Omega} = \frac{
\sum_{c} \tr{\omega_c}
}{\sum_{c\in\Omega} \tr{\omega_c}} \rho_{\land\Omega} .
\end{align}
We informally refer to these states as the subnormalized and normalized conditional states respectively (the latter is perhaps a slight misnomer if $\tr{\rho}<1$, but this situation does not arise in our proofs). The process of taking subnormalized conditional states is commutative and ``associative'', in the sense that for any events $\Omega,\Omega'$ we have $(\rho_{\land\Omega})_{\land\Omega'} = (\rho_{\land\Omega'})_{\land\Omega} = \rho_{\land(\Omega\land\Omega')}$; hence for brevity we will denote all of these expressions as
\begin{align}
\rho_{\land\Omega\land\Omega'} \defvar (\rho_{\land\Omega})_{\land\Omega'} = (\rho_{\land\Omega'})_{\land\Omega} = \rho_{\land(\Omega\land\Omega')}.
\end{align}
On the other hand, some disambiguating parentheses are needed when combined with taking normalized conditional states.
\end{definition}

\begin{definition}
(2-universal hashing) A \emph{2-universal family of hash functions} is a set $\mathcal{H}$ of functions from a set $\mathcal{X}$ to a set $\mathcal{Y}$, such that if $h$ is drawn uniformly at random from $\mathcal{H}$, then
\begin{align}
\pr{h(x) = h(x')} \leq \frac{1}{|\mathcal{Y}|} \qquad \forall x\neq x'.
\end{align}
\end{definition}

To ensure consistency of definitions, we now state the definitions of smoothed entropies relevant for this work, though we will not need to use the explicit expressions. We follow the presentation in~\cite{DFR20,DF19}, which can be shown to be equivalent to the definitions in~\cite{Tom16,TL17}.

\begin{definition}
For $\rho,\sigma \in \dop{\leq}(A)$, the \emph{generalized fidelity} is
\begin{align}
F(\rho,\sigma) \defvar \norm{
\sqrt{\rho}\sqrt{\sigma}
}_1 + \sqrt{(1-\tr{\rho})(1-\tr{\sigma})},
\end{align}
and the \emph{purified distance} is $\pd(\rho,\sigma)\defvar\sqrt{1-F(\rho,\sigma)^2}$.
\end{definition}

\begin{definition}
For $\rho\in\dop{\leq}(AB)$, the \emph{min- and max-entropies of $A$ conditioned on $B$} are
\begin{align}
\Hmin(A|B)_\rho &\defvar 
-\log 
\min_{\substack{\sigma \in \dop{\leq}(B) \suchthat\\ \ker(\rho_B)\subseteq\ker(\sigma_B)}} 
\norm{\rho_{AB}^\frac{1}{2}
(\id_A \otimes \sigma_{B})
^{-\frac{1}{2}}}_\infty^2,
\\ 
\Hmax(A|B)_\rho &\defvar \log 
\max_{\sigma \in \dop{\leq}(B)} 
\norm{\rho_{AB}^\frac{1}{2}
(\id_A \otimes \sigma_{B})
^\frac{1}{2}}_1^2, 
\end{align}
where in the first equation the 
$(\id_A \otimes \sigma_{B})
^{-\frac{1}{2}}$ term
should be understood in terms of the Moore-Penrose generalized inverse.
(Note that the optimum is indeed attained in both equations~\cite{Tom16}, and it can be attained by a normalized state, so $\dop{\leq}(B)$ can be replaced by $\dop{=}(B)$ without loss of generality.)
\end{definition}

\begin{definition}
For $\rho\in\dop{\leq}(AB)$ and $\eps\in\left[0,\sqrt{\tr{\rho_{AB}}}\right)$, the \emph{$\eps$-smoothed min- and max-entropies of $A$ conditioned on $B$} are
\begin{align}
\Hmin^\eps(A|B)_\rho \defvar
\max_
{\substack{\tilde{\rho} \in \dop{\leq}(AB) \suchthat\\ \pd(\tilde{\rho},\rho)\leq\eps}}
\Hmin(A|B)_{\tilde{\rho}}, 
\qquad
\Hmax^\eps(A|B)_\rho \defvar
\min_
{\substack{\tilde{\rho} \in \dop{\leq}(AB) \suchthat\\ \pd(\tilde{\rho},\rho)\leq\eps}} 
\Hmax(A|B)_{\tilde{\rho}}.
\end{align}
\end{definition}

\subsection{Security definitions}
\label{sec:secdef}

The question of formalizing an appropriate security definition for the device-independent setting has not been definitively resolved yet~\cite{ARV19}, due to considerations regarding the device-reuse attacks described in~\cite{BCK13}. However, we shall proceed by following the security definitions used in~\cite{ARV19}, which were based on strong\footnote{In the sense that the definitions imply \emph{composable security}, which roughly speaking means that we can safely use this protocol in place of a simpler, more idealized functionality that generates an ideal secret key whenever it does not abort --- see~\cite{PR14} for details.} security definitions~\cite{PR14} for standard QKD, and should be sufficient under suitable constraints on the nature of the device memories (we shall briefly discuss these in the next section, when listing the assumptions). 

Qualitatively, the concepts involved in the security definition we use here are: \emph{completeness}, meaning that the honest devices will accept with high probability, and \emph{soundness}, meaning that the devices remain ``secure'' (possibly by aborting) even in the presence of dishonest behaviour. Note that the completeness concept relies on having some description of the honest devices, which should be understood to mean the device behaviour in the situation where they perform ``according to specifications'', without manipulation or eavesdropping attempts from Eve (we give examples of such descriptions in Sec.~\ref{sec:hon}).
The following definition formalizes these concepts:

\begin{definition} \label{def:secure}
Consider a DIQKD protocol such that at the end, the honest parties either \emph{accept} (producing keys $K_A$ and $K_B$ of length $\lkey$ for Alice and Bob respectively) or \emph{abort} (producing an abort symbol $\perp$ for all parties).
It is said to be 
$\ecom$-complete and $\esound$-sound if the following properties hold:
\begin{itemize}
\item (Completeness) The honest protocol implementation aborts with probability at most $\ecom$.
\item (Soundness) For any implementation of the protocol, we have 
\begin{align}\label{eq:sound}
\pr{\mathrm{accept}} \frac{1}{2} \norm{\sigma_{K_A K_B E'} - \left(\frac{1}{2^{\lkey}} \sum_k \pure{kk}_{K_A K_B} \right) \otimes \sigma_{E'}}_1 \leq \esound,
\end{align}
where $\sigma$ denotes the normalized state conditioned on the protocol accepting, and $E'$ denotes all side-information registers available to the adversary at the end of the protocol.
\end{itemize}
\end{definition}

In the security proof, it is convenient to use the fact that the soundness property is implied by a pair of slightly simpler conditions, as shown in~\cite{PR14}. Specifically, to prove a DIQKD protocol is $\esound$-sound, it suffices to find $\ecorr,\esecr$ such that $\esound\geq\ecorr+\esecr$ and the protocol is both $\ecorr$-correct and $\esecr$-secret, defined as follows:

\begin{definition}
A DIQKD protocol as described above is said to be 
$\ecorr$-correct and $\esecr$-secret if
the following properties hold:
\begin{itemize}
\item (Correctness) For any implementation of the protocol, we have\footnote{In~\cite{ARV19}, this definition is stated slightly differently since the accept condition is implicitly absorbed into the $K_A,K_B$ condition by defining $K_A=K_B=\perp$ when the protocol aborts.}
\begin{align}
\pr{K_A \neq K_B \land \mathrm{accept}} \leq \ecorr.
\end{align}
\item (Secrecy) For any implementation of the protocol, we have 
\begin{align}
\pr{\mathrm{accept}} \frac{1}{2} \norm{\sigma_{K_A E'} - \idk_{K_A} \otimes \sigma_{E'}}_1 \leq \esecr,
\label{eq:secret}
\end{align}
where $\sigma$ is as described in Definition~\ref{def:secure}, and $\idk_{K_A}$ denotes the maximally mixed state (i.e.~a uniformly random key for Alice).
\end{itemize}
\end{definition}

\section{Main protocol}
\label{sec:prot}

The overall structure of our main protocol is stated as Protocol~\ref{prot:DIQKD} below, with the details of some steps to be specified in the following subsections. 
We make the following fairly standard assumptions, following the presentation in~\cite{ARV19}:
\begin{itemize}
\item Alice and Bob can prevent unwanted information from leaking outside of their respective locations (for instance, the inputs and outputs to the devices remain private for each party until/unless they are revealed in the public communication steps).
\item Alice and Bob can generate trusted (local) randomness.
\item Alice and Bob have trusted post-processing units to perform classical computations.
\item Alice and Bob perform all classical communication using an authenticated public channel.
\item All systems can be modelled as finite-dimensional quantum registers (though we do not impose any bounds on the dimensions unless otherwise specified).
\end{itemize}
Regarding the first point in particular, we remark that while the loophole-free Bell tests in~\cite{HBD+15,SMC+15,GVW+15,RBG+17} used spacelike separation to motivate the assumption that the devices do not reveal their inputs to each other, this may not be strictly necessary for a DIQKD implementation. It could be possible, and perhaps more reasonable, to instead justify this assumption by implementing some ``shielding'' measures on the devices, to prevent them from leaking unwanted information. In any case, ``shielding'' measures of this nature are likely necessary to prevent the raw \emph{outputs} of the devices from leaking to the adversary, so it may be expedient to use these measures to prevent the inputs from leaking as well.

The above assumptions will be sufficient for us to show that the protocol satisfies the completeness and soundness definitions stated above. However, when considering whether these formal definitions yield security in an intuitive (or composable) sense, some additional assumptions are needed to address the memory attack of~\cite{BCK13}. One approach would be to impose the condition that the devices do not access any registers storing ``private'' data from previous protocols, and that this condition continues to hold if the devices are reused in the future. (Note that this condition is always implicit in standard QKD, because the register being measured is inherently specified when describing the trusted measurements the devices perform.) It remains to fully formalize this condition in a suitable framework for composable security, but this would be beyond the scope of this work.
\begin{savenotes} 
\begin{breakablealgorithm}
\caption{} 
\label{prot:DIQKD}
The protocol is defined in terms of the following parameters (chosen before the protocol begins), which we qualitatively describe:
{\setlist[description]{leftmargin=2cm,labelindent=2cm,itemsep=0mm}
\begin{description}
\item $n$: Total number of rounds
\item $\gamma$: Probability of a test round
\item $\p$: Noisy-preprocessing bias
\item $\cmax$: Bound on number of bits used for error correction
\item $\wexp$: Expected winning probability for the (IID) honest devices
\item $\dtol$: Tolerated deviation from expected winning probability $\wexp$
\item $\lkey$: Length of final key
\end{description}}
\noindent The honest behaviour consists of $n$ IID copies of a device characterized by $\wexp$ and an error-correction parameter $h_\mathrm{hon}$ (details in Sec.~\ref{sec:hon}).\\
\begin{algorithmic}[1]
\State\label{step:1rnd}\textbf{Measurement:} For each $j \in \upto{n}$, perform the following steps: 
\begin{algsubstates}
\State Alice and Bob's devices each receive some share of a quantum state.
\State Alice chooses a uniform input $X_j \in \inX$. With probability $\gamma$, Bob chooses a uniform input $Y_j \in \inYt$, otherwise Bob chooses a uniform input $Y_j \in \inYg$.
\State Alice and Bob supply their inputs to their devices, and record the outputs as $A_j$ and $B_j$ respectively. 
\end{algsubstates}
\State Alice and Bob publicly announce their input strings $\str{X}$ and $\str{Y}$. 
\State \textbf{Sifting:} For all rounds such that $Y_j \in \inYg$ and $X_j \neq Y_j$, Alice and Bob overwrite their outputs with $A_j = B_j = 0$.
\State\label{step:NPP}\textbf{Noisy preprocessing:} For all rounds such that $Y_j \in \inYg$ and $X_j = Y_j$, Alice generates a biased random bit $F_j$ with $\pr{F_j=1} = \p$, and overwrites her output $A_j$ with $A_j \oplus F_j$.
\State\label{step:EC}\textbf{Error correction:} Alice and Bob publicly communicate some bits $\str{L}=(\str{L}_\mathrm{EC},\str{L}_\mathrm{h})$ 
for error correction as follows (see Sec.~\ref{sec:EC}): 
\begin{algsubstates}
\State\label{step:guess}Alice and Bob communicate some bits $\str{L}_\mathrm{EC}$ to allow Bob to produce a guess $\tilde{\str{A}}$ for $\str{A}$. If at some point the number of communicated bits reaches $\cmax$, Alice and Bob immediately cease communication and proceed to the next step, with Bob producing $\tilde{\str{A}}$ using only the information he has at that point.
\State\label{step:hash}Alice computes a 2-universal hash $\str{L}_\mathrm{h} = \hash(\str{A})$ of length $\ceil{\log(1/\eh)}$. She sends $\str{L}_\mathrm{h}$ (and the choice of hash function) to Bob.
\end{algsubstates} 
\State\label{step:PA}\textbf{Parameter estimation:} 
For all $j\in\upto{n}$, Bob sets 
$C_j=\perp$ if $Y_j \in \inYg$; otherwise he sets $C_j=0$ if $\tilde{A}_j \oplus B_j \neq X_j \cdot (Y_j-2)$ and $C_j=1$ if $\tilde{A}_j \oplus B_j = X_j \cdot (Y_j-2)$.
\State\label{step:abort}Bob checks whether $\str{L}_\mathrm{h}=\hash(\tilde{\str{A}})$, as well as whether the value $\str{c}$ on registers $\str{C}$ satisfies $\freq_\str{c}(1)\geq (\wexp-\dtol)\gamma$ and $\freq_\str{c}(0)\leq (1-\wexp+\dtol)\gamma$. If all those conditions hold, Alice and Bob proceed to the next step. Otherwise, the protocol aborts.
\State \textbf{Privacy amplification:} Alice and Bob apply privacy amplification (see Sec.~\ref{sec:PA}) on $\str{A}$ and $\tilde{\str{A}}$ respectively to obtain final keys $K_A$ and $K_B$ of length $\lkey$.
\end{algorithmic}
\end{breakablealgorithm}
\end{savenotes}
The rounds in which $Y_j \in \inYt$ will be referred to as test rounds, and the rounds in which $Y_j \in \inYg$ will be referred to as generation rounds (though strictly speaking, the final key in this protocol is obtained from all the rounds, not merely the generation rounds alone).
In each round,
Eve is allowed to hold some extension of the state distributed to the devices. We will use $E$ to denote the collection of all such quantum side-information she retains over the entire protocol. (We do not denote this using separate registers $E_j$ for individual rounds, because in a general scenario, Eve's side-information may not necessarily ``factorize'' into a tensor product across the individual rounds.) 

We briefly highlight some aspects of this protocol that may differ slightly as compared to more commonly used QKD protocols. Firstly, we do not choose a random subset of fixed size as test rounds, but rather, each round is independently chosen to be a test or generation round, following~\cite{ARV19}. This was in order to apply the entropy accumulation theorem, which holds for processes that can be described using a sequence of maps. Furthermore, the parameter-estimation check is performed on \emph{both} $\freq_\str{c}(1)$ and $\freq_\str{c}(0)$. This was required in order to derive a critical inequality in the security proof (following~\cite{BRC20}), though in some cases it is possible to omit the $\freq_\str{c}(1)$ check (see Eq.~\eqref{eq:fminPEineq} and the subsequent discussion).

We now describe some of the individual steps in more detail.

\subsection{Error correction}\label{sec:EC}

We first discuss Step~\ref{step:hash}, because it will have an impact on our discussion of Step~\ref{step:guess}. Given any $\eh \in (0,1]$, 
if we consider a 2-universal family of hash functions where the output is a bitstring of length $\ceil{\log(1/\eh)}$, then the defining property of 2-universal hashing guarantees that
\begin{align}
\pr{\hash(\str{A})=\hash(\tilde{\str{A}})\middle|\str{A}\neq\tilde{\str{A}}} \leq \eh. \label{eq:hash}
\end{align}
In other words, the probability of getting matching hashes from different strings can be made arbitrarily small, by using sufficiently long hashes. Informally speaking, this gives us some laxity in Step~\ref{step:guess}, because regardless of how much the devices deviate from the honest behaviour, the guarantee~\eqref{eq:hash} will still hold, providing a final ``check'' on how bad the guess $\tilde{\str{A}}$ could be. 
Importantly, our later proof of the \emph{soundness} of the protocol will not rely on any guarantees regarding the procedure in Step~\ref{step:guess} --- only the \emph{completeness} of the protocol (i.e.~the probability that the honest devices mistakenly abort) requires such guarantees. 

We now study Step~\ref{step:guess}. $\cmax$ is defined as follows: it is the length of $\str{L}_\mathrm{EC}$ required such that given the honest devices, Bob can use $\str{L}_\mathrm{EC}$ and $\str{B}$ to produce a guess $\tilde{\str{A}}$ 
satisfying
\begin{align}
\pr{\str{A} \neq \tilde{\str{A}}}_\mathrm{hon} \leq \ecEC.
\label{eq:comEC}
\end{align}
(In this section, we will use the subscript $_\mathrm{hon}$ to emphasize quantities computed with respect to an honest behaviour.)
We stress that while some preliminary characterization of the devices can be performed beforehand to choose a suitable $\cmax$, this parameter \emph{must not} be changed once the protocol begins. 

This immediately raises the question of what value should be chosen for $\cmax$ in order to achieve a desired $\ecEC$. Theoretically, there exists a protocol~\cite{RR12}
with one-way communication that achieves~\eqref{eq:comEC} as long as we choose $\cmax$ 
such that
\begin{align}
\cmax \geq 
\Hmax^{\ez}(\str{A}|\str{B}\str{X}\str{Y})_\mathrm{hon} + 2\log\frac{1}{\ecEC-\ez} + 4,
\label{eq:optEC}
\end{align}
where $\ez\in[0,\ecEC)$ is a parameter that can be optimized over.
(This bound is essentially tight for one-way protocols.)
Since the honest behaviour is IID, 
the max-entropy can be bounded by using the asymptotic equipartition property (AEP) in the form stated as Corollary~4.10 of~\cite{DFR20}, which yields\footnote{In this analysis, we deviated slightly from~\cite{ARV19} by using the error-correction protocol from~\cite{RR12} instead of~\cite{RW05}, and the AEP stated in~\cite{DFR20} rather than~\cite{TCR09}. Both of these yield slight improvements in the bounds (the former at large $n$, and the latter when
$\dim(A_j)$ is not too large).
}
\begin{align}
\Hmax^{\ez}(\str{A}|\str{B}\str{X}\str{Y})_\mathrm{hon} \leq n h_\mathrm{hon} + 
\sqrt{n} \, (2\log5)\sqrt{\log\frac{2}{\ez^2}},
\label{eq:HmaxAEP}
\end{align}
where (using the decomposition $H(Q|Q'C)=\sum_c \pr{c} H(Q|Q';C=c)$ for classical $C$)
\begin{align}
h_\mathrm{hon} \defvar H(A_j|B_jX_jY_j)_\mathrm{hon} 
&= \frac{1-\gamma}{4} 
\sum_{z\in\inX} H(A_j | B_j;X_j = Y_j = z)_\mathrm{hon}  
\nonumber \\
& \qquad 
+ \frac{\gamma}{4} 
\sum_{x\in\inX,y\in\inYt} 
H(A_j| B_j ;X_j = x, Y_j = y)_\mathrm{hon},
\label{eq:ECrate}
\end{align}
where the terms in the summation over $z$ should be understood to refer to the $A_jB_j$ values \emph{after} the noisy-preprocessing step. (Any value of $j$ can be used in the above equation since the honest behaviour is IID.)

However, the protocol achieving the bound~\eqref{eq:optEC} may not be easy to implement. In practice, error-correction protocols typically achieve performance described by
\begin{align}
\cmax \approx \xi(n,\ecEC) n h_\mathrm{hon}, 
\end{align}
where $\xi(n,\ecEC)$ lies between $1.05$ and $1.2$ for ``typical'' values of $n$ and $\ecEC$. (More precise characterizations can be found in~\cite{TMP+17}, which gives for instance an estimate
\begin{align}
\cmax \approx \xi_1 n h_\mathrm{hon} + \tilde{\xi}(\ecEC,h_\mathrm{hon})\sqrt{n},
\end{align}
for a constant $\xi_1$ and a specific function $\tilde{\xi}$.) Furthermore, some protocols used in practice do not have a theoretical bound on $\ecEC$ (for a given $\cmax$), only heuristic estimates. 

Fortunately, as mentioned earlier, the choice of error-correction procedure in Step~\ref{step:guess} will have no effect in our proof of the soundness of Protocol~\ref{prot:DIQKD} (as long as $\cmax$ is a fixed parameter), only its completeness. This means that as long as we are willing to accept heuristic values for $\ecom$, we can use the heuristic values of $\ecEC$ provided by using some ``practical'' error-correction procedure in Step~\ref{step:guess}, and the value of $\esound$ (i.e.~how ``secure'' the protocol is) will be completely unaffected. The critical point to remember is that $\cmax$ is a value to be fixed before the protocol begins, and Alice and Bob \emph{must} stop Step~\ref{step:guess} once they have reached that number of communicated bits. With this in mind, we remark that while we mainly focus on protocols using one-way error correction, this is not quite a strict requirement --- in principle, one could use a procedure involving two-way communication (such as Cascade), as long as $\cmax$ includes all the communicated bits, not just those sent from Alice to Bob. Another possibility worth considering might be adaptive procedures that adjust to the noise level encountered during execution of the protocol, rather than the expected noise level (again, making sure to halt once $\cmax$ bits are communicated, where $\cmax$ is defined beforehand based on the expected behaviour). 

We remark that in our situation, the registers $\str{A}\str{B}\str{X}\str{Y}$ 
have some substructure in the sense that they can be naturally divided into the substrings where $X_j\neq Y_j\in\inYg$, $X_j=Y_j\in\inYg$, and $Y_j\in\inYt$, so the error-correction procedure should ideally take advantage of this substructure (for instance, no error-correction data needs to be sent regarding the rounds where $X_j\neq Y_j\in\inYg$). Also, if we assume that $\pr{A_j = B_j |X_j = x, Y_j = y}_\mathrm{hon}$ is the same for all $x\in\inX,y\in\inYt$ (in which case it must equal $\wexp$), then for the $y\in\inYt$ terms in Eq.~\eqref{eq:ECrate} we have 
\begin{align}
H(A_j| B_j ;X_j = x, Y_j = y)_\mathrm{hon} = \binh\left(\pr{A_j \neq B_j |X_j = x, Y_j = y}_\mathrm{hon}\right) = \binh(\wexp),
\end{align}
which lies in approximately $[0.600,0.811]$ for $\wexp\in[3/4,{(2+\sqrt{2})}/{4}]$.
If the protocol parameters are such that $\xi(n,\ecEC)\binh(\wexp)$ turns out to be fairly close to $1$,
there is not much loss incurred by simply sending the outputs of the test rounds directly rather than expending the effort to compute appropriate error-correction data.

\subsection{Privacy amplification}\label{sec:PA}

Privacy amplification is essentially centred around the \emph{Leftover Hashing Lemma}, which we state below in the form described in~\cite{TL17} (obtained via a small modification of the proof in~\cite{rennerthesis}):
\begin{proposition}
(Leftover Hashing Lemma) Consider any $\sigma \in \dop{\leq}(CQ)$ where $C$ is a classical $n$-bit register. Let 
$\mathcal{H}$ be 
a 2-universal family of hash functions from $\mathbb{Z}_2^n$ to $\mathbb{Z}_2^\ell$, 
and let $H$ be a register of dimension $|\mathcal{H}|$.
Define the state
\begin{align}
\omega_{KCQH} \defvar \mathcal{E}\!\left(\sigma_{CQ} \otimes 
\idk_H
\right),
\end{align}
where the map $\mathcal{E}$ represents the (classical) process of applying the hash function specified in the register $H$ to the register $C$, and recording the output in register $K$.
Then for any $\eps \in \left[0,\sqrt{\tr{\sigma_{CQ}}}\right)$, we have
\begin{align}
\frac{1}{2}\norm{\omega_{KQH} - \idk_K \otimes\omega_{QH}}_1 \leq 2^{-\frac{1}{2}(\Hmin^\eps(C|Q)_\sigma - \ell + 2)} + 2\eps.
\label{eq:LHL}
\end{align}
\end{proposition}

Practically speaking, the privacy amplification step simply consists of Alice choosing a random function from the 2-universal family and publicly communicating it to Bob, followed by Alice and Bob applying that function to $\str{A}$ and $\tilde{\str{A}}$ respectively. The Leftover Hashing Lemma then ensures that the output of this process is close to an ideal key, as long as the conditional min-entropy of the original state was sufficiently large. (Notice that the register $H$ is included in the ``side-information'' term in Eq.~\eqref{eq:LHL}, so it can be publicly communicated.)

\subsection{Honest behaviour}
\label{sec:hon}

For this protocol, the honest implementation consists of $n$ IID copies of a device characterized by 2 parameters, $\wexp$ and $h_\mathrm{hon}$. The first parameter $\wexp$ is the probability with which the device wins the CHSH game when supplied with uniformly random inputs $X_j \in \inX, Y_j \in \inYt$, while the second parameter $h_\mathrm{hon}$ is defined in Eq.~\eqref{eq:ECrate}. While $h_\mathrm{hon}$ does not explicitly appear in the protocol description, it is implicitly used to define the parameter $\cmax$, as was described in Sec.~\ref{sec:EC}. (Since $h_\mathrm{hon}$ has a dependence on $\gamma$, strictly speaking it may be more precise to instead view the honest device behaviour as being parametrized by a tuple specifying all the individual entropies in Eq.~\eqref{eq:ECrate}, but for brevity we shall summarize this as the honest behaviour being parametrized by $h_\mathrm{hon}$. In the more specific models of honest devices described below, these entropies are expressed in terms of some simpler parameters.) 

When computing the keyrates shown in Figs.~\ref{fig:experiments} and~\ref{fig:protIID} later for honest devices corresponding to the Bell tests in~\cite{HBD+15} (resp.~\cite{RBG+17}), we used the following model of the honest devices: following the estimates given in~\cite{MvDR+19}, we characterize them via the parameters $\wexp = 0.797$ (resp.~$0.777$) and $\perr = 0.06$ (resp.~$0.035$), where $\perr$ is a parameter such that the probabilities \emph{before} noisy preprocessing satisfy\footnote{It can be shown that this is equivalent to taking $\pr{A_j B_j |X_j = Y_j = z}_\mathrm{hon}$ to be independent of $z$ and taking the marginal distributions $\pr{A_j |X_j = Y_j = z}_\mathrm{hon}$, $\pr{B_j |X_j = Y_j = z}_\mathrm{hon}$ to be uniform, then defining $\perr \defvar \pr{A_j \neq B_j |X_j = Y_j = z}_\mathrm{hon}$.}
\begin{align}
\pr{A_j B_j |X_j = Y_j = z}_\mathrm{hon} = 
(1-2\perr)\frac{\delta_{A_j,B_j}}{2} + 2\perr\, \frac{1}{4}
& \quad \text{ for all } z\in\inYg,
\end{align}
where $\delta_{j,k}$ is the Kronecker delta. Furthermore, we take $\pr{A_j = B_j |X_j = x, Y_j = y}_\mathrm{hon}$ to be the same for all $x\in\inX,y\in\inYt$ (in which case it must equal $\wexp$). Under this model, the expression~\eqref{eq:ECrate} for $h_\mathrm{hon}$ can be simplified:
\begin{align}
h_\mathrm{hon} 
& = \frac{1-\gamma}{2}\binh(\p+(1-2\p)\perr) + \gamma \binh(\wexp),
\end{align}
where the $\p+(1-2\p)\perr$ term is obtained by an explicit computation~\cite{WAP21}.

When computing the keyrates shown in Figs.~\ref{fig:depol}--\ref{fig:entbnds} later for a depolarizing-noise scenario, we modelled the honest devices as being described by a parameter $\q \in [0,1/2]$, such that the devices hold the two-qubit Werner state $(1-2\q)\pure{\Phi^+} + 2\q\, \id/4$ (where $\ket{\Phi^+}=(\ket{00}+\ket{11})/\sqrt{2}$). The measurements corresponding to Alice and Bob's inputs $X_j\in\inX,Y_j\in\inYt$ are the ideal CHSH measurements (see e.g.~\cite{PAB+09}), and the measurements corresponding to Bob's inputs $Y_j\in\inYg$ are measurements in the same bases as Alice's measurements. In terms of $\wexp$ and $h_\mathrm{hon}$, this yields
\begin{align}
\wexp&=(1-2\q)\frac{2+\sqrt{2}}{4} + 2\q\,\frac{1}{2}, \\
h_\mathrm{hon} & = \frac{1-\gamma}{2}\binh(\p+(1-2\p)\q) + \gamma \binh(\wexp),
\end{align}
where the second expression follows by the same computation as above, noting that essentially we have $\perr=\q$ in this case.

Finally, for modelling photonic experiments in Fig.~\ref{fig:keyrateseta} later, we follow~\cite{Ebe93} and use a highly simplified model with a single parameter $\eta\in[0,1]$, which is intended to represent the overall detection efficiency (grouping together various effects such as fibre-optic losses and photodetector efficiency into this single parameter). In this model, we take the honest devices to be able to implement arbitrary two-qubit states and measurements perfectly well, but then with probability $1-\eta$ the outcome is replaced with a no-detection symbol $\phi$. 
In order to apply our security proof, which requires the test-round measurements to have binary outcomes, we impose that for inputs $X_j \in \inX$ by Alice and inputs $Y_j \in \inYt$ by Bob, the no-detection outcome $\phi$ is deterministically mapped to the output value $0$ (this is a common approach for Bell tests and/or DIQKD using this model~\cite{Ebe93,PAB+09}). However, for inputs $Y_j \in \inYg$ by Bob, we preserve the no-detection outcome, as it slightly improves the keyrates by reducing the error-correction term $h_\mathrm{hon}$~\cite{ML12}.
Unlike the depolarizing-noise model, in this case we do not stick to a fixed choice of states and measurements for all $\eta$, but rather we (heuristically) optimize the states and measurements to maximize the keyrate for each $\eta$ --- as is typical in this model, this makes a significant difference in the threshold $\eta$ value required to achieve e.g.~Bell violation~\cite{Ebe93} or nonzero keyrates~\cite{BFF21}.

\subsection{Finite-size keyrates}
\label{sec:finiteresults}

We now present our main theorem, giving the length of the final key as a function of the number of rounds and the desired security parameters. To do so, we need to introduce some notation and ancillary functions. 
First, we require an function $\lin_\p$ satisfying some properties we now describe. Consider 
any tripartite quantum state $\rho_{\bar{A}\bar{B}\bar{E}}$, 
and suppose there are two possible binary-outcome measurements (indexed by $x$) on register $\bar{A}$, and similarly two binary-outcome measurements (indexed by $y$) on register $\bar{B}$. Let $w$ be the probability of winning the CHSH game (with uniform inputs) for these measurements on the state $\rho_{\bar{A}\bar{B}}$.
Let $\hat{A}_x$ be a register that stores the result if measurement $x$ is performed on register $\bar{A}$ and noisy preprocessing (Step~\ref{step:NPP}) with bias $\p$ is then applied to the outcome.
We require $\lin_\p$ to be an affine function such that for any choice of $\rho_{\bar{A}\bar{B}\bar{E}}$ and measurements, we have
\begin{align}
\sum_{x\in\{0,1\}} \frac{1}{2} H(
\hat{A}_x
|\bar{E}) \geq \lin_\p(w).
\label{eq:linbnd}
\end{align}
(The factor of $1/2$ arises from 
the input distributions in the protocol.)
We give details on how to obtain such a bound in Sec.~\ref{sec:1rndbnd} (in principle, the $\p=0$ case could also be obtained from the results of~\cite{SGP+21}).
Given such a function $\lin_\p$, we then define the affine function
\begin{align}
\g(w) \defvar \frac{1-\gamma}{2} \lin_\p(w) + \gamma \lin_0(w).
\end{align}
Informally, $\g$ can be interpreted as a lower bound on the entropy ``accumulated'' in one round of the protocol.

Finally, for an affine function $f$ defined on all probability distributions on some register $C$, and any subset $\mathcal{S}$ of its domain,
we will define
\begin{align}
\begin{gathered}
\Max(f) \defvar \max_q f(q), 
\qquad 
\Min_{\mathcal{S}}(f) \defvar \inf_{q\in\mathcal{S}} f(q), 
\\
\Var_\mathcal{S}(f) \defvar \sup_{q\in\mathcal{S}} \left(\sum_c q(c) f(\delta_c)^2 - \left(\sum_c q(c) f(\delta_c)\right)^2\right),
\end{gathered}
\end{align}
where $\max_q$ is taken over all distributions on $C$, and $\delta_c$ denotes the distribution with all its weight on the symbol $c$. 

With these definitions, we can state the security guarantees of the protocol. The following theorem involves various parameters in addition to those listed at the start of Protocol~\ref{prot:DIQKD}, which we first qualitatively describe (note that several descriptions are somewhat informal, not meant to be entirely precise on their own): 
{
\setlist[description]{leftmargin=2.9cm,rightmargin=1cm,labelindent=5mm,itemsep=0mm}
\begin{description}
\item \makebox[15mm]{$\ecEC$}: Bound on the probability for the honest implementation that Bob's guess~$\tilde{\str{A}}$ for $\str{A}$ is wrong
\item \makebox[15mm]{$\ecPE$}: Bound on the probability for the honest implementation that Bob's guess~$\tilde{\str{A}}$ for $\str{A}$ is correct but $\freq_\str{c}$ does not satisfy the parameter-estimation checks
\item \makebox[15mm]{$\eEA$}: Informally, a bound on the probability that a ``virtual'' parameter estimation step (in a related virtual protocol) accepts when given devices that produce insufficient min-entropy 
\item \makebox[15mm]{$\ePA$}: Informally, a bound on the secrecy parameter~\eqref{eq:secret} of the keys after privacy amplification (in a related virtual protocol) when given devices that produce sufficient min-entropy, up to some smoothing corrections
\item \makebox[15mm]{$\eh$}: Bound on the probability that $\hash(\tilde{\str{A}})$ matches $\hash(\str{A})$ even when Bob's guess~$\tilde{\str{A}}$ for $\str{A}$ is wrong
\item \makebox[15mm]{$\es,\es',\es''$}: Smoothing parameters for several entropic terms
\item \makebox[15mm]{$\alpha,\alpha'$}: R\'{e}nyi entropy parameters arising in the EAT
\item \makebox[15mm]{$\cperp$}: Free parameter in constructing min-tradeoff function (see Eq.~\eqref{eq:fmin})
\end{description}
}
\noindent These parameters, together with the protocol parameters $\gamma,\p,\dtol$ described earlier, can be considered to be variational parameters that should be chosen to optimize the keyrate as much as possible. The formal theorem statement is as follows:
\begin{theorem}\label{th:DIQKD}
Take any 
$\ecEC,\ecPE,\eEA,\ePA,\eh,\es,\es',\es''\in (0,1]$ 
such that $\es>\es'+2\es''$, and any $\alpha\in(1,2)$, $\alpha'\in(1,1+2/V')$, $\cperp\in[\g(0),\g(1)]$, $\gamma\in(0,1)$, $\p\in[0,1/2]$, where $V'\defvar2\log5$.
Protocol~\ref{prot:DIQKD} is $(\ecEC + \ecPE)$-complete and $(\max\{\eEA, \ePA + 2\es\} + 2\eh)$-sound when performed with any choice of $\cmax$ such that Eq.~\eqref{eq:comEC} holds, and $\dtol,\lkey$ satisfying
\begin{align}
\ecPE &\geq \cdfBin{n}{\gamma\wexp}{\floor{(\wexp-\dtol)\gamma n}} + \cdfBin{n}{
1-\gamma+\gamma\wexp
}{\floor{(1-\gamma+\wexp\gamma-\dtol\gamma)n}},
\label{eq:ecPE}\\
\lkey &\leq n\g(\wexp-\dtol) - n\frac{(\alpha-1)\ln 2}{2}V^2 - n(\alpha-1)^2K_\alpha - n\gamma - n\left(\frac{\alpha'-1}{4}\right)V'^2 
\nonumber \\ &\qquad 
- \frac{\smf{\es'}}{\alpha-1} - \frac{\smf{\es''}}{\alpha'-1} - \left(\frac{\alpha}{\alpha-1}+\frac{\alpha'}{\alpha'-1}-2\right)\log\frac{1}{\eEA} - 3\smf{\es-\es'-2\es''} \nonumber\\&\qquad
- \cmax - \ceil{\log\left(\frac{1}{\eh}\right)} - 2\log\frac{1}{\ePA} + 2,
\label{eq:keylength}
\end{align}
where $\cdfBin{n}{p}{k}$ is the cumulative distribution function of a binomial distribution (Definition~\ref{def:binom}), and
\begin{align}
\begin{aligned}
\smf{\eps} &\defvar \log\frac{1}{1-\sqrt{1-\eps^2}} \leq \log\frac{2}{\eps^2}, 
\\
V &\defvar \sqrt{\Var_{\mathcal{Q}_{f}}(\fmin)+2} + \log
33,
\\
K_\alpha &\defvar \frac{2^{(\alpha-1)(2\log4 + \Max(\fmin)-\Min_{\mathcal{Q}_{f}} (\fmin))} }{6(2-\alpha)^3\ln2}
\ln^3\left(2^{2\log4 + \Max(\fmin)-\Min_{\mathcal{Q}_{f}} (\fmin)} + e^2\right),
\end{aligned}
\label{eq:VK}
\end{align}
with $\fmin$ and $\mathcal{Q}_{f}$ being a function and a set (defined explicitly in Sec.~\ref{sec:fmin}) that satisfy
\begin{align}
\begin{gathered}
\Max(\fmin)
= \frac{1}{\gamma}\g(1) + \left(1-\frac{1}{\gamma}\right)\cperp, 
\qquad 
\Min_{\mathcal{Q}_{f}} (\fmin) = \g\!\left(\frac{2-\sqrt{2}}{4}\right), \\
\Var_{\mathcal{Q}_{f}}(\fmin) \leq 
\frac{2-\sqrt{2}}{4\gamma} 
\min \left\{\Delta_0^2, \Delta_1^2 \right\}
+ \frac{2+\sqrt{2}}{4\gamma} 
\max \left\{\Delta_0^2, \Delta_1^2 \right\}
, \text{ where } \Delta_w \defvar \cperp - \g(w).
\end{gathered}
\label{eq:minmaxvarbounds}
\end{align}
\end{theorem}
\noindent Qualitatively, the $\smf{\eps}$ terms in the above theorem arise from various properties of smoothed entropies, while the $V,V'$ terms are measures of ``variance'' of some functions considered in the the EAT, and the $K_\alpha$ term is an additional correction that essentially depends on the range of these functions.

\begin{figure}
\centering
\subfloat[\cite{HBD+15} parameters]{
\includegraphics[width=0.48\textwidth]{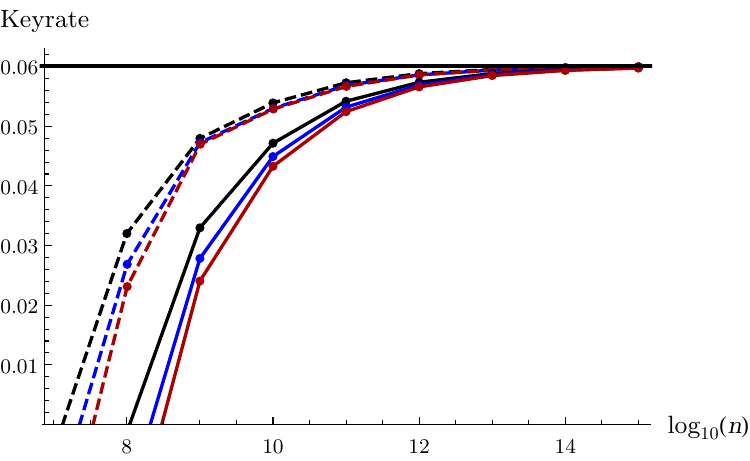}
} 
\subfloat[\cite{RBG+17} parameters]{
\includegraphics[width=0.48\textwidth]{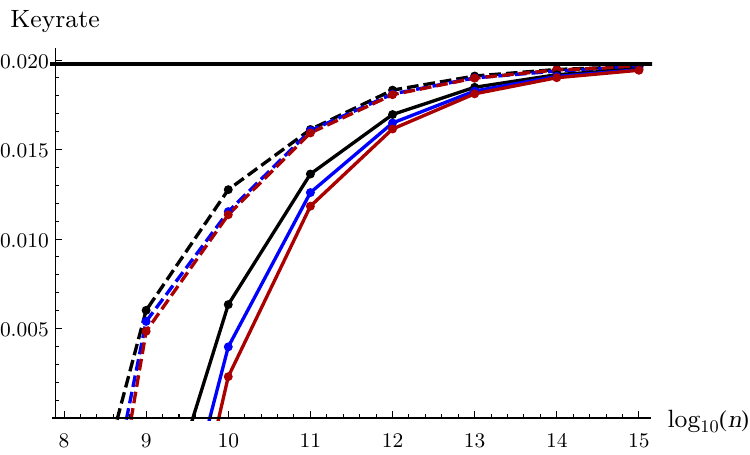}
}
\caption{Finite-size keyrates as a function of number of rounds in Protocol~\ref{prot:DIQKD}
without noisy preprocessing ($\p=0$), 
for honest devices following the estimated parameters in~\cite{MvDR+19} for the Bell tests in~\cite{HBD+15} and~\cite{RBG+17}.
The solid curves show the results for general attacks (Theorem~\ref{th:DIQKD}), while the dashed curves show the results under the assumption of collective attacks (Theorem~\ref{th:collective}), with the error-correction protocol taken to satisfy Eqs.~\eqref{eq:optEC}--\eqref{eq:ECrate}. The colours correspond to soundness parameters (informally, a measure of how ``insecure'' the key is) of $\esound=10^{-3}$, $10^{-6}$, and $10^{-9}$ for black, blue, and red respectively, while the completeness parameter (the probability that the honest devices abort) is $\ecom=10^{-2}$ in all cases. The horizontal line denotes the asymptotic keyrate. 
All other parameters in Theorems~\ref{th:DIQKD} and \ref{th:collective} were numerically optimized, except $\cperp$. 
}
\label{fig:experiments}
\vspace{2cm}
\includegraphics[width=0.6\textwidth]{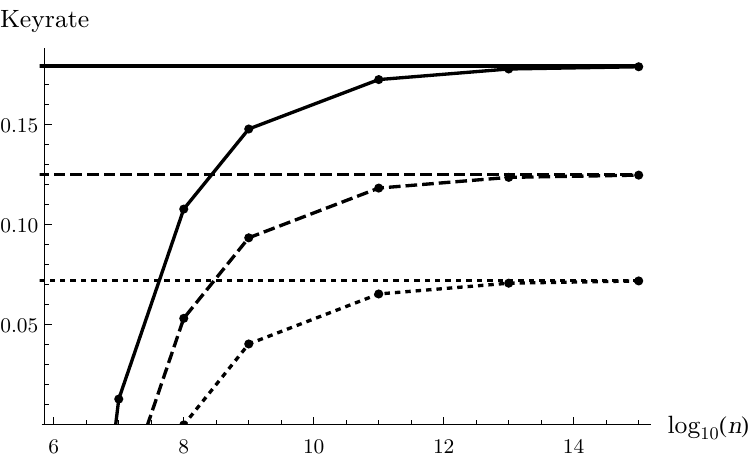}
\caption{Finite-size keyrates secure against general attacks (Theorem~\ref{th:DIQKD}) as a function of number of rounds in Protocol~\ref{prot:DIQKD}
without noisy preprocessing ($\p=0$), where the honest devices are described by depolarizing noise $\q$ (see Sec.~\ref{sec:hon}). The solid, dashed and dotted curves denote $\q=5\pct$, $6\pct$ and $7\pct$ respectively, with the error-correction protocol taken to satisfy Eqs.~\eqref{eq:optEC}--\eqref{eq:ECrate}. The soundness parameter is $\esound=10^{-6}$ and the completeness parameter is $\ecom=10^{-2}$. The horizontal lines denote the asymptotic keyrates. All other parameters were numerically optimized.
}
\label{fig:depol}
\end{figure}

To compute the finite-size keyrates presented in this work, for instance in Figs.~\ref{fig:experiments} and~\ref{fig:depol} later, we optimized the parameter choices by using the inbuilt (heuristic) constrained-optimization functions in Mathematica or MATLAB to maximize $\lkey$ while imposing the constraint~\eqref{eq:ecPE} on the parameters $\ecPE,\gamma,\dtol$.\footnote{
To enforce that for instance the parameters $\ecEC,\ecPE$ satisfy the condition $\ecEC + \ecPE = \ecom$ for the desired completeness value $\ecom$, rather than imposing this condition as a constraint in the optimization, we instead introduced a reparametrization $\ecEC=\ecom\sin^2\theta$, $\ecPE=\ecom\cos^2\theta$ and optimized over $\theta \in (0,\pi/2)$. This ensures that $\ecEC + \ecPE = \ecom$ automatically holds without imposing it as an explicit constraint. A similar approach (with more parameters) was taken for conditions such as $\max\{\eEA, \ePA + 2\es\} + 2\eh = \esound$ and so on.}
We highlight that the exact expression for $\smf{\eps}$ is numerically unstable, and hence we replaced it with the upper bound of $\log(2/\eps^2)$ (this bound is basically tight at small $\eps$, so it makes little difference). Furthermore, the optimization over $\cperp$ also appears to be somewhat unstable. We observed heuristically that the optimal value of $\cperp$ appears to typically be very close to $\g(1)$, and hence for simplicity in some cases we did not optimize over it but instead simply fixed $\cperp=\g(1)$ (or slightly below it, to avoid some instabilities at $\cperp-\g(1)=0$). Finally, we found that direct computation of $\cdfBin{n}{p}{k}$ (e.g.~via the regularized beta function) could sometimes be slow or unstable, and in such cases we followed~\cite{LLR+21} and replaced it with the upper bound in the following theorem:
\begin{proposition}\label{prop:ZSbound}\cite{ZS13}
Let
\begin{align}
\fZS{n}{p}{k} \defvar \Phi\left(\operatorname{sign}\left(\frac{k}{n}-p\right)\sqrt{2n \,D_e\!\left(\frac{k}{n} \;\middle\Vert\; p\right)}\right),
\end{align}
where $D_e(q \Vert p)\defvar q\ln\frac{q}{p} + (1-q)\ln\frac{1-q}{1-p}$ is the base-$e$ relative entropy between the distributions $\{q,1-q\}$ and $\{p,1-p\}$, while $\Phi(z)\defvar \int_{-\infty}^z 
(2\pi)^{-1/2}
e^{-u^2/2} \mathrm{d}u 
=\left(1/2\right)\operatorname{erfc}\!\left(-{z}/{\sqrt{2}}\right)
$ is the cumulative distribution function of the standard normal distribution.
Then for any $k\in\upto{n-1}$,
\begin{align}
\fZS{n}{p}{k} \leq \cdfBin{n}{p}{k} \leq \fZS{n}{p}{k+1}. 
\label{eq:ZSbound}
\end{align}
\end{proposition}
\noindent Replacing $\cdfBin{n}{p}{k}$ with $\fZS{n}{p}{k+1}$ and computing the latter (which is a Gaussian integral) appeared to be faster and more stable than computing $\cdfBin{n}{p}{k}$ directly. There is little loss incurred by performing this replacement --- the inequalities~\eqref{eq:ZSbound} imply $\fZS{n}{p}{k+1} \leq \cdfBin{n}{p}{k+1}$, so the effect is no larger than replacing $\cdfBin{n}{p}{k}$ by $\cdfBin{n}{p}{k+1}$, which is basically negligible in the parameter regimes studied in this work.

Note that in general, the optimal parameter values (especially for $\alpha$ and $\alpha'$) would depend heavily on $n$.
To get an estimate for the asymptotic scaling of $\lkey$, we can choose all the $\eps$ parameters to take some constant values satisfying the desired completeness and soundness bounds, set $\cperp=\g(1)$ (this ensures the formula~\eqref{eq:Khat} later is independent of $\gamma$), then choose~\cite{DFR20,DF19}
\begin{align}
\begin{aligned}
\alpha-1&= \frac{1}{\sqrt{n}} \sqrt{\frac{2}{V^2 \ln 2} \left(\smf{\es'} + 2\log\frac{1}{\eEA}\right)} , \\
\alpha'-1&= \frac{1}{\sqrt{n}} \sqrt{\frac{4}{V'^2} \left(\smf{\es''} + 2\log\frac{1}{\eEA} \right)} ,
\end{aligned}
\end{align}
taking $n$ to be large enough that $\alpha \in (1,3/2)$ and $\alpha'\in(1,1+2/V')$. Also, introduce the constant
\begin{align}
\hat{K} \defvar \frac{2^{(2\log4 + \Max(\fmin)-\Min_{\mathcal{Q}_{f}} (\fmin))} }{(3/4)\ln2}
\ln^3\left(2^{2\log4 + \Max(\fmin)-\Min_{\mathcal{Q}_{f}} (\fmin)} + e^2\right) ,
\label{eq:Khat}
\end{align}
which satisfies $\hat{K} \geq K_\alpha$ for $\alpha\in(1,3/2)$. 
With these choices, Eq.~\eqref{eq:keylength} can be satisfied by taking
\begin{align}
\lkey &= \bigg\lfloor
n\g(\wexp-\dtol) - n\frac{(\alpha-1)\ln 2}{2}V^2 - n(\alpha-1)^2 \hat{K} - n\gamma - n\left(\frac{\alpha'-1}{4}\right)V'^2 
\nonumber \\ &\qquad 
- \frac{1}{\alpha-1}\left(\smf{\es'} + 2\log\frac{1}{\eEA} \right) - \frac{1}{\alpha'-1}\left(\smf{\es''} + 2\log\frac{1}{\eEA} \right) + 2\log\frac{1}{\eEA} - 3\smf{\es-\es'-2\es''} \nonumber\\&\qquad
- \cmax - \ceil{\log\left(\frac{1}{\eh}\right)} - 2\log\frac{1}{\ePA} + 2 \bigg\rfloor \nonumber \\
&= \bigg\lfloor n\g(\wexp-\dtol) - \sqrt{n}\,2\sqrt{\frac{V^2 \ln 2}{2} \left(\smf{\es'} + 2\log\frac{1}{\eEA}\right)} - \frac{2}{V^2 \ln 2} \left(\smf{\es'} + 2\log\frac{1}{\eEA}\right)\hat{K} 
- n\gamma \nonumber \\ &\qquad 
- \sqrt{n} \, 2\sqrt{\frac{V'^2}{4} \left(\smf{\es''} + 2\log\frac{1}{\eEA} \right)} + 2\log\frac{1}{\eEA} - 3\smf{\es-\es'-2\es''} \nonumber\\&\qquad 
- \cmax - \ceil{\log\left(\frac{1}{\eh}\right)} - 2\log\frac{1}{\ePA} + 2 \bigg\rfloor \nonumber \\
&= \left\lfloor n\g(\wexp-\dtol) - n\gamma - O(\sqrt{n}) - O(1) - \cmax \right\rfloor.
\label{eq:estkeylength}
\end{align}
Furthermore, Eq.~\eqref{eq:ecPE} can be satisfied by choosing $\gamma=3 n^{-1} \dtol^{-2} \log(2/\ecPE)$ (see Eq.~\eqref{eq:chernoffs}), 
in which case by taking $\dtol \propto n^{-1/3}$ we have $\dtol,\gamma \to 0$ as $n\to\infty$, and Eq.~\eqref{eq:estkeylength} then yields (strictly speaking, here we glossed over the dependency of the $V$ terms on $\gamma$; see~\cite{DF19} for more details): 
\begin{align}
\lim_{n\to\infty} \frac{\lkey}{n} = 
\frac{1}{2}\left(\lin_\p(\wexp) - 
\sum_{z\in\inX} \frac{1}{2}H(A_j | B_j;X_j = Y_j = z)_\mathrm{hon}\right)
,
\label{eq:asympt}
\end{align}
taking $\cmax$ according to Eqs.~\eqref{eq:optEC}--\eqref{eq:ECrate}. This is the expected asymptotic result, e.g.~according to the Devetak-Winter bound~\cite{DW05} (with the prefactor of $1/2$ being due to the sifting).

Given the scaling behaviour shown in Eq.~\eqref{eq:estkeylength}, it can be seen that the optimal values of the various $\eps$ parameters (given some desired values of $\ecom$ and $\esound$) may be of rather different orders of magnitude. This is because some of them appear in $O(1/\sqrt{n})$ corrections to the finite-size keyrate while others appear in $O(1/n)$ corrections. Intuitively speaking, the $\eps$ parameters in the latter can be chosen to be substantially smaller than the former, since the $O(1/n)$ scaling reduces their contribution to the finite-size effects.

With Theorem~\ref{th:DIQKD}, we compute the achievable finite-size keyrates for the various types of honest devices described in Sec.~\ref{sec:hon}, showing the results in Figs.~\ref{fig:experiments} and~\ref{fig:depol}. As mentioned previously, the former shows the results regarding the Bell tests in \cite{HBD+15} and \cite{RBG+17}, while the latter shows the results for a depolarizing-noise model. We remark that for the computations in Fig.~\ref{fig:experiments}, noisy preprocessing was not applied because it appears to only slightly improve the keyrates for those experimental parameters; see Fig.~\ref{fig:protIID} later. For reference, in Fig.~\ref{fig:experiments} we also include plots of the finite-size keyrates against collective attacks as derived in Theorem~\ref{th:collective} later --- comparing them to the keyrates against general attacks given by Theorem~\ref{th:DIQKD}, we see that there is indeed some difference, but the threshold value of $n$ to achieve positive keyrates only differs by about one order of magnitude. Furthermore, the various choices we considered for the soundness parameter all yielded fairly similar keyrate curves, indicating that changing the soundness requirements would not significantly change the $n$ threshold.

\section{Finite-size security proof}
\label{sec:finite}

We now prove that Protocol~\ref{prot:DIQKD} indeed satisfies the security properties claimed in Theorem~\ref{th:DIQKD}.
To do so, we first introduce a virtual protocol that is more convenient to analyze.
For the purposes of understanding this construction, it may be helpful to think of it as being based on a \emph{specific} set of states and measurements that could be occurring in a run of Protocol~\ref{prot:DIQKD} (as opposed to simultaneously considering all possible states and measurements that could be occurring). In particular, this virtual protocol
(and the channels $\map_j$ in Sec.~\ref{sec:EAT} later) should be understood as being constructed in terms of this specific set of states and measurements. Since we will not impose any additional assumptions on these states and measurements beyond those specified by the protocol, this will still yield a valid way for us to prove the desired security properties (in particular the soundness property, which has to be proven for all possible states and measurements that could occur in a run of the protocol).

Consider the state at the end of Step~\ref{step:PA} in Protocol~\ref{prot:DIQKD}. We now describe a virtual protocol\footnote{We stress that this ``protocol'' is not performed in practice 
(and in fact cannot be, since the ``virtual parameter estimation'' step cannot be performed locally by either party). 
However, it produces exactly the same state as Protocol~\ref{prot:DIQKD} on all relevant registers, and can hence be used for the security analysis.} that produces exactly the same state (when it is implemented using the same input state and measurements as those used in a run of Protocol~\ref{prot:DIQKD}), apart from the introduction of two additional registers $\Btstr\Ctstr$.

\let\oldthealgorithm\thealgorithm 
\renewcommand{\thealgorithm}{$1'$}
\begin{savenotes}
\begin{algorithm}[H]
\caption{A virtual protocol}\label{prot:virtual}
\begin{algorithmic}[1]
\State Alice and Bob's devices each receive and store all quantum states that they will subsequently measure.
\State\label{step:EATmap}For each $j \in \upto{n}$, perform the following steps: 
\begin{algsubstates}
\State Alice chooses a uniform input $X_j \in \inX$. With probability $\gamma$, Bob chooses a uniform input $Y_j \in \inYt$, otherwise Bob chooses a uniform input $Y_j \in \inYg$.
\State Alice and Bob supply their inputs to their devices, and record the outputs as $A_j$ and $B_j$ respectively. 
\State Alice and Bob publicly announce their inputs.
\State\textbf{Sifting:} If $Y_j \in \inYg$ and $X_j \neq Y_j$, Alice and Bob overwrite their outputs with $A_j = B_j = 0$.
\State\textbf{Noisy preprocessing:} If $Y_j \in \inYg$ and $X_j = Y_j$, Alice generates a biased random bit $F_j$ with $\pr{F_j=1} = \p$, and overwrites her output $A_j$ with $A_j \oplus F_j$.
\State 
If $Y_j \in \inYg$, Bob sets $\Bt_j = 0$, otherwise Bob sets $\Bt_j = B_j$. 
\State\textbf{Virtual parameter estimation:} 
Set $\Ct_j=\perp$ if $Y_j \in \inYg$; otherwise set $\Ct_j=0$ if $A_j \oplus \Bt_j \neq X_j \cdot (Y_j-2)$ and $\Ct_j=1$ if $A_j \oplus \Bt_j = X_j \cdot (Y_j-2)$.
\end{algsubstates}
\State\textbf{Error correction:} Alice and Bob publicly communicate some bits $\str{L}$ for error correction as previously described, allowing Bob to construct a guess $\tilde{\str{A}}$ for $\str{A}$.
\State\textbf{Parameter estimation:} 
For all $j\in\upto{n}$, Bob sets 
$C_j=\perp$ if $Y_j \in \inYg$; otherwise he sets $C_j=0$ if $\tilde{A}_j \oplus \Bt_j \neq X_j \cdot (Y_j-2)$ and $C_j=1$ if $\tilde{A}_j \oplus \Bt_j = X_j \cdot (Y_j-2)$.
\end{algorithmic}
\end{algorithm}
\end{savenotes}
\let\thealgorithm\oldthealgorithm 
\addtocounter{algorithm}{-1} 
\noindent The key changes as compared to Protocol~\ref{prot:DIQKD} are as follows:
\begin{itemize}
\item All the states that the devices will measure are distributed immediately at the start (note that this is possible because in Protocol~\ref{prot:DIQKD}, the measurement choices $\str{X},\str{Y}$ are not disclosed until all measurements have been performed, and hence the distributed states cannot behave adaptively with respect to the inputs). 
\item The sifting and noisy preprocessing steps are now performed immediately after each measurement, instead of after all measurements are performed. This is to allow us to subsequently apply the EAT.
\item Two additional registers were introduced: $\Btstr$, which is equal to $\str{B}$ on all the test rounds but is otherwise set to $0$, and $\Ctstr$, which is analogous to $\str{C}$ but computed using $\str{A}$ in place of $\tilde{\str{A}}$. These registers were used in a \emph{virtual parameter estimation} step.
\item All parameter estimation is performed with $\Btstr$ instead of $\str{B}$ (this substitution has no physical effect since $\Bt_j = B_j$ 
in all rounds used for parameter estimation).
\end{itemize}

Let $\rho$ denote the state on registers ${\str{A}\tilde{\str{A}}\str{B}\Btstr\str{X}\str{Y}\str{L}\str{C}\Ctstr E}$ (as well as the choice of hash function in the error-correction step) at the end of Protocol~\ref{prot:virtual}. As mentioned above, the reduced state after tracing out $\Btstr \Ctstr$ is exactly the same as that at the end of Step~\ref{step:PA} in Protocol~\ref{prot:DIQKD}. 
Since all subsequent steps in Protocol~\ref{prot:DIQKD} (i.e.~simply the accept/abort check and the privacy amplification) only involve this reduced state, to analyze Protocol~\ref{prot:DIQKD} it suffices to consider the equivalent (apart from $\Btstr\Ctstr$) process where all the steps up to Step~\ref{step:PA} are replaced by this virtual protocol, and then the remaining steps in Protocol~\ref{prot:DIQKD} are performed. 
With this in mind, let us define the following events on the state $\rho$:
{\setlist[description]{leftmargin=2cm,labelindent=2cm}
\begin{description} 
\item $\Og$: $\str{A} = \tilde{\str{A}}$ 
(i.e.~Bob correctly guesses $\str{A}$)
\item $\Oh$: $\hash(\str{A}) = \hash(\tilde{\str{A}})$
\item $\OPE$: $\freq_\str{c}(1)\geq (\wexp-\dtol)\gamma$ and $\freq_\str{c}(0)\leq (1-\wexp+\dtol)\gamma$
\item $\OpPE$: $\freq_{\ctstr}(1)\geq (\wexp-\dtol)\gamma$ and $\freq_{\ctstr}(0)\leq (1-\wexp+\dtol)\gamma$
\end{description} 
}
\noindent Note that in terms of these events, the accept condition of the protocol is $\Oh \land \OPE$. With the virtual protocol and the above events in mind, we now turn to proving completeness and soundness of Protocol~\ref{prot:DIQKD}.

\subsection{Completeness}
\label{sec:com}

Completeness is defined entirely with respect to the honest behaviour of the devices, hence all discussion in this section is with respect to the situation where the state $\rho$ described above is the one produced by the honest states and measurements. 
To prove completeness, we simply need to obtain an upper bound on the probability that this honest behaviour yields an abort, i.e.~$\pr{\no{\Oh} \lor \no{\OPE}}_\mathrm{hon}$ (recall we use $\no{\Omega}$ to denote the complement of an event). 
However, we encounter a slight inconvenience here because the event $\no{\OPE}$ involves the register $\str{C}$ produced using Bob's guess $\tilde{\str{A}}$ rather than Alice's actual string $\str{A}$, and there is some small probability that his guess was wrong. To cope with this, we shall break down $\pr{\no{\Oh} \lor \no{\OPE}}_\mathrm{hon}$ into simpler terms that can be bounded in terms of probabilities involving only the ``virtual'' string $\Ctstr$ rather than $\str{C}$, where the former is easier to handle since it is produced from the actual value of $\str{A}$.

We begin by noting that the hashes of $\str{A}$ and $\tilde{\str{A}}$ can only differ if $\str{A}\neq\tilde{\str{A}}$, which is to say that the event $\no{\Oh}$ implies the event $\no{\Og}$. With this, we write
\begin{align}
\pr{\no{\Oh} \lor \no{\OPE}}_\mathrm{hon} 
&\leq \pr{\no{\Og} \lor \no{\OPE}}_\mathrm{hon} \nonumber\\
&= \pr{\no{\Og}}_\mathrm{hon} + \pr{\Og \land \no{\OPE}}_\mathrm{hon}
\end{align}
where in the second line we have partitioned the event $\no{\Og} \lor \no{\OPE}$ into the disjoint events $\no{\Og}$ and $\Og \land \no{\OPE}$.
We shall now upper bound the probabilities of each of these events. 

The $\pr{\no{\Og}}_\mathrm{hon}$ term is straightforward to handle, since by construction the error-correction step ensures that this probability is at most $\ecEC$.
As for the $\pr{\Og \land \no{\OPE}}_\mathrm{hon}$ term, we now make the critical observation that $\Og \land \no{\OPE} = \Og \land \OpPEnot$ (because the event $\Og$ implies that $\str{C}=\Ctstr$). Therefore, we have
\begin{align}
\pr{\Og \land \no{\OPE}}_\mathrm{hon} = \pr{\Og \land \OpPEnot}_\mathrm{hon} \leq \pr{\OpPEnot}_\mathrm{hon},
\end{align}
which is the desired reduction to a term involving $\Ctstr$ rather than $\str{C}$. To explicitly upper bound $\pr{\OpPEnot}_\mathrm{hon}$, observe that under the honest behaviour, the string $\Ctstr$ consists of $n$ IID rounds such that
$\mathrm{Pr}[{\Ct_j=1}]_\mathrm{hon} = \gamma \wexp$ and $\mathrm{Pr}[{\Ct_j=0}]_\mathrm{hon} = \gamma (1-\wexp)$ in each round. Therefore, we have
\begin{align}
\pr{\freq_{\ctstr}(1) < (\wexp-\dtol)\gamma}_\mathrm{hon}
&\leq 
\pr{\freq_{\ctstr}(1)n \leq \floor{(\wexp-\dtol)\gamma n}}_\mathrm{hon} \nonumber\\
&= \cdfBin{n}{\gamma\wexp}{\floor{(\wexp-\dtol)\gamma n}}, \label{eq:ecPEbound1}\\
\pr{\freq_{\ctstr}(0) > (1-\wexp+\dtol)\gamma}_\mathrm{hon} 
&= \pr{\freq_{\ctstr}(
\neg 0
) \leq 1-(1-\wexp+\dtol)\gamma}_\mathrm{hon} \nonumber\\
&= \pr{\freq_{\ctstr}(\neg 0) n \leq \floor{
(1-\gamma+\wexp\gamma-\dtol\gamma)
n}}_\mathrm{hon} \nonumber\\
&= \cdfBin{n}{
1-\gamma+\gamma\wexp
}{\floor{(1-\gamma+\wexp\gamma-\dtol\gamma)n} },
\label{eq:ecPEbound0}
\end{align}
where $\neg 0$ represents all symbols other than $0$. 
Hence by the union bound, $\pr{\OpPEnot}_\mathrm{hon}$ is upper bounded by $\ecPE$ is as specified in Eq.~\eqref{eq:ecPE} (i.e.~the sum of the expressions~\eqref{eq:ecPEbound1} and~\eqref{eq:ecPEbound0}). This yields a final upper bound of $\ecEC + \ecPE$ on the probability of the honest behaviour aborting, as desired.

In principle, various somewhat simpler expressions could be obtained using the (multiplicative) Chernoff bound (where in the following formulas we briefly use the notation $t \defvar \dtol/\wexp$ and $t' \defvar \dtol/(1-\wexp)$ for brevity, and suppose that $\dtol<\min\{\wexp,1-\wexp\}$ so $t,t'<1$):
\begin{align}
\begin{gathered}
\pr{\freq_{\ctstr}(1) < (\wexp-\dtol)\gamma}_\mathrm{hon}
\leq \left(\frac{e^{-t}}{(1-t)^{1-t}}\right)^{n\gamma\wexp}
\leq e^{-\frac{n \gamma \dtol^2}{2\wexp}}
, 
\\
\pr{\freq_{\ctstr}(0) > (1-\wexp+\dtol)\gamma}_\mathrm{hon} 
\leq \left(\frac{e^{t'}}{(1+{t'})^{1+{t'}}}\right)^{n\gamma(1-\wexp)}
\leq e^{-\frac{n \gamma \dtol^2}{3(1-\wexp)}}
.
\end{gathered}
\label{eq:chernoffs}
\end{align}
However, it was observed in~\cite{LLR+21} that these bounds are weaker than Prop.~\ref{prop:ZSbound} by a significant amount.
As yet another alternative, Hoeffding's inequality yields a bound of $e^{-2n \gamma^2 \dtol^2}$, 
but this is worse than the above Chernoff bounds whenever e.g.~$\gamma \leq 1/6$; furthermore, it does not yield nontrivial bounds if we choose $\gamma\propto1/n$ (for constant $\ecEC,\dtol$), which ought to be possible in principle according to the scaling analysis in~\cite{DF19}.

\subsection{Soundness}
\label{sec:soundness}

Soundness has to be proven against all possible dishonest behaviours (subject to the protocol assumptions as listed in Sec.~\ref{sec:prot}). 
In this section, we shall consider any particular state $\rho$ (as defined after the Protocol~\ref{prot:virtual} description) that would be produced by a particular choice of such dishonest behaviour, and prove that it satisfies the soundness condition~\eqref{eq:sound} (for a specific value of $\esound$) regardless of which dishonest behaviour was considered. 
All probabilities are to be understood as being defined with respect to that state $\rho$.

We first note that it is straightforward to show that Protocol~\ref{prot:DIQKD} is $\eh$-correct --- recalling that the accept condition can be written as $\Oh \land \OPE$, we obtain the desired upper bound on the probability that $K_A \neq K_B$ and the protocol accepts: 
\begin{align}
\pr{K_A \neq K_B \land \Oh \land \OPE} 
\leq \pr{K_A \neq K_B \land \Oh} 
\leq \pr{\no{\Og} \land \Oh} 
\leq \pr{\Oh | \no{\Og}}
\leq \eh,
\label{eq:corrproof}
\end{align}
where the last inequality holds by the defining property of 2-universal hashing.

It remains to prove secrecy.
Denote the privacy amplification step in Protocol~\ref{prot:DIQKD} as the map $\mPA$, so the subnormalized state conditioned on the event of Protocol~\ref{prot:DIQKD} accepting can be written as $\mPA(\rho_{\land \Oh \land \OPE})$. Let $H$ be the register storing the choice of hash function used for privacy amplification, and denote $E' \defvar \str{X}\str{Y}\str{L}EH$ for brevity. We can now rewrite the secrecy condition as the requirement
\begin{align}
\frac{1}{2}\norm{\mPA(\rho_{\land \Oh \land \OPE})_{K_A E'} - \idk_{K_A} \otimes \mPA(\rho_{\land \Oh \land \OPE})_{E'}}_1 \leq \esecr.
\label{eq:esecrnew}
\end{align}
Now, somewhat similarly to the completeness analysis, we shall find a way to upper bound the left-hand-side of the above expression in terms of $\OpPE$ rather than $\OPE$, as the former is easier to handle.
Specifically, by noting that $\rho_{\land \Oh \land \OPE} = \rho_{\land \Og \land \Oh \land \OPE} + \rho_{\land \no{\Og} \land \Oh \land \OPE}$, we find the following bound using norm subadditivity:
\begin{align}
&\frac{1}{2}\norm{\mPA(\rho_{\land \Oh \land \OPE})_{K_A E'} - \idk_{K_A} \otimes \mPA(\rho_{\land \Oh \land \OPE})_{E'}}_1 \nonumber\\
\leq{}& \frac{1}{2}\norm{\mPA(\rho_{\land \Og \land \Oh \land \OPE})_{K_A E'} - \idk_{K_A} \otimes \mPA(\rho_{\land \Og \land \Oh \land \OPE})_{E'}}_1 \nonumber\\ 
&\qquad + \frac{1}{2}\norm{\mPA(\rho_{\land \no{\Og} \land \Oh \land \OPE})_{K_A E'} - \idk_{K_A} \otimes \mPA(\rho_{\land \no{\Og} \land \Oh \land \OPE})_{E'}}_1 \nonumber\\
\leq{}& \frac{1}{2}\norm{\mPA(\rho_{\land \Og \land \Oh \land \OpPE})_{K_A E'} - \idk_{K_A} \otimes \mPA(\rho_{\land \Og \land \Oh \land \OpPE})_{E'}}_1 
+ \eh, \label{eq:secsplit}
\end{align}
where in the last line we have 
performed the substitution $\Og \land \OPE = \Og \land \OpPE$ in the first term (again, because the event $\Og$ implies that $\str{C}=\Ctstr$),
and bounded the second term using
\begin{align}
\pr{\no{\Og} \land \Oh \land \OPE} \leq \pr{\no{\Og} \land \Oh} \leq \eh,
\end{align}
as noted in Eq.~\eqref{eq:corrproof}.

We now aim to bound the first term in Eq.~\eqref{eq:secsplit}. To do so, we study three exhaustive possibilities for the state $\rho$ (the first two are not mutually exclusive, but this does not matter):
{\setlist[enumerate]{leftmargin=2cm,labelindent=2cm}
\begin{enumerate}[label*=Case \arabic*:,ref=\arabic*]
\item \label{case:ECsmall} $\pr{\Og \land \Oh|\OpPE} \leq \es^2$.
\item \label{case:PEsmall} $\pr{\OpPE} \leq \eEA$.
\item \label{case:neither} Neither of the above are true.
\end{enumerate}
}
\noindent In case~\ref{case:ECsmall}, that term is bounded by
\begin{align}
\pr{\Og \land \Oh \land \OpPE} = \pr{\Og \land \Oh |\OpPE} \pr{\OpPE} \leq \es^2.
\label{eq:caseECsmall}
\end{align}
In case~\ref{case:PEsmall}, that term is bounded by
\begin{align}
\pr{\Og \land \Oh \land \OpPE} \leq \pr{\OpPE} \leq \eEA.
\end{align}
The main challenge is case~\ref{case:neither}. To study this case, we first focus on the conditional state $\rho_{|\OpPE}$. Importantly, the relevant smoothed min-entropy of this state can be bounded by using the following theorem, which we prove in Sec.~\ref{sec:EAT} using entropy accumulation:
\begin{theorem}\label{th:rawHmin}
For all parameter values as specified in Theorem~\ref{th:DIQKD}, the state at the end of Protocol~\ref{prot:virtual} satisfies
\begin{align}
\Hmin^{\es}(\str{A}|\str{X}\str{Y}E)_{\rho_{|\OpPE}} &>
n\g(\wexp-\dtol) - n\frac{(\alpha-1)\ln 2}{2}V^2 - n(\alpha-1)^2K_\alpha - n\gamma - n\left(\frac{\alpha'-1}{4}\right)V'^2 
\nonumber \\ &\qquad 
- \frac{\smf{\es'}}{\alpha-1} - \frac{\smf{\es''}}{\alpha'-1} - \left(\frac{\alpha}{\alpha-1}+\frac{\alpha'}{\alpha'-1}\right)\log\frac{1}{\pr{\OpPE}} - 3\smf{\es-\es'-2\es''}.
\label{eq:rawHmin}
\end{align}
where $V',\smf{\eps},V,K_\alpha$ are as defined in Theorem~\ref{th:DIQKD}.
\end{theorem}

To relate this to our state of interest, 
we use~\cite{TL17} Lemma~10, which states that for a state $\sigma\in\dop{\leq}(ZZ'QQ')
$ that is classical on registers $ZZ'$, an event $\Omega$ on the registers $ZZ'$, and any $\eps \in [0,\sqrt{\tr{\sigma_{\land\Omega}}})$,
we have
\begin{align}
\Hmin^\eps(ZQ|Q')_{\sigma_{\land\Omega}} \geq \Hmin^\eps(ZQ|Q')_{\sigma} .
\label{eq:condlemma}
\end{align}
In our context, we observe that the probability of the event $\Og \land \Oh$ on the normalized state $\rho_{|\OpPE}$ is $\pr{\Og \land \Oh|\OpPE}$, which is greater than $\es^2$ since we are in case~\ref{case:neither}. Hence the conditions
of the lemma are satisfied (identifying $\str{A}$ with $Z$, $\tilde{\str{A}}$ (and the error-correction hash choice) with $Z'$, $\str{X}\str{Y}\str{L}E$ with $Q'$, and leaving $Q$ empty), allowing us to obtain the bound
\begin{align}
\Hmin^{\es}(\str{A}|\str{X}\str{Y}\str{L}E)_{\big(\rho_{|\OpPE}\big) {}_{\land \Og \land \Oh} } &\geq \Hmin^{\es}(\str{A}|\str{X}\str{Y}\str{L}E)_{\rho_{|\OpPE}} \nonumber\\
&\geq \Hmin^{\es}(\str{A}|\str{X}\str{Y}E)_{\rho_{|\OpPE}} 
- \len(\str{L}) \nonumber\\
&\geq \Hmin^{\es}(\str{A}|\str{X}\str{Y}E)_{\rho_{|\OpPE}} 
- \cmax - \ceil{\log\left(\frac{1}{\eh}\right)}
,\label{eq:entPA}
\end{align}
where in the second line we have applied a chain rule for smoothed min-entropy (see e.g.~\cite{WTHR11} Lemma~11 or~\cite{Tom16} Lemma~6.8).

Putting together Eq.~\eqref{eq:entPA} and Theorem~\ref{th:rawHmin}, we find that for a key of length $\lkey$ satisfying Eq.~\eqref{eq:keylength}, we have\footnote{Keeping the $\log\left(1/\pr{\OpPE}\right)$ term here yields a slightly tighter result as compared to~\cite{ARV19,MvDR+19}, which instead used $\pr{\OpPE}\leq1$ to write an inequality in place of the equality in the second-last line of Eq.~\eqref{eq:finalPA}.}
\begin{align}
&\frac{1}{2}\left(\Hmin^{\es}(\str{A}|\str{X}\str{Y}\str{L}E)_{\big(\rho_{|\OpPE}\big) {}_{\land \Og \land \Oh} } - \lkey  + 2\right) + \log\frac{1}{\pr{\OpPE}} \nonumber\\
\geq{}&
\frac{1}{2} \left(- \left(\frac{\alpha}{\alpha-1}+\frac{\alpha'}{\alpha'-1}\right)\log\frac{1}{\pr{\OpPE}} +
2\log\frac{1}{\ePA} + 
\left(\frac{\alpha}{\alpha-1}+\frac{\alpha'}{\alpha'-1} - 2\right)\log\frac{1}{\eEA} \right) + \log\frac{1}{\pr{\OpPE}} \nonumber\\
={}& \log\frac{1}{\ePA} +\frac{1}{2} \left(\frac{\alpha}{\alpha-1}+\frac{\alpha'}{\alpha'-1}-2\right) \left(\log\frac{1}{\eEA} - \log\frac{1}{\pr{\OpPE}}\right) \nonumber\\
\geq{}& \log\frac{1}{\ePA}, 
\end{align}
where the last line holds because $\pr{\OpPE} \geq \eEA$ in case~\ref{case:neither} (and also $\frac{\alpha}{\alpha-1}+\frac{\alpha'}{\alpha'-1} -2 \geq 2 
> 0
$ since $\alpha,\alpha' \in (1,2)$).

This finally allows us to bound Eq.~\eqref{eq:secsplit}, since
we have $\es^2 \leq \pr{\Og \land \Oh|\OpPE} = \operatorname{Tr}[(\rho_{|\OpPE})_{\land \Og \land \Oh}]$ in case~\ref{case:neither}, and hence we can apply the Leftover Hashing Lemma to see that\footnote{We remark that the events $\OpPE$ and $\Og \land \Oh$ were treated somewhat differently in this analysis, in that 
while we introduced a parameter $\eEA$ to divide the analysis of $\OpPE$ into separate cases, there is no analogous condition for $\Og \land \Oh$ (the condition for case~\ref{case:ECsmall} is not quite analogous; it is merely a technicality to allow us to apply Eq.~\eqref{eq:condlemma} and the Leftover Hashing Lemma).
This is fundamentally because conditioning on $\OpPE$ via Eq.~\eqref{eq:rawHmin} has a ``larger'' effect on min-entropy compared to conditioning on $\Og \land \Oh$ (via Eq.~\eqref{eq:condlemma} modified for normalized conditional states). When substituting these effects into the final security parameter, one finds that the former worsens the security parameter by an amount that depends on $\log(1/\pr{\OpPE})$, while the latter does not, and hence we required a bound on $\pr{\OpPE}$ in the former.}
\begin{align}
&\frac{1}{2}\norm{\mPA(\rho_{\land \Og \land \Oh \land \OpPE})_{K_A E'} - \idk_{K_A} \otimes \mPA(\rho_{\land \Og \land \Oh \land \OpPE})_{E'}}_1 \nonumber\\
={}& \pr{\OpPE} \frac{1}{2}\norm{\mPA((\rho_{|\OpPE})_{\land \Og \land \Oh})_{K_A E'} - \idk_{K_A} \otimes \mPA((\rho_{|\OpPE})_{\land \Og \land \Oh})_{E'}}_1 \nonumber\\
\leq{}& \pr{\OpPE} \left(2^{-\frac{1}{2}\left(\Hmin^{\es}(\str{A}|\str{X}\str{Y}\str{L}E)_{\big(\raisebox{1.5\depth}{$\mathsmaller{\rho}$}|\OpPE\big) {\land \Og \land \Oh} } - \lkey + 2\right)} + 2\es \right) \nonumber\\
={}& 2^{-\frac{1}{2}\left(\Hmin^{\es}(\str{A}|\str{X}\str{Y}\str{L}E)_{\big(\raisebox{1.5\depth}{$\mathsmaller{\rho}$}|\OpPE\big)\land \Og \land \Oh } - \lkey + 2\right) - \log\left(1/\pr{\OpPE}\right)} + 2\es\pr{\OpPE} 
\nonumber\\
\leq{}& \ePA + 2\es. \label{eq:finalPA}
\end{align}

Since the three possible cases are exhaustive, we conclude that the secrecy condition is satisfied by choosing
\begin{align}
\esecr = \max\{\es^2, \eEA, \ePA + 2\es\} + \eh = \max\{\eEA, \ePA + 2\es\} + \eh.
\end{align}
Recalling that we have already shown the protocol is $\eh$-correct, we finally conclude that it is $(\max\{\eEA, \ePA + 2\es\} + 2\eh)$-sound.

\subsection{Entropy accumulation}\label{sec:EAT}

This section is devoted to the proof of Theorem~\ref{th:rawHmin}. The key theoretical tool in this proof is the entropy accumulation theorem, which we shall now briefly outline in the form stated in~\cite{DF19}. To do so, we shall first introduce \emph{EAT channels} and \emph{tradeoff functions}.

\begin{definition}\label{def:EATchann}
A \emph{sequence of EAT channels} is a sequence 
$\{\map_j\}_{j\in\upto{n}}$ 
where each $\map_j$ is a channel from a register $R_{j-1}$ to registers $D_j S_j T_j R_j$, 
which satisfies the following properties:
\begin{itemize}
\item All $D_j$ are classical registers with a common alphabet $\mathcal{D}$, and all $S_j$ have the same finite dimension.
\item 
For each $\map_j$, the value of $D_j$ is determined from the registers $S_j T_j$ alone. Formally, this means 
$\map_j$ is of the form $\mathcal{P}_j \circ \map'_j$, where $\map'_j$ is a channel from $R_{j-1}$ to $S_j T_j R_j$, and $\mathcal{P}_j$ is a channel from $S_j T_j$ to $D_j S_j T_j$ of the form
\begin{align}
\mathcal{P}_j(\rho_{S_j T_j}) = \sum_{s\in\mathcal{S},t\in\mathcal{T}} (\Pi_{S_j,s} \otimes \Pi_{T_j,t}) \rho_{S_j T_j} (\Pi_{S_j,s} \otimes \Pi_{T_j,t}) \otimes \pure{d(s,t)}_{D_j},
\label{eq:nodisturb}
\end{align}
where $\{\Pi_{S_j,s}\}_{s\in\mathcal{S}}$ and $\{\Pi_{T_j,t}\}_{t\in\mathcal{T}}$ are families of orthogonal projectors on $S_j$ and $T_j$ respectively, and $d: \mathcal{S} \times \mathcal{T} \to \mathcal{D}$ is a deterministic function.
\end{itemize}
\end{definition}

\begin{definition}\label{def:fminfmax}
Let $\fmin$ be a real-valued affine function defined on probability distributions over the alphabet $\mathcal{D}$. It is called a \emph{min-tradeoff function} for a sequence of EAT channels $\{\map_j\}_{j\in\upto{n}}$ if for any distribution $q$ on $\mathcal{D}$, we have 
\begin{align}
\fmin(q) \leq \inf_{\sigma \in \Sigma_j(q)} H(S_j|T_jR)_\sigma \qquad \forall j\in\upto{n},
\end{align}
where $\Sigma_j(q)$ denotes the set of states of the form $(\map_j\otimes\idmap_{R})(\omega_{R_{j-1}R})$ such that the reduced state on $D_j$ has distribution $q$. 
Analogously, a real-valued affine function $\fmax$ defined on probability distributions over $\mathcal{D}$ is called a \emph{max-tradeoff function} if 
\begin{align}
\fmax(q) \geq \sup_{\sigma \in \Sigma_j(q)} H(S_j|T_jR)_\sigma \qquad \forall j\in\upto{n}.
\end{align}
The infimum and supremum of an empty set are defined as $+\infty$ and $-\infty$ respectively.
\end{definition}
\noindent With these definitions, we can now state the theorem:
\begin{proposition}\label{prop:EAT}
(Entropy accumulation theorem~\cite{DF19,LLR+21}) 
Consider a sequence of EAT channels $\{\map_j\}_{j\in\upto{n}}$ and a state of the form $\rho = (\map_n \circ\dots \circ\map_1 \otimes \idmap_E)\left(\rho^0_{R_0E}\right)$ satisfying the Markov conditions
\begin{align}
I(S_{\upto{j-1}} : T_j | T_{\upto{j-1}} E)_\rho 
= 0 \qquad \forall j\in\upto{n}.
\label{eq:markov}
\end{align}

\noindent Let $\fmin$ be a min-tradeoff function for $\{\map_j\}_{j\in\upto{n}}$ and consider any $h\in\mathbb{R}, \eps\in(0,1), \alpha\in(1,2)$. Then for any event $\Omega \subseteq \mathcal{D}^n$ such that 
$\fmin(\freq_{\str{d}}) \geq h$ for all $\str{d}\in\Omega$, we have\footnote{This expression differs slightly from those in~\cite{DF19,LLR+21} because we have not performed the simplifications based on $\smf{\eps} \leq \log(2/\eps^2)$ and $\alpha < 2$ (though the former is only a very small improvement for typical values of $\eps$).
We also remark that in~\cite{LLR+21}, a modification was made to the EAT to improve its dependence on the $\Var$ term, which we have not included here as it involves a further optimization that would significantly increase the complexity of our keyrate computations. 
(In an earlier version of this work, we stated that this modification did not make a difference here as our choice of affine $\fmin$ is essentially equal to the tight bound on $H(S_j|T_jR)$ in our scenario; however, this was in error as this equality does not hold on e.g.~any parts of the domain where $\fmin$ is negative.)
Still, we would expect that such an optimization would not yield better keyrates than the collective-attacks scenario (shown in Fig.~\ref{fig:experiments}) at least, so the potential improvement is limited to some extent.}
\begin{align}
\Hmin^{\eps}(\str{S}|\str{T}E)_{\rho_{|\Omega}}  
>nh - n\frac{(\alpha-1)\ln 2}{2}V^2 - n(\alpha-1)^2K_\alpha 
- \frac{\smf{\eps}}{\alpha-1} - \frac{\alpha}{\alpha-1}\log\frac{1}{\pr{\Omega}}, 
\end{align}
where $\smf{\eps}$ is as defined in Eq.~\eqref{eq:VK}, and
\begin{align}
\begin{aligned}
V &\defvar \sqrt{\Var_{\mathcal{Q}}(\fmin)+2} + \log
(2\dim(S_j)^2+1),
\\
K_\alpha &\defvar \frac{2^{(\alpha-1)(2\log\dim(S_j) + \Max(\fmin)-\Min_{\mathcal{Q}} (\fmin))} }{6(2-\alpha)^3\ln2}
\ln^3\left(2^{2\log\dim(S_j) + \Max(\fmin)-\Min_{\mathcal{Q}} (\fmin)} + e^2\right),
\end{aligned}
\end{align}
with $\mathcal{Q}$ being the set of all distributions on $\mathcal{D}$ that could be produced by applying some EAT channel to some state.
\end{proposition}
\noindent Informally, the Markov conditions
impose the requirement that the register $T_j$ does not ``leak any information'' about the previous registers $S_{\upto{j-1}}$ beyond what is already available from $T_{\upto{j-1}} E$. (Without this Markov condition, one could for instance have channels such that $T_j$ simply contains a copy of $S_{j-1}$, in which case there could be situations where a nontrivial min-tradeoff function holds but the conclusion of the entropy accumulation theorem would be completely false.) There is also an analogous EAT statement regarding the max-entropy~\cite{DFR20}, using a max-tradeoff function, though we will be using it slightly differently and will elaborate further on it at that point.

We now describe how the EAT can be used to prove Theorem~\ref{th:rawHmin}. First, note that to prove Theorem~\ref{th:rawHmin} it would be sufficient to consider only the registers $\str{A}\Btstr\str{X}\str{Y}\Ctstr E$ of the state $\rho$ (the conditioning event $\OpPE$ is determined by $\Ctstr$ alone). The reduced state on these registers is the same as that at the point when Step~\ref{step:EATmap} of Protocol~\ref{prot:virtual} has finished looping over $j$, since the subsequent steps do not change these registers, and thus we can equivalently study that state in place of $\rho_{|\OpPE}$.
From this point onwards, all smoothed min- or max-entropies refer to that state conditioned on $\OpPE$ (and normalized),
hence for brevity we will omit the subscript specifying the state.

Each iteration of Step~\ref{step:EATmap} of Protocol~\ref{prot:virtual} can be treated as a channel $\map_j$ in a sequence of EAT channels, by considering it to be a channel performing the following operations:
\begin{enumerate}
\item Alice generates $X_j$ as specified in Step~\ref{step:EATmap}.
Conditioned on the value of $X_j$, Alice's device performs some measurement on its share of the stored quantum state $R_{j-1}$ (which includes any memory retained from previous rounds), then performs sifting and noisy preprocessing on the outcome, storing the final result in register $A_j$. 
\item Bob's device behaves analogously, producing the registers $Y_j$ and $\Bt_j$ (we will not need to consider $B_j$).
\item The value of $\Ct_j$ is computed from $A_j \Bt_j X_j Y_j$.
\end{enumerate}
We highlight that in the above description of $\map_j$, the only ``unknowns'' are the measurements it performs on the input state on $R_{j-1}$ --- all other operations are taken to be performed in trusted fashion. 
(This is reasonable because these measurements and the stored states are the only untrusted aspects in the true protocol.) If we had simply considered completely arbitrary channels $\map_j$ producing the respective registers, it would not be possible to 
make a nontrivial security statement about the output.

Identifying $\Ct_j$ with $D_j$, $A_j\Bt_j$ with $S_j$, and $X_jY_j$ with $T_j$ in Definition~\ref{def:EATchann}, we see that these channels $\map_j$ indeed form a valid sequence of EAT channels: $\Ct_j$ is determined from $A_j \Bt_j X_j Y_j$ in the manner specified by Eq.~\eqref{eq:nodisturb}. Additionally, the state they produce always fulfills the Markov conditions, because the values of $X_jY_j$ in each round are generated independently of all preceding registers.

Intuitively, it seems that we could now use the EAT to bound $\Hmin^{\es}(\str{A}|\str{X}\str{Y}E)$.
However, there is a technical issue: to apply the EAT, the event $\OpPE$ must be defined entirely in terms of the (classical) registers that appear in the smoothed min-entropy term that we are bounding, which is not a condition satisfied by the registers $\str{A}\str{X}\str{Y}$ alone. This is where the register $\Btstr$ comes into play, following the same approach as~\cite{ARV19}: by a chain rule for the min- and max-entropies (\cite{VDT13} or \cite{Tom16} Eq.~(6.57)), 
we have for any $\es' + 2\es'' < \es$:
\begin{align}
\Hmin^{\es}(\str{A}|\str{X}\str{Y}E) &\geq \Hmin^{\es'}(\str{A}\Btstr|\str{X}\str{Y}E) - \Hmax^{\es''}(\Btstr|\str{A}\str{X}\str{Y}E) - 3\smf{\es-\es'-2\es''} \nonumber\\
&\geq \Hmin^{\es'}(\str{A}\Btstr|\str{X}\str{Y}E) - \Hmax^{\es''}(\Btstr|\str{X}\str{Y}E) - 3\smf{\es-\es'-2\es''}.
\label{eq:chain}
\end{align}

The $\Hmax^{\es''}(\Btstr|\str{X}\str{Y}E)$ term admits a fairly simple bound, as follows: consider a sequence of EAT channels $\widetilde{\map}_j$ that are identical to $\map_j$ except that they do not produce the registers $A_j\Ct_j$. As before, these maps obey the required Markov conditions.
In addition, recall that for every round the register $\Bt_j$ is deterministically set to $0$ whenever $y\in\inYg$ (which happens with probability $1-\gamma$), hence we always have
\begin{align}
H(\Bt_j|X_j Y_j R)_{(\widetilde{\map}_j\otimes\idmap_{R})(\omega_{R_{j-1}R})} 
= \sum_{y} \pr{Y_j=y} H(\Bt_j|X_j R;Y_j =y)_{(\widetilde{\map}_j\otimes\idmap_{R})(\omega_{R_{j-1}R})} 
\leq \gamma .
\label{eq:fmax}
\end{align}
This means we can apply the max-entropy version of the EAT\footnote{Here we shall use the results from~\cite{DFR20}, because for a constant tradeoff function, this turns out to yield a slightly better bound as compared to the version of the EAT~\cite{DF19} stated here.
Strictly speaking, the reasoning used here is not a direct application of the EAT, because once again, the event $\OpPE$ is not defined on the registers $\Btstr\str{X}\str{Y}$ alone (attempting to address this by including $\str{A}$ in the conditioning registers could result in the Markov conditions not being fulfilled). Fortunately, the bound \eqref{eq:fmax} holds for our maps $\widetilde{\map}_j$ even without a constraint on the output distribution. Hence the reasoning is as follows, in terms of the equations and lemmas in~\cite{DFR20}: first apply Eq.~(32) \emph{without} the event-conditioning term (this is valid since \eqref{eq:fmax} holds without constraints), {then} condition on $\OpPE$ using Lemma~B.6 (noting that $\rho_{\Btstr\str{X}\str{Y}E} 
= \pr{\OpPE}(\rho_{|\OpPE})_{\Btstr\str{X}\str{Y}E} + \pr{\OpPEnot}(\rho_{|\OpPEnot})_{\Btstr\str{X}\str{Y}E}$),
and finally apply Lemma~B.10 to obtain Eq.~\eqref{eq:Hmaxbnd}. 
(Alternatively, one could use $\Ctstr$ instead of $\Btstr$. This was done in~\cite{MvDR+19} to slightly improve the bound in the block analysis, but it does not make a difference in our analysis. However, using $\Ctstr$ would seem to make it harder to sharpen the slightly crude bound used to obtain Eq.~\eqref{eq:gproof}.)
} with a \emph{constant} max-tradeoff function of value $\gamma$. Letting $V'=2\log(1+2
\dim(\Bt)
) = 2\log5$, 
this yields the following bound for any $\alpha' \in (1,1+2/V')$:
\begin{align}
\Hmax^{\es''}(\Btstr|\str{X}\str{Y}E) < n\gamma + n\left(\frac{\alpha'-1}{4}\right)V'^2 + \frac{\smf{\es''}}{\alpha'-1} +  \frac{\alpha'}{\alpha'-1}\log\frac{1}{\pr{\OpPE}}.
\label{eq:Hmaxbnd}
\end{align}

The bulk of our task is to bound the $\Hmin^{\es'}(\str{A}\Btstr|\str{X}\str{Y}E)$ term. To do so, we will need an appropriate min-tradeoff function, which we shall now construct.

\subsubsection{Min-tradeoff function}
\label{sec:fmin}

Consider an arbitrary state of the form $(\map_j\otimes\idmap_{R})(\omega_{R_{j-1}R})$.
In this section, all entropies will be computed with respect to this state, and hence for brevity we will omit the subscript specifying the state.

We first note that
\begin{align}
H(A_j \Bt_j | X_j Y_j \Fj R) 
&= \frac{1-\gamma}{4} 
\sum_{z\in\inX} H(A_j | \Fj R;X_j = Y_j = z)  \nonumber \\
& \qquad + \frac{\gamma}{4} 
\sum_{y\in\inYt} \sum_{x\in\inX} 
H(A_j \Bt_j | \Fj R;X_j = x, Y_j = y), \label{eq:1rndmix}
\end{align}
where we have used the fact that $H(A_j \Bt_j | \Fj R;X_j = x, Y_j = y) = 0$ when $(x,y)=(0,1) \text{ or } (1,0)$, and $H(A_j \Bt_j | \Fj R;X_j = x, Y_j = y) = H(A_j | \Fj R;X_j = x, Y_j = y)$ when $(x,y)=(0,0) \text{ or } (1,1)$.

Let $w$ denote the probability that the state wins the CHSH game, conditioned on the game being played. 
Then by applying the simple but somewhat crude bound\footnote{This marks a point where the analysis could be slightly sharpened, in that if we had a tight bound on the ``two-party entropies'' $H(A_j \Bt_j | \Fj R;X_j = x, Y_j = y)$ rather than just the ``one-party entropies'' $H(A_j | \Fj R;X_j = x, Y_j = y)$, we could improve the second term in~\eqref{eq:gproof}. However, note that it would only improve the keyrate by $O(\gamma)$, because of the $\gamma$ prefactor on that term.} $H(A_j \Bt_j | \Fj R;X_j = x, Y_j = y) \geq H(A_j | \Fj R;X_j = x, Y_j = y)$ to the terms in the second sum in Eq.~\eqref{eq:1rndmix}, we get the bound\footnote{Recall that noisy preprocessing is not applied to the rounds with $Y_j \in \inYt$.}
\begin{align}
H(A_j \Bt_j | X_j Y_j \Fj R) 
\geq \frac{1-\gamma}{2} \lin_\p(w) + \frac{\gamma}{2} \sum_{y\in\inYt} \lin_0(w) 
= \g(w).
\label{eq:gproof}
\end{align}

We can now use the function $\g$ to construct a min-tradeoff function $\fmin$, with the domain of $\fmin$ being distributions on $\Ct_j$ (recall that this register is set to $\perp$ if $Y_j\in\inYg$,
and otherwise is set to $0$ or $1$ if the CHSH game is lost or won respectively). First observe that the channel is an \emph{infrequent-sampling channel} in the sense described in~\cite{DF19,LLR+21}. By the argument in Appendix A.7 of~\cite{LLR+21}, a valid min-tradeoff function $\fmin$ for the channel is given by 
the (unique) affine function specified by the following values (in a minor abuse of notation, here we interpret $\g$ as a function of a distribution instead of a winning probability):
\begin{align}
\fmin(\delta_c) = 
\begin{cases} 
\frac{1}{\gamma} \g(\delta_c) + \left(1-\frac{1}{\gamma}\right)\cperp & \text{ if } c\neq\perp \\
\cperp & \text{ if } c = \perp
\end{cases}
\, ,
\label{eq:fmin}
\end{align}
where $\cperp\in[\g\left(\delta_0\right),\g\left(\delta_1\right)]$ is a constant that can be chosen to optimize the keyrate. (Intuitively, this function is constructed simply by noting that the maps $\map_j$ can only produce distributions that lie in the slice of the probability simplex specified by the constraint 
$\mathrm{Pr} 
[\Ct_j = \perp] = 1-\gamma$, and hence the min-tradeoff function is free to take any value for distributions outside of this slice, recalling that we take the infimum of an empty set to be $+\infty$. For distributions within this slice, we know that $\g$ is an affine lower bound on the entropy as a function of the winning probability, and hence we can just set $\fmin$ equal to $\g$ (up to a domain rescaling) on this slice. Any $\fmin$ constructed this way is precisely of the form described in Eq.~\eqref{eq:fmin}, with $\cperp$ being a constant determining its value on all distributions outside of the $\mathrm{Pr}[\Ct_j = \perp] = 1-\gamma$ slice.)

As shown in~\cite{LLR+21}, the min-tradeoff function constructed this way satisfies
\begin{align}
\begin{gathered}
\Max(\fmin) =
\max\left\{
\frac{1}{\gamma}\Max(\g) + \left(1-\frac{1}{\gamma}\right)\cperp
, \cperp \right\},
\qquad
\Min_{\mathcal{Q}_{f}} (\fmin) = \Min_{\mathcal{Q}_{\g}} (\g), \\
\Var_{\mathcal{Q}_{f}}(\fmin) \leq \sup_{q\in\mathcal{Q}_{\g}} \sum_{c\in\{0,1\}} \frac{q(c)}{\gamma} \left(\cperp - \g(\delta_c)\right)^2,
\end{gathered}
\end{align}
where $\mathcal{Q}_{f}$ denotes
the set of distributions on $\Ct_j$ such that $\mathrm{Pr}[\Ct_j = \perp] = 1-\gamma$ and $\mathrm{Pr}[\Ct_j = 1] \in [\gamma(2-\sqrt{2})/4, \gamma(2+\sqrt{2})/4]$, while $\mathcal{Q}_{\g}$ denotes
the set of all distributions on 
the alphabet $\{0,1\}$
such that $\mathrm{Pr}[1] \in [(2-\sqrt{2})/4, (2+\sqrt{2})/4]$.

For the specific $\g$ and range of $\cperp$ that we consider, 
these expressions simplify to Eq.~\eqref{eq:minmaxvarbounds}, where 
we have solved the optimization $\sup_{q\in\mathcal{Q}_{\g}}$ in the bound on $\Var_{\mathcal{Q}_{f}}(\fmin)$ by observing that it is an affine function of the distribution $q$, and the set $\mathcal{Q}_{\g}$ we use here is essentially
a line segment (in a $1$-dimensional probability simplex).

\subsubsection{Final min-entropy bound}

The event $\OpPE$ is defined by the conditions
$\freq_\ctstr(1)\geq (\wexp-\dtol)\gamma$ and 
$\freq_\ctstr(0)\leq (1-\wexp+\dtol)\gamma$. Hence for all 
$\ctstr \in \OpPE$,
we have (since $\fmin$ is affine)
\begin{align}
\fmin(\freq_\ctstr)
&= \freq_\ctstr(0) \left(\frac{1}{\gamma} \g\left(\delta_0\right) +  \left(1-\frac{1}{\gamma}\right)\cperp\right)
+ \freq_\ctstr(1) \left(\frac{1}{\gamma} \g\left(\delta_1\right) +  \left(1-\frac{1}{\gamma}\right)\cperp\right)
+ \freq_\ctstr(\perp) \cperp \nonumber\\
&= \freq_\ctstr(0) \left(\frac{1}{\gamma} \g\left(\delta_0\right) -\frac{1}{\gamma}\cperp\right)
+ \freq_\ctstr(1) \left(\frac{1}{\gamma} \g\left(\delta_1\right) -\frac{1}{\gamma}\cperp\right)
+ \cperp \nonumber\\
&\geq (1-\wexp+\dtol) \left(\g\left(\delta_0\right) - \cperp\right)
+ (\wexp-\dtol) \left(\g\left(\delta_1\right) - \cperp\right)
+ \cperp \nonumber\\
&= (1-\wexp+\dtol)\g\left(\delta_0\right)
+ (\wexp-\dtol)\g\left(\delta_1\right)
\nonumber\\
&= \g(\wexp-\dtol),
\label{eq:fminPEineq}
\end{align}
where the inequality holds because $\cperp\in[\g\left(\delta_0\right),\g\left(\delta_1\right)]$, and in the last line we use the fact that $\g$ is affine and revert to interpreting it as a function of winning probability. (We remark that if we fix $\cperp=\g(1)$, then in fact this inequality can be derived using only the $\freq_\str{c}(0)$ condition, following~\cite{DF19}. Hence in principle one could sacrifice the option of optimizing $\cperp$ in exchange for reducing the number of checks to perform in the protocol, which improves the completeness parameters.)

Therefore, we can choose $h=\g(\wexp-\dtol)$ in the EAT statement (Prop.~\ref{prop:EAT}) to conclude that the state conditioned on $\OpPE$ satisfies
\begin{align}
\Hmin^{\es'}(\str{A}\Btstr | \str{X}\str{Y}E) &> n
\g(\wexp-\dtol) 
- n\frac{(\alpha-1)\ln 2}{2}V^2 - n(\alpha-1)^2K_\alpha 
\nonumber \\
&\qquad 
- \frac{\smf{\es'}}{\alpha-1} - \frac{\alpha}{\alpha-1}\log\frac{1}{\pr{\OpPE}}, \label{eq:EATbound}
\end{align}
where $\smf{\eps},V,K_\alpha$ are as defined in Eq.~\eqref{eq:VK}. Putting this together with Eqs.~\eqref{eq:chain} and~\eqref{eq:Hmaxbnd}, we finally obtain the bound in Theorem~\ref{th:rawHmin}.

\subsection{Scope of applicability}
\label{sec:nonCHSH}

In the above analysis, we have focused on a protocol that only uses the CHSH game. However, it would be possible to modify the analysis to account for arbitrary Bell inequalities, as was done in~\cite{BRC20}. Essentially, Alice and Bob would simply need to choose a different distribution of their input settings, corresponding to a different game being played. Furthermore, it is not strictly necessary for Alice to choose uniformly random inputs in the generation rounds --- as noted in~\cite{SGP+21}, she could instead choose some biased distribution of inputs. It would even be possible to consider applying different amounts of noisy preprocessing for the different inputs in generation rounds. All of these modifications would essentially correspond to finding an appropriate min-tradeoff function, which we describe how to do in the subsequent section. However, we show in Sec.~\ref{sec:bestbnd} that for the depolarizing-noise model at least, there is little to be gained by considering these modifications.

There is a technical issue that in order to implement the above modifications, Alice and Bob may need to know which rounds are test rounds and which rounds are generation rounds, if they need to choose different input distributions in the two cases. (In Protocol~\ref{prot:DIQKD}, this was not an issue because the CHSH game is played with uniformly random inputs, and we also used uniformly random inputs for the generation rounds, so there is no difference in the input distributions between test and generation rounds.) However, this could in principle be addressed by using a short pre-shared key in order to choose the test rounds --- we describe this in more detail in Sec.~\ref{sec:pubchann}. 

\section{Single-round bound}
\label{sec:1rndbnd}

We now explain how to derive the function $\lin_\p$ as described in Eq.~\eqref{eq:linbnd}. To reduce clutter, we will use slightly different notation in this section, which should not be confused with the earlier notation. In particular, we will omit the~~$\bar{ }$~~accents from the registers $\bar{A}\bar{B}\bar{E}$, and we will be using hermitian operators $A_x,B_y$ that should not be confused with the registers $A_j,B_j$ in the earlier notation.

As described previously, consider an arbitrary state $\rho_{ABE}$ and possible measurements on the $A$ and $B$ subsystems, indexed by $x$ and $y$ respectively. 
For the purpose of bounding the entropy, all measurements can be assumed projective by considering a suitably chosen simultaneous Stinespring dilation; see e.g.~\cite{TSG+21}.
Let $\pvm_{a|x}$ for Alice (resp.~$\pvmB_{b|y}$ for Bob) denote the projector corresponding to outcome $a$ (resp.~$b$) from measurement $x$ (resp.~$y$). We first consider a somewhat more general setting than that required for the above security proof; namely, we aim to find a lower bound on the conditional entropies, knowing only that the state and measurements produce the values $\constr_j$ for some observables $\Gamma_j(\pvm_{a|x},\pvmB_{b|y})\defvar \sum_{abxy} c^{(j)}_{abxy} \pvm_{a|x}\otimes\pvmB_{b|y}$ defined by fixed coefficients $c^{(j)}_{abxy}\in\mathbb{R}$. (For instance, these could be the values of some Bell expressions, or even simply the entire set of output probabilities.) Furthermore, we allow for the possibility that the inputs are not uniformly random.
Making this precise, we aim to find the function 
\begin{align}
\breve{\lin}_\p(\vec{\constr}) \defvar \,
\begin{gathered}
\inf_{\rho_{ABE},\pvm_{a|x},\pvmB_{b|y}} \sum_{x\in\{0,1\}} \keyw_x H(
\hat{A}_x
|{E}) \\ 
\suchthat\; \tr{\Gamma_j(\pvm_{a|x},\pvmB_{b|y}) \rho_{AB}} = \constr_j \quad \forall j
\end{gathered},
\label{eq:mainopt}
\end{align}
where $\keyw_x \geq 0$ are coefficients 
that depend on the input distributions~\cite{SGP+21}.
(For the security proof of Protocol~\ref{prot:DIQKD}, we would only need to consider a single $\Gamma_j(\pvm_{a|x},\pvmB_{b|y})$, which describes the probability of winning the CHSH game. Also, we simply have $\keyw_0=\keyw_1=1/2$, since we want a bound of the form~\eqref{eq:linbnd}.)
Without loss of generality, we can restrict the optimization to pure $\rho_{ABE}$.

By considering the effect of Eve using mixtures of strategies, it is easily seen that $\breve{\lin}_\p$ must be convex. It can thus be shown that for any $\vec{\constr}$ in the interior of the set of values achievable by quantum theory, this optimization is in fact equal to its Lagrange dual (see e.g.~\cite{TSG+21} for a detailed explanation):
\begin{align}
\breve{\lin}_\p(\vec{\constr}) = \sup_{\vec{\lambda}}
\inf_{\rho_{ABE},\pvm_{a|x},\pvmB_{b|y}} \left( \sum_{x\in\{0,1\}} \keyw_x H(\hat{A}_x|{E})\right) - \vec{\lambda} \cdot \left(\tr{\vec{\Gamma}(\pvm_{a|x},\pvmB_{b|y}) \rho_{AB}} - \vec{\constr} \right) .
\label{eq:dual}
\end{align}
Since the optimization over $\vec{\lambda}$ is a supremum, it follows that for any value of $\vec{\lambda}$, we have a lower bound on $\breve{\lin}_\p(\vec{\constr})$ of the form
\begin{align}
\breve{\lin}_\p(\vec{\constr}) \geq
\vec{\lambda}\cdot\vec{\constr} + c_{\vec{\lambda}}, 
\label{eq:duallin}
\end{align}
where
\begin{align}
c_{\vec{\lambda}} \defvar \inf_{\rho_{ABE},\pvm_{a|x},\pvmB_{b|y}} \left(\sum_{x\in\{0,1\}} \keyw_x H(\hat{A}_x|{E})\right) - \vec{\lambda} \cdot \tr{\vec{\Gamma}(\pvm_{a|x},\pvmB_{b|y}) \rho_{AB}}.
\label{eq:intercept}
\end{align}
Importantly, such a lower bound is automatically affine with respect to $\vec{\constr}$, and hence we can take it as a possible choice of $\lin_\p$ for the security proof.
We thus see that in order to find an affine $\lin_\p$ for use in the security proof, it suffices to choose some $\vec{\lambda}$ and compute (or lower-bound) the value of $c_{\vec{\lambda}}$ as defined by Eq.~\eqref{eq:intercept}.
In addition, this approach yields essentially tight bounds for the asymptotic rates\footnote{There is a technicality here that was overlooked in earlier versions of this work. Specifically, even though $\breve{\lin}_\p$ is basically affine over a large range of values (see Sec.~\ref{sec:1rndresults}), the optimal choice of $\lin_\p$ for the asymptotic rates still may not yield the optimal finite-size keyrates at each $n$ when applying the EAT --- this is because choosing a bound $\lin_\p$ that is suboptimal with respect to the asymptotic rates can yield a min-tradeoff function with smaller variance and range, reducing the finite-size corrections. However, given the computationally intensive nature of the algorithm we describe here, we did not attempt to optimize the choice of $\lin_\p$ as a function of $n$, instead sticking to the optimal choice for the asymptotic limit. Still, as mentioned previously in Sec.~\ref{sec:EAT}, we would expect that any potential improvements by optimizing $\lin_\p$ would at best bring the keyrates somewhat closer to the collective-attacks values shown in Fig.~\ref{fig:experiments}.}, in the sense that taking the supremum of the bounds given by all possible $\vec{\lambda}$ returns the value of $\breve{\lin}_\p(\vec{\constr})$, by Eq.~\eqref{eq:dual}. We also note that when there are multiple constraints, for any given $\vec{\constr}$ the corresponding optimal $\vec{\lambda}$ yields a single Bell expression that certifies the same bound on the entropy as all the original constraints. When applying this bound in the EAT, the task of choosing $\vec{\lambda}$ here is equivalent to the task in~\cite{ARV19,LLR+21} of choosing a tangent point to the rate curves. 

The $H(\hat{A}_x|E)$ term in the optimization can be rewritten based on the approach in~\cite{WLC18}. Specifically, we observe that the state produced on $\hat{A}_x E$ by performing Alice's measurement $x$ on $\rho_{ABE}$ and then applying noisy preprocessing could also be obtained by the following process: append an ancilla $T$ in the state
\begin{align}
\ket{\phi_\p}_T \defvar \sqrt{1-\p} \ket{0}_T + \sqrt{\p} \ket{1}_T,
\end{align}
apply a pinching channel $\mathcal{P}(\sigma_T)\defvar\sum_t \pure{t}_T \sigma_T \pure{t}_T$ to $T$, then perform a measurement on $AT$ described by the projectors
\begin{align}
\tilde{\pvm}_{a|x} \defvar \pvm_{a|x}\otimes\pure{0}_T + \pvm_{a\oplus 1|x}\otimes\pure{1}_T,
\label{eq:NPPprojs}
\end{align}
and store the outcome in $\hat{A}_x$ without further processing. However, 
the pinching channel on $T$ can in fact be omitted, because the subnormalized conditional states produced on $E$ would still be the same without it (in the following, we leave some tensor factors of $\id$ implicit for brevity, and consider an arbitrary state $\sigma_{ABET}$ before applying the pinching channel):
\begin{align}
&\tr[ABT]{\tilde{\pvm}_{a|x} 
\,\mathcal{P}_T (\sigma_{ABET})\,
\tilde{\pvm}_{a|x}}\nonumber\\
=&\tr[ABT]{\tilde{\pvm}_{a|x} \left(\sum_{t} 
\pure{t}_T
\sigma_{ABET} 
\pure{t}_T
\right) \tilde{\pvm}_{a|x}}\nonumber\\
=& \sum_{t} \tr[ABT]{(\pvm_{a\oplus t|x}\otimes\pure{t}_T)  \sigma_{ABET}(\pvm_{a\oplus t|x}\otimes\pure{t}_T) }\nonumber\\
=& \sum_{t,t'} \tr[ABT]{(\pvm_{a\oplus t|x}\otimes\pure{t}_T)  \sigma_{ABET}\left(\pvm_{a\oplus t'|x}\otimes\pure{t'}_T\right) }\nonumber\\
=& \tr[ABT]{\tilde{\pvm}_{a|x} \sigma_{ABET}\tilde{\pvm}_{a|x} }.
\end{align}
Hence we no longer consider the pinching channel on $T$, i.e.~we simply study the situation where we immediately perform the projective measurement~\eqref{eq:NPPprojs} on the state $\hat{\rho}_{ABET} \defvar \rho_{ABE} \otimes \pure{\phi_\p}_T$ and store the outcome in register $\hat{A}_x$. As mentioned previously, we can take $\rho_{ABE}$ to be pure without loss of generality, in which case $\hat{\rho}_{ABET}$ is pure as well\footnote{It was important to remove the pinching channel because if $\hat{\rho}_{ABET}$ were a mixed state, then Eq.~\eqref{eq:Hduality} would be replaced by $D(\hat{\rho}_{ABT} \Vert \calZ_x(\hat{\rho}_{ABT})) = H(\hat{A}_x|EE')$, where $E'$ purifies $\hat{\rho}_{ABET}$. While $H(\hat{A}_x|EE')$ is a valid lower bound on $H(\hat{A}_x|E)$, it also turns out to be a trivial one, because the value of $T$ is ``copied'' into $E'$, 
removing the advantage of noisy preprocessing (in which Eve is not supposed to know whether the bit-flip has occurred).}, and must hence obey the following relation (derived in e.g.~\cite{Col12}):
\begin{align}
H(\hat{A}_x|E) = D(\hat{\rho}_{ABT} \Vert \calZ_x(\hat{\rho}_{ABT})) = D(\calG({\rho}_{AB}) \Vert \calZ_x(\calG({\rho}_{AB}))),
\label{eq:Hduality}
\end{align}
where
\begin{align}
\calG(\sigma_{AB}) &\defvar \sigma_{AB} \otimes \pure{\phi_\p}_T, \\
\calZ_x(\sigma_{ABT}) &\defvar \sum_a (\tilde{\pvm}_{a|x}\otimes\id_{B}) \sigma_{ABT} (\tilde{\pvm}_{a|x}\otimes\id_{B}) .
\end{align}
Importantly, this expression for $H(\hat{A}_x|E)$ is entirely in terms of the reduced state $\rho_{AB}$, and is convex with respect to $\rho_{AB}$. (We give an alternative expression in Appendix~\ref{app:altNPP}, which may be of use in other situations.)

We remark that the above analysis was fairly general, in that in fact it applies to DI scenarios with arbitrary numbers of inputs and outputs, as long as Alice's key-generating measurements still only have 2 outcomes.
In particular, the approaches described in~\cite{TSG+21,BFF21} are able to yield bounds on these optimizations when there is no noisy preprocessing, and it may be useful to study whether the above map for describing noisy preprocessing allows one to apply those approaches to this scenario. However, we leave this question for future work.

We now specialize to 2-input 2-output scenarios. In such cases, 
we have the following ``qubit reduction''~\cite{PAB+09,HST+20}: if 
one finds
a convex function that lower-bounds the right-hand-side of Eq.~\eqref{eq:mainopt} with its optimization restricted to states $\rho_{ABE}$ of dimension $2\times2\times4$ and Pauli measurements, 
then this function is also a lower bound on 
$\breve{\lin}_\p$.
We note (by following the same arguments as before, but with the optimization domain restricted) that picking some $\vec{\lambda}$ and solving the optimization~\eqref{eq:intercept} over such states and measurements yields such a lower bound via Eq.~\eqref{eq:duallin}, which is trivially convex since it is affine. Hence it suffices to consider the optimization~\eqref{eq:intercept} restricted to such states and measurements. Furthermore, the resulting bounds are still tight in the same sense as before, in that taking the supremum over choices of $\vec{\lambda}$ yields the convex envelope of this restricted version of the optimization~\eqref{eq:mainopt} (except possibly for $\vec{\constr}$ not in the interior of the quantum-achievable values).

Since we have reduced the analysis to Pauli measurements, it is convenient to interpret the measurements as producing values in $\pm1$ instead of $\mathbb{Z}_2$, and define the corresponding hermitian observables $A_x \defvar \sum_a a \pvm_{a|x}$ and $B_y \defvar \sum_b b \pvmB_{b|y}$. For 2-input 2-output scenarios, the full probability distribution $\pr{ab|xy}$ (subject to the no-signalling constraints) is completely parametrized by the 4 correlators $\expvaltr{(A_x \otimes B_y)}$ and the 4 marginals $\expvaltr{(A_x \otimes \id)}$, $\expvaltr{(\id \otimes B_y)}$. To simplify the analysis, one can focus on the case where the marginals are zero\footnote{While more detailed explanations can be found in e.g.~\cite{PAB+09,HST+20}, the outline is as follows: consider a virtual \emph{symmetrization step} in which Alice and Bob jointly flip their outputs using a uniformly random public bit, forcing their marginals to be zero while leaving the correlators unchanged. Some calculation shows~\cite{SR08,PAB+09,HST+20} that the entropy of their original outputs conditioned on Eve's side-information is equal to the entropy of their symmetrized outputs conditioned on Eve's side-information and the publicly communicated bit (and this still holds true with noisy preprocessing). Absorbing the publicly communicated bit into Eve's side-information, this implies that if the original optimization~\eqref{eq:mainopt} had constraints on the marginals $\expvaltr{(A_x \otimes \id)}$, $\expvaltr{(\id \otimes B_y)}$ in addition to the correlators $\expvaltr{(A_x \otimes B_y)}$, we could bound it by instead considering the optimization where the marginal constraints are replaced by the condition that they are zero (with the correlator constraints unchanged). (Note that the virtual symmetrization does not need to be physically performed; it merely serves as an intermediate construction in this analysis.)}, in which case only the correlators $\expvaltr{(A_x \otimes B_y)}$ remain to be considered. In terms of the optimization~\eqref{eq:mainopt}, this means that it suffices to consider situations where there are 4 constraints, corresponding to the observables
\begin{align}
\Gamma_{xy}(\pvm_{a|x},\pvmB_{b|y}) = A_x \otimes B_y, 
\quad \text{ for } x,y\in\{0,1\}.
\end{align}
Considering other forms of constraints in 2-input 2-output scenarios is essentially equivalent to making specific choices of Lagrange multipliers $\vec{\lambda}$ for this 4-constraint formulation. For instance, in Protocol~\ref{prot:DIQKD} we only impose a constraint based on the CHSH value\footnote{Here, by CHSH value we mean the quantity $\constr = \tr{(A_0\otimes B_0 + A_0\otimes B_1 + A_1\otimes B_0 - A_1\otimes B_1) \rho_{AB}}$. This is related to the probability $w$ of winning the CHSH game by the simple equation $\constr = 8w-4$, so an affine function of $\constr$ is easily converted to an affine function of $w$.}, which is equivalent to restricting to Lagrange-multiplier combinations of the form $(\lambda_{00},\lambda_{01},\lambda_{10},\lambda_{11}) = (\lambda,\lambda,\lambda,-\lambda)$ for some $\lambda\in\mathbb{R}$.

We still have the freedom to choose the basis in which to express the optimization. Following~\cite{SGP+21}, we can use the measurement axes of Alice's two Pauli measurements to define the $X$-$Z$ plane on her system, taking $A_0=Z$ and $A_1 = \cos(\thA)Z + \sin(\thA)X$ for some $\thA \in [0,\pi]$ 
(values of $\thA$ in $[\pi,2\pi]$ can be brought into this range by rotating our axis choice by $\pi$ around the $Z$-axis). 
Analogously, we can choose a basis for Bob such that $B_0=Z$ and $B_1 = \cos(\thB)Z + \sin(\thB)X$ with $\thB \in [0,\pi]$. (This is a different basis choice from the one in~\cite{PAB+09,HST+20} that allows a reduction to \emph{Bell-diagonal} 
$\rho_{AB}$. That choice involves more parameters for the measurements, but fewer parameters for the state.)

With this in mind, we rewrite the optimization~\eqref{eq:intercept} as 
\begin{align}
\begin{gathered}
\min_{\thA} \min_{\thB} 
\min_{\rho_{AB}} 
F_\mathrm{obj}(\thA,\thB,\rho_{AB}),
\text{ where } \\ 
F_\mathrm{obj}(\thA,\thB,\rho_{AB}) \defvar \left( \sum_{x\in\{0,1\}} \keyw_x D(\calG({\rho}_{AB}) \Vert \calZ_x(\calG({\rho}_{AB})))\right)
- \vec{\lambda} \cdot \tr{\vec{\Gamma}(
\thA,\thB
) \rho_{AB}}.
\end{gathered}
\end{align}
(The 
minima are 
attained because the objective function is continuous and the domain is compact.)
Furthermore, the objective function is invariant under the substitutions $\rho_{AB} \to (Y \otimes Y) \rho_{AB} (Y \otimes Y)$ and $\rho_{AB} \to \rho_{AB}^*$, which implies~\cite{PAB+09} we can restrict the optimization to $\rho_{AB}$ that are ``almost'' Bell-diagonal --- specifically, with respect to the Bell basis $\{\ket{\Phi^+}, \ket{\Psi^-}, \ket{\Phi^-}, \ket{\Psi^+}\}$ where $\ket{\Phi^\pm} = (\ket{00}\pm\ket{11})/\sqrt{2}$ and $\ket{\Psi^\pm} = (\ket{01}\pm\ket{10})/\sqrt{2}$, we can take
\begin{align}
\rho_{AB} = 
\begin{pmatrix}
L_{\Phi^+} & \ell_1 & 0 & 0 \\
\ell_1 & L_{\Psi^-} & 0 & 0 \\
0 & 0 & L_{\Phi^-} & \ell_2 \\
0 & 0 & \ell_2 & L_{\Psi^+}
\end{pmatrix},
\label{eq:almostdiag}
\end{align}
for some $L_{\Phi^+} , L_{\Psi^-} , L_{\Phi^-} , L_{\Psi^+} , \ell_1 , \ell_2 \in \mathbb{R}$.

We now describe how each minimization can be tackled
when treating the parameters in the other minimizations as constants, then summarize how all these algorithms can be put together in a consistent manner, and argue that this approach indeed yields arbitrarily tight bounds.

\subsection{Minimization over Alice's measurement}
We tackle this minimization simply by applying a (uniform) continuity bound for $\thA$.
Specifically, for $\delta\in[0,\pi]$ we describe a monotone increasing function $\cont(\delta)$ that bounds the change in the objective function when $\thA$ is replaced by $\thA+\delta$ (treating $\thB$ and $\rho_{AB}$ as constants). 
Then for any set of intervals 
of the form $\{[\theta_j-\delta_j,\theta_j+\delta_j]\}_j$ that covers the interval $[0,\pi]$, we would have 
\begin{align}
\min_{\thA} F_\mathrm{obj}(\thA,\thB,\rho_{AB}) \geq \min_j 
F_\mathrm{obj}(\theta_j,\thB,\rho_{AB}) - \cont(\delta_j)
.
\end{align}
We apply this in practice by starting with a fairly ``coarse'' choice of intervals, then iteratively applying the process of deleting the interval that currently achieves the minimization over $j$, and replacing it with smaller intervals that cover the deleted interval.

To derive such a continuity bound, we first analyze the entropic term in the objective function, following~\cite{SBV+21} (with a minor modification to slightly improve the bound).
Take any pure initial state $\ket{\rho}_{ABE}$, and let $\sigma_{\hat{A}_1 BE}$ be the state obtained by performing a Pauli measurement along angle $\thA$ in the $X$-$Z$ plane on the $A$ register of $\ket{\rho}_{ABE}$, applying noisy preprocessing, then storing the result in the classical register $\hat{A}_1$ and tracing out $A$. 
Let $\sigma'_{\hat{A}_1 BE}$ be the analogous state with $\thA$ replaced by $\thA+\delta$ for some $\delta \in [0,\pi]$. Our goal would be to bound $\left|H(\hat{A}_1| E)_{\sigma} - H(\hat{A}_1| E)_{\sigma'}\right|$. 

We have $H(\hat{A}_1| E)_{\sigma} 
= \sum_{\hat{a}_1} \pr{\hat{a}_1} H(\sigma_{E|\hat{A}_1=\hat{a}_1}) + H(\hat{A}_1)_{\sigma} - H(E)_{\sigma}
$, and analogously for $\sigma'$. Since the operations on $A$ do not affect $E$, we have $H(E)_{\sigma}=H(E)_{\sigma'}$. Also, for states of the form~\eqref{eq:almostdiag}, we have $\rho_{A}=\id/2$ and hence 
$H(\hat{A}_1)_{\sigma} = H(\hat{A}_1)_{\sigma'} = 1$. This gives us
\begin{align}
\left|H(\hat{A}_1| E)_{\sigma} - H(\hat{A}_1| E)_{\sigma'}\right| 
\leq \max_{\hat{a}_1} \left| \left(H(\sigma_{E|\hat{A}_1=\hat{a}_1}) - H(\sigma'_{E|\hat{A}_1=\hat{a}_1})\right)\right|,
\end{align}
so it suffices to bound the difference in entropies of the conditional states on $E$.

Now observe that exactly the same state $\sigma_{\hat{A}_1 B E}$ would have been produced if the initial state had been $\left(e^{i\thA Y_A/2} \otimes \id_{BE}\right)\ket{\rho}_{ABE}$ and the initial Pauli measurement were replaced by a $Z$ measurement.
Furthermore, the fact that $\rho_A=\id/2$ implies we can write $\ket{\rho}_{ABE} = \sum_a \ket{a}_A \ket{a}_{BE} / \sqrt{2}$, where $\{\ket{a}_A\}_a$ is the $Z$-eigenbasis of $A$ and $\{\ket{a}_{BE}\}_a$ are two orthonormal states on $BE$. This implies 
\begin{align}
\left(e^{i\thA Y_A/2} \otimes \id_{BE}\right)\ket{\rho}_{ABE} 
= \left(\id_A \otimes \left(e^{i\thA Y_{BE}/2}\right)^T\right)\ket{\rho}_{ABE} 
= \frac{1}{\sqrt{2}}\sum_a \ket{a}_A \otimes 
\rot{\thA}
\ket{a}_{BE},
\end{align}
where 
$Y_{BE} \defvar i\ketbra{-1}{+1}_{BE}-i\ketbra{+1}{-1}_{BE}$ 
and $\rot{\thA}\defvar\left(e^{i\thA Y_{BE}/2}\right)^T$.
Performing the analogous analysis for
$\sigma'$,
we conclude that 
\begin{align}
\begin{aligned}
\sigma_{BE|\hat{A}_1=\hat{a}_1} &= 
(1-\p) \rot{\thA}\pure{\hat{a}_1}_{BE}\rot{\thA}^\dagger + \p \rot{\thA}\pure{-\hat{a}_1}_{BE}\rot{\thA}^\dagger, \\
\sigma'_{BE|\hat{A}_1=\hat{a}_1} &= 
(1-\p) \rot{\thA+\delta}\pure{\hat{a}_1}_{BE}\rot{\thA+\delta}^\dagger + \p \rot{\thA+\delta}\pure{-\hat{a}_1}_{BE}\rot{\thA+\delta}^\dagger.
\end{aligned}
\end{align}
Therefore, we have
\begin{align}
F(\sigma_{E|\hat{A}_1=\hat{a}_1},\sigma'_{E|\hat{A}_1=\hat{a}_1}) &\geq F(\sigma_{BE|\hat{A}_1=\hat{a}_1},\sigma'_{BE|\hat{A}_1=\hat{a}_1}) \nonumber \\
&\geq (1-\p) \left|\bra{\hat{a}_1}\rot{\thA}^\dagger\rot{\thA+\delta}\ket{\hat{a}_1}\right| + \p \left|\bra{-\hat{a}_1}\rot{\thA}^\dagger\rot{\thA+\delta}\ket{-\hat{a}_1}\right| \nonumber \\
&\geq \left|\cos\frac{\delta}{2}\right|,
\end{align}
where the second inequality holds by concavity of fidelity, 
and the third inequality is given by explicit calculation (see~\cite{SBV+21}).

This lets us apply a fidelity-based continuity bound~\cite{SBV+21}:
\begin{align}
\max_{\hat{a}_1} \left| \left(H(\sigma_{E|\hat{A}_1=\hat{a}_1}) - H(\sigma'_{E|\hat{A}_1=\hat{a}_1})\right)\right| \leq 4.023 \acos F(\sigma_{E|\hat{A}_1=-1},\sigma'_{E|\hat{A}_1=-1}) \leq 2.012 \delta,
\label{eq:Hcontbnd}
\end{align}
where in the second inequality we used the condition $\delta \in [0,\pi]$. 
(Numerical heuristics suggest that the true bound in Eq.~\eqref{eq:Hcontbnd} may simply be $\delta$, so there is some potential for improvement here, though the effect would be fairly small.
In Appendix~\ref{app:altcont} we present an approach based on trace distance instead of fidelity, but it appears to scale poorly at small $\delta$.)

As for the $\vec{\Gamma}$ term,
we note that 
\newcommand{\cs}[1]{\cos(#1)}
\newcommand{\sn}[1]{\sin(#1)}
\begin{align}
&\left|\tr{\left(\vec{\lambda}\cdot\vec{\Gamma}(\thA+\delta,\thB)-\vec{\lambda}\cdot\vec{\Gamma}(\thA,\thB)\right)\rho_{AB}}\right| 
\nonumber\\
\leq& \norm{\vec{\lambda}\cdot\vec{\Gamma}(\thA+\delta,\thB) - \vec{\lambda}\cdot\vec{\Gamma}(\thA,\thB)}_\infty \nonumber\\
=& \norm{\left((\cs{\thA+\delta}Z + \sn{\thA+\delta}X) - (\cs{\thA}Z + \sn{\thA}X) \right) \otimes (\lambda_{10} B_0+\lambda_{11} B_1)}_\infty \nonumber\\
\leq& (|\lambda_{10}| + |\lambda_{11}|)\norm{
(\cs{\thA+\delta}Z + \sn{\thA+\delta}X) - (\cs{\thA}Z + \sn{\thA}X)
}_\infty \nonumber\\
=& (|\lambda_{10}| + |\lambda_{11}|)\sqrt{2-2\cos(\delta)},
\end{align}
where the last line follows from an explicit eigenvalue calculation.

Overall, this means that for $\delta\in[0,\pi]$ we can choose 
\begin{align}
\cont(\delta) = 
2.012 \keyw_1
\delta + (|\lambda_{10}| + |\lambda_{11}|)\sqrt{2-2\cos(\delta)},
\label{eq:contbnd}
\end{align}
accounting for the $\keyw_1$ factor on the $H(\hat{A}_1| E)$ term. This bound is monotone increasing for $\delta\in[0,\pi]$, as required (so that it also bounds the change in entropy when the measurement angle is changed from $\thA$ to any value in the interval $[\thA,\thA+\delta]$).

\subsection{Minimization over Bob's measurement}

The entropic term in the objective function has no dependence on Bob's measurement, so we only need to consider the $\vec{\Gamma}$ term.
In principle, this could be approached using the same argument as above, where we would arrive at the continuity bound
\begin{align}
&\left|\tr{\left(\vec{\lambda}\cdot\vec{\Gamma}(\thA,\thB+\delta)-\vec{\lambda}\cdot\vec{\Gamma}(\thA,\thB)\right)\rho_{AB}}\right| 
\leq
(|\lambda_{01}| + |\lambda_{11}|)\sqrt{2-2\cos(\delta)}.
\end{align}

However, some heuristic experiments indicate that the following approach (used in~\cite{SGP+21}) is more efficient: we can let $r_Z\defvar\cos(\thB)$ and $r_X\defvar\sin(\thB)$ and write
\begin{align}
\sum_{xy} \lambda_{xy} A_x \otimes B_y 
&= \left(\sum_{x} \lambda_{x0} A_x\right) \otimes Z + \left(\sum_{x} \lambda_{x1} A_x\right) \otimes (r_Z Z + r_X X),
\end{align}
in which case the minimization over $\thB\in[0,\pi]$ is equivalent to minimizing over $(r_Z,r_X)$ that lie on the set $\semicset \defvar \{(r_Z,r_X)\mid r_Z^2+r_X^2=1 \text{ and } r_X \geq 0\}$ (i.e.~a semicircular arc). Crucially, the objective function is affine with respect to the vector $(r_Z, r_X)$. Hence if $V$ is any finite set of points such that $\semicset$ is contained in their convex hull $\operatorname{Conv}(V)$, we immediately have 
\begin{align}
\min_{(r_Z,r_X)\in \semicset} F_\mathrm{obj}(\thA,(r_Z,r_X),\rho_{AB}) 
&\geq \min_{(r_Z,r_X)\in \operatorname{Conv}(V)} F_\mathrm{obj}(\thA,(r_Z,r_X),\rho_{AB}) \nonumber\\
&= \min_{(r_Z,r_X)\in V} F_\mathrm{obj}(\thA,(r_Z,r_X),\rho_{AB}),
\end{align}
because the minimum of an affine function over the convex hull of a finite set $V$ is always attained at an extremal point (which will be a point in $V$). 
To apply this result, we start with a simple choice of the set $V$ (for instance,
in our code we use 
$V = \{(1,0),(1,1),(-1,1),(-1,0)\}$) and find the point in $V$ that yields the minimum value. We then delete this point and replace it with two other points such that $\semicset$ is still contained in the convex hull, and iterate this process until a sufficiently tight bound is obtained (for instance, by checking that there is a feasible point of the optimization that is sufficiently close to the lower bound we have obtained).

\subsection{Minimization over states}

The minimization over states can be tackled by expressing it as a convex optimization and applying the \emph{Frank-Wolfe algorithm}~\cite{FW56}, as was observed in~\cite{WLC18}. For completeness, we now describe the method, with minor 
modifications and clarifications for our specific scenario. 
We emphasize that while this procedure is numerical, the bounds it returns are \emph{secure}, in the sense that it will never over-estimate the true value of the minimization problem. 

We observe that (for fixed measurements, parametrized as described above) the minimization over states is the minimization of the convex function 
\begin{align}
f_\mathrm{obj}(\rho) \defvar
\left(\sum_{x\in\{0,1\}} \keyw_x D(\calG({\rho}) \Vert \calZ_x(\calG({\rho})))\right) - \vec{\lambda}\cdot \tr{\vec{\Gamma}(\thA,(r_Z,r_X)) \rho}.
\end{align}
This function is differentiable for all $\rho$ such that $\calG(\rho) > 0$, with~\cite{WLC18}\footnote{The notation here follows that used in~\cite{WLC18}, in which for a function $f$ defined on matrices parametrized by their matrix entries ($\rho=\sum_{jk}\rho_{jk}\ketbra{j}{k}$), its derivative is $\nabla f(\rho) \defvar \sum_{jk} (\partial f/ \partial\rho_{jk}) \ketbra{j}{k}$. 
Denoting the tangent to the graph of $f$ at $\rho$ as $t_{\rho}$, this definition of $\nabla f$ satisfies $t_{\rho}(\rho+\Delta) = f(\rho) + \tr{(\nabla f(\rho))^T \Delta}$.
}
\begin{align}
(\nabla f_\mathrm{obj}(\rho))^T = \left( \sum_{x\in\{0,1\}} \keyw_x \calG^\dagger \!\left(\log\calG(\rho) - \log\calZ_x(\calG(\rho))\right)\right) - \vec{\lambda}\cdot\vec{\Gamma}(\thA,(r_Z,r_X)),
\label{eq:fobjder}
\end{align}
where $\calG^\dagger$ is the adjoint channel of $\calG$. In practice, we will not need to explicitly compute $\calG^\dagger$, because all subsequent arguments rely only on ``inner products'' $\tr{(\nabla f_\mathrm{obj}(\rho))^T\sigma}$, which can be rewritten as
\begin{align}
\tr{(\nabla f_\mathrm{obj}(\rho))^T\sigma} = \tr{\left( \sum_{x\in\{0,1\}} \keyw_x \left(\log\calG(\rho) - \log\calZ_x(\calG(\rho))\right)\calG(\sigma)\right) - \vec{\lambda}\cdot\vec{\Gamma}(\thA,(r_Z,r_X))\sigma}.
\end{align}

We also note that the optimization domain is a set defined by PSD constraints (namely, $\rho$ of the form~\eqref{eq:almostdiag} with $\rho\geq0$ and $\tr{\rho}=1$), which means that optimizing any affine function over this set is an SDP, which can be efficiently solved and bounded~\cite{BV04v8} via its dual value. Together with the explicit expression~\eqref{eq:fobjder} for the derivative of the objective function, this makes the optimization problem here a prime candidate for the {Frank-Wolfe algorithm}~\cite{FW56}, which yields \emph{secure} lower bounds on the true minimum value of the optimization. 

\newcommand{\epsd}{\eps}
The Frank-Wolfe algorithm is based on a simple geometric insight: for any point in the domain of a convex function $f_\mathrm{obj}$, the tangent hyperplane (or a supporting hyperplane, if $f_\mathrm{obj}$ is not differentiable at that point) to the graph of $f_\mathrm{obj}$ at that point yields an affine lower bound on $f_\mathrm{obj}$. Hence if we can minimize affine functions over the optimization domain, then any such tangent hyperplane lets us obtain a lower bound on the minimum of $f_\mathrm{obj}$ on the domain. In addition, it intuitively seems that taking the tangent at points closer to the true optimal solution should yield tighter lower bounds (though there are some technical caveats). 
Thus in principle, one could simply perform some heuristic computations to get an estimate of where the true minimum lies, then take the tangent at that estimate and obtain the corresponding lower bound. For this optimization, however, we found that the results were fairly sensitive to deviations from the true optimum. To get the bounds to converge, we found it more efficient to use the standard Frank-Wolfe algorithm,
which (under mild assumptions) can be proven to converge in order $O(1/k)$ after $k$ iterations: 
\newpage
\begin{savenotes}
\begin{algorithm}[htp]
\caption*{\textbf{Frank-Wolfe algorithm} (As presented in~\cite{WLC18})}
Let the domain of optimization be $\optdom$
and the acceptable gap between the feasible values and the lower bounds be $\eps_\mathrm{tol}>0$.
\begin{algorithmic}[1]
\State Set $k=0$
and heuristically find $\rho_0 \defvar \arg\min_{\rho\in\optdom}f_\mathrm{obj}(\rho)$.
\State\label{step:FWSDP}Solve the SDP
\begin{align}
\min_{\Delta} \tr{(\nabla f_\mathrm{obj}(\rho_k))^T \Delta} \text{ s.t. } \rho_k + \Delta \in \optdom, \label{eq:FWSDP}
\end{align}
which returns a feasible point $\Delta^*$
as well as a dual value $-\epsd \leq 0$ 
that lower-bounds the true minimum of the SDP.
\State If $\epsd \leq \eps_\mathrm{tol}$ then stop and return $f_\mathrm{obj}(\rho_k)-\epsd$ as a lower bound on the optimization.
\State Otherwise, heuristically find 
$\mu^* \defvar \arg\min_{\mu\in(0,1)}f_\mathrm{obj}(\rho_k + \mu\Delta^*)$.
\State Set $\rho_{k+1} = \rho_k + \mu^* \Delta^*$, $k\leftarrow k+1$ and return to Step~\ref{step:FWSDP}.
\end{algorithmic}
\end{algorithm}
\end{savenotes}

\noindent Geometrically, the SDP in Eq.~\eqref{eq:FWSDP} corresponds to considering the tangent to $f_\mathrm{obj}$ at $\rho_k$ and computing the maximum amount by which it can decrease (as compared to its value at $\rho_k$) over the domain $\optdom$. 
As described earlier,
this bounds the maximum amount by which $f_\mathrm{obj}$ can decrease from $f_\mathrm{obj}(\rho_k)$ over $\optdom$.

In our application of the Frank-Wolfe algorithm, there is the technical issue that if $\calG(\rho_k)$ is singular (or has negative eigenvalues, from numerical imprecision), then the derivative at $\rho_k$ (Eq.~\eqref{eq:fobjder}) is ill-behaved. 
To cope with this, we used
the heuristic solution of simply replacing $\rho_k$ with $(1-\delta)\rho_k + \delta \idk$ in Step~\ref{step:FWSDP}, where $\delta 
\defvar 
\max(\operatorname{eigenvalues}(-\calG(\rho_k)) \cup \{ 10^{-14} \})
$. 
Note that this does not affect the \emph{security} of the result, since it merely corresponds to taking a tangent at a slightly different point, which still yields a valid lower bound. On the other hand, it could possibly affect the theoretical convergence rates, but in practice this did not appear to pose a significant problem in our setting. (In~\cite{WLC18}, this issue is addressed by analyzing a ``perturbed'' version of the optimization and applying a continuity bound, but we found that for the level of accuracy we desired in this work, the admissible perturbation values are too small to cope with the negative eigenvalues that occur.)

\subsection{Overall algorithm}
\label{sec:algorithmsummary}

Putting the above results together, we see that for any set of intervals $[\theta_j-\delta_j,\theta_j+\delta_j]$ such that $[0,\pi]\subseteq \bigcup_j[\theta_j-\delta_j,\theta_j+\delta_j]$, and set $V$ such that $\semicset \subseteq \operatorname{Conv}(V)$, we have 
\begin{align}
\min_{\rho_{AB}} \min_{\thB} \min_{\thA}
F_\mathrm{obj}(\thA,\thB,\rho_{AB}) 
&\geq \min_{\rho_{AB}} \min_{\thB} \min_j F_\mathrm{obj}(\theta_j,\thB,\rho_{AB}) - \cont(\delta_j) \nonumber \\
&= \min_j \min_{\rho_{AB}} \min_{(r_Z,r_X)\in \semicset} F_\mathrm{obj}(\theta_j,(r_Z,r_X),\rho_{AB}) - \cont(\delta_j) \nonumber \\
&\geq \min_j \min_{\rho_{AB}} \min_{(r_Z,r_X)\in V} F_\mathrm{obj}(\theta_j,(r_Z,r_X),\rho_{AB}) - \cont(\delta_j) \nonumber \\
&= \min_j \min_{(r_Z,r_X)\in V} \min_{\rho_{AB}}  F_\mathrm{obj}(\theta_j,(r_Z,r_X),\rho_{AB}) - \cont(\delta_j).
\label{eq:mincommute}
\end{align}
We can refine the intervals $[\theta_j-\delta_j,\theta_j+\delta_j]$ and set $V$ by the iterative processes described above, with the innermost minimization over $\rho_{AB}$ being handled by the Frank-Wolfe algorithm. It is clearly possible to swap the order of $\min_j$ and $\min_{(r_Z,r_X)\in V}$ in the last line; however, some heuristic plots of the objective function suggest that performing the optimizations in the order shown here is slightly faster. 

To see that this expression can converge to a tight bound, observe that the first inequality in~\eqref{eq:mincommute} becomes arbitrarily tight as we choose smaller values of $\delta_j$. For the second inequality, note that 
the described algorithm chooses $V$ in such a way that the minimum over $V$ approaches the minimum over $\semicset$, hence this inequality also becomes arbitrarily tight.

In general, the above approach faces the difficulty that it needs to optimize over the choice of Lagrange multipliers $\vec{\lambda}$. Given that our approach for solving~\eqref{eq:mincommute} for a specific choice of Lagrange multipliers is already highly computationally intensive (requiring about $5000$ core-hours to achieve the level of accuracy in the bounds~\eqref{eq:certbnds} below for each value of $\p$), it would be impractical to also optimize over the Lagrange multipliers while doing so. It is more feasible to first optimize the Lagrange multipliers while using a simple heuristic algorithm to estimate the minimizations, then certify the final result using our approach for solving~\eqref{eq:mincommute}. (This is essentially the same perspective as presented in~\cite{SBV+21}.)

We remark that this approach can also yield arbitrarily tight bounds for DIRE, where the goal would typically be~\cite{ARV19,BRC20} to find lower bounds on (weighted sums of) ``two-party entropies'' $H(A_x B_y|E)$. This is because by the same arguments as above, we have $H(A_x B_y|E) = D({\rho}_{AB} \Vert \calZ_{xy}({\rho}_{AB}))$~\cite{Col12,TSG+21},
where
\begin{align}
\calZ_{xy}(\sigma_{AB}) &\defvar \sum_{ab} (\pvm_{a|x}\otimes\pvmB_{b|y}) \sigma_{AB} (\pvm_{a|x}\otimes\pvmB_{b|y}) .
\end{align}
(Here we omit the parts corresponding to noisy preprocessing, since it is not applied in randomness expansion.) This expression can be bounded in the same way as we have just described above, though the objective function would no longer be affine with respect to $(r_Z,r_X)$, and hence the optimization over Bob's measurements would also have to be approached using a continuity bound. Our approach should yield a substantial improvement over previous results, which are restricted to the CHSH inequality and only bound the entropy of one party's outputs~\cite{LLR+21}, or which consider the full distribution and bound the entropy of both outputs but use inequalities that are not tight~\cite{BRC20,TSG+21,BFF21}. In addition, the fact that it allows for the random-key-measurement approach could yield further improvements, though there are some technicalities that we address at the end of Sec.~\ref{sec:preshared}.

It would be convenient for future analysis if it were possible to develop closed-form expressions for $\breve{\lin}_\p$ (as was done in~\cite{HST+20,WAP21,SBV+21}), rather than the computationally intensive numerical approaches shown above. However, this appears to be rather challenging, as noted in~\cite{WAP21}. In particular, we found numerical evidence against the conjecture proposed in~\cite{SGP+21} that the minimum in~\eqref{eq:mainopt} (when restricted to qubit strategies) can always be attained by states such that $\rho_{AB}$ is of rank $2$. (A similar observation was reported in~\cite{WAP21}.) More precisely, in the process of heuristically solving the optimizations in order to estimate suitable choices of $\lambda$, we discovered that if we imposed the additional restriction that $\rho_{AB}$ has rank $2$, there was a small but numerically significant difference as compared to the results without this restriction. (The states that heuristically approach the minimum in the latter indeed tend to have two very small eigenvalues, but it appears that these eigenvalues cannot be reduced exactly to zero.) This suggests that the aforementioned conjecture is not true after all, which poses a challenge for closed-form analysis because the eigenvalues involved in computing the entropy are genuinely roots of a fourth-degree polynomial (in~\cite{HST+20,SBV+21}, a key element of the analysis was to argue that it suffices to consider rank-$2$ $\rho_{AB}$, simplifying the expression for the eigenvalues). 

\subsection{Resulting bounds}
\label{sec:1rndresults}

\begin{figure}
\centering
\subfloat[$\p=0$]{
\includegraphics[width=0.49\textwidth]{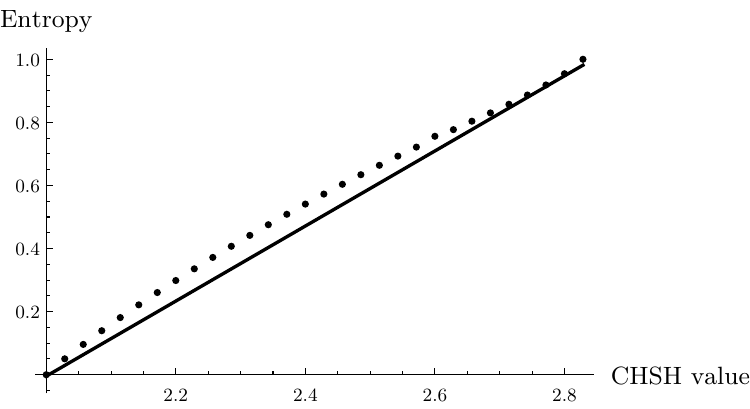}
} 
\subfloat[$\p=0.2$]{
\includegraphics[width=0.49\textwidth]{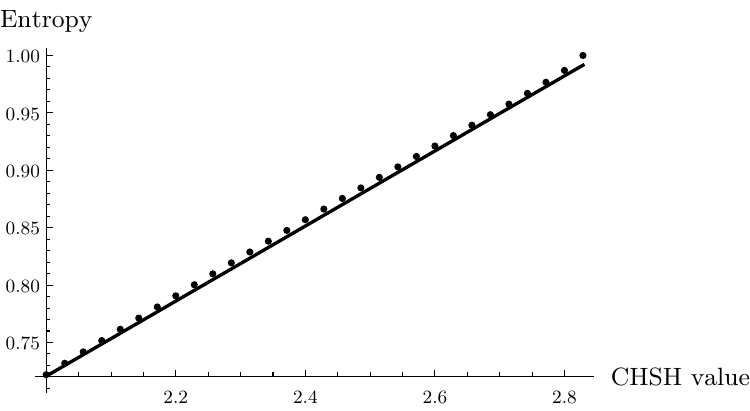}
}
\caption{The solid lines are the certified lower bounds we obtained (Eq.~\eqref{eq:certbnds}), while the points indicate the results of heuristically solving the optimization~\eqref{eq:mainopt} over qubit states and measurements (with just the CHSH value as the constraint). As previously discussed, the tight bound in each case would be given by the convex envelope of the curve traced out by the points, assuming that the heuristics have found the true minimum. However, we can see that in each case that curve appears to be nonconvex over the interval $[2,2.75]$ (approximately), and its convex envelope would be affine over that interval --- specifically, it would be given by the linear interpolation between the feasible points at the ends of that interval. The certified bound is almost flush with this linear interpolation, indicating that it is basically tight over this interval.}
\label{fig:entbnds}
\end{figure}

For the protocols in this work, the relevant bound to compute is for the case where the only constraint imposed is the CHSH value. More precisely, we consider the described optimization with a single constraint corresponding to the operator
\begin{align}
\Gamma(\thA,\thB) = A_0\otimes B_0 + A_0\otimes B_1 + A_1\otimes B_0 - A_1\otimes B_1,
\end{align}
with the constraint value $\constr$ being the CHSH value. (As previously mentioned, this can be implemented in the formulation where are 4 constraint operators $\Gamma_{x,y}(\thA,\thB)$ by simply restricting to Lagrange-multiplier choices of the form $(\lambda_{00},\lambda_{01},\lambda_{10},\lambda_{11}) = (\lambda,\lambda,\lambda,-\lambda)$.) Each choice of the associated Lagrange multiplier $\lambda$ yields an affine lower bound of the form ${\lambda}{\constr} + c_{{\lambda}}$, as noted in Eq.~\eqref{eq:duallin}.
Importantly, some heuristic computations (also observed in~\cite{SGP+21} for the $\p=0$ case) suggest that the true bound $\breve{\lin}_\p({\constr})$ in this situation is in fact affine over a wide range of CHSH values --- we show this in Fig.~\ref{fig:entbnds}, which displays the results of heuristic minimizations compared to our certified bound in some cases. In particular, this range on which the bound is affine covers all currently experimentally reasonable values. 
This has the implication that there is a single ${\lambda}$ that yields an affine lower bound which is equal to $\breve{\lin}_\p({\constr})$ (i.e.~it is tight) over this entire range; specifically, it is simply the value of ${\lambda}$ corresponding to the gradient of $\breve{\lin}_\p({\constr})$ in this range. This greatly simplifies our task since we only need to solve the optimization for this specific value of ${\lambda}$.

\begin{figure}
\centering
\includegraphics[width=0.6\textwidth]{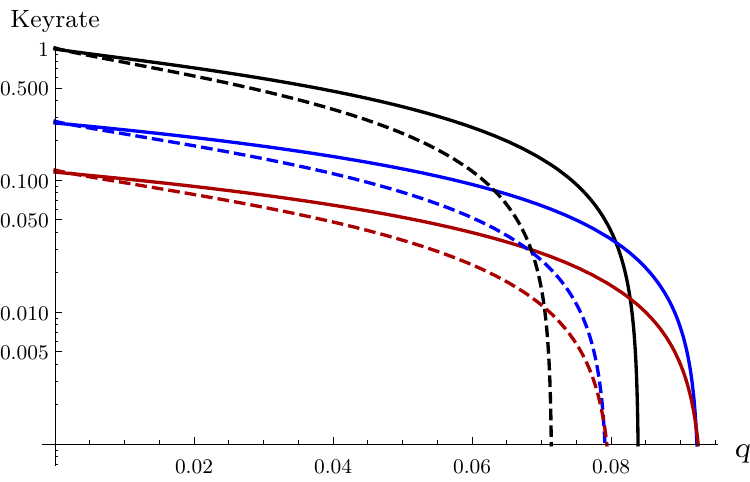}
\caption{
Lower bounds on asymptotic keyrates under depolarizing noise $\q$ (defined in Sec.~\ref{sec:hon}), on a vertical log scale. The black, blue, and red solid curves show the (net) asymptotic keyrate of Protocol~\ref{prot:preshared} (defined in Sec.~\ref{sec:preshared})\protect\footnotemark, for noisy-preprocessing values of $\p=0$, $0.2$ and $0.3$ respectively, based on the bounds we computed (Eq.~\eqref{eq:certbnds}). 
For comparison, the dashed curves show the corresponding asymptotic keyrates of the protocol in~\cite{HST+20}, which does not use the random-key-measurement method (the $\p=0$ case is equivalent to the~\cite{PAB+09} protocol). The solid curves intersect the horizontal axis at $\q=8.39\pct$, $9.26\pct$ and $9.33\pct$, in order of increasing $\p$. The first value is a minor improvement over~\cite{SGP+21} (despite being effectively the same protocol), likely because our algorithm provably converges to a tight keyrate bound (for fixed $\p$). The last value exceeds all previous bounds~\cite{HST+20,SGP+21,WAP21,SBV+21} for depolarizing-noise tolerance, by an amount comparable to their respective improvements over the original~\cite{PAB+09} protocol. It is also close to the upper bound of $9.57\pct$ that we derive in Sec.~\ref{sec:bestbnd}.}
\label{fig:keyrates}
\end{figure}
\footnotetext{The asymptotic keyrates of Protocol~\ref{prot:DIQKD} would be given by simply halving these keyrates, which does not change their horizontal intercepts.}

We focused on several values of noisy preprocessing, ranging from $\p=0$ to $\p=0.45$.
In each case, we first solved the minimization~\eqref{eq:intercept} heuristically for some selection of values of ${\lambda}$, in order to estimate the choice of $\lambda$ that yields a tight bound over the range of CHSH values in which $\breve{\lin}_\p({\constr})$ is affine. Then using our algorithm to get certified bounds on the corresponding $c_{\lambda}$ in~\eqref{eq:intercept}, we arrived at the final bounds
\begin{align}
\begin{gathered}
\breve{\lin}_{0}({\constr}) \geq 1.190(\constr-2) - 0.00454, \qquad
\breve{\lin}_{0.2}({\constr}) \geq 0.327(\constr-2) + 0.72063, \\
\breve{\lin}_{0.3}({\constr}) \geq 0.139(\constr-2) + 0.88051, \qquad
\breve{\lin}_{0.4}({\constr}) \geq 0.0341(\constr-2) + 0.97055, \\
\breve{\lin}_{0.45}({\constr}) \geq 0.00855(\constr-2) + 0.992487.
\end{gathered}
\label{eq:certbnds}
\end{align}
(To express the bounds as functions of winning probability $w$ instead, as we required for our various keyrate computations, simply replace $\constr$ in each expression with $8w-4$.)
With these bounds, we computed the asymptotic keyrates for these noisy-preprocessing values, and we show the results up to $\p=0.3$ in Fig.~\ref{fig:keyrates}. From the $\p=0.3$ case, we obtain the depolarizing-noise threshold of $9.33\pct$ mentioned in the introduction. However, we were unable to obtain better thresholds using the higher values of $\p$, for reasons we shall soon discuss (in Sec.~\ref{sec:bestbnd} below).

\begin{figure}
\centering
\includegraphics[width=0.6\textwidth]{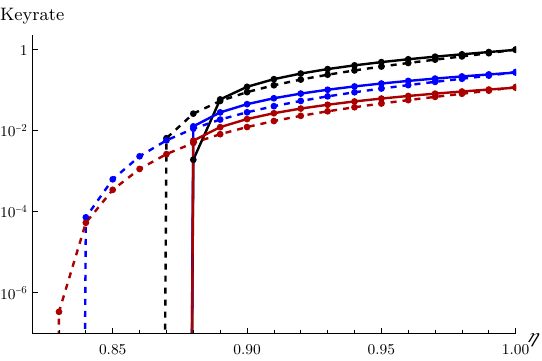}
\caption{Lower bounds on asymptotic keyrates for limited detection efficiency $\eta$ (defined in Sec.~\ref{sec:hon}), on a vertical log scale. The black, blue, and red solid curves show the (net) asymptotic keyrate of Protocol~\ref{prot:preshared} (defined in Sec.~\ref{sec:preshared}), for noisy-preprocessing values of $\p=0$, $0.2$ and $0.3$ respectively, based on the bounds we computed (Eq.~\eqref{eq:certbnds}). 
At each value of $\eta$, the states and measurements were optimized to maximize the keyrate.
For comparison, the dashed curves show the corresponding optimized asymptotic keyrates of the protocol in~\cite{HST+20}, which does not use the random-key-measurement method (the $\p=0$ case is equivalent to the~\cite{PAB+09} protocol). In contrast to the depolarizing-noise case (Fig.~\ref{fig:keyrateseta}), the threshold $\eta$ value required for nonzero keyrates is significantly worse for our protocol than the~\cite{HST+20} protocol (although our protocol does yield slightly better keyrates at higher values of $\eta$).
}
\label{fig:keyrateseta}
\end{figure}

For photonic experiments, we consider the~\cite{Ebe93} model described in Sec.~\ref{sec:hon} earlier, 
and the resulting keyrates for noisy-preprocessing values up to $\p=0.3$ are shown in Fig.~\ref{fig:keyrateseta} (as in the depolarizing-noise case, higher values of $\p$ gave somewhat worse results, for similar reasons that will be discussed in Sec.~\ref{sec:bestbnd}). As mentioned previously, in this scenario, for each value of $\eta$ we have to optimize the honest states and measurements to maximize the keyrates. Since this optimization was performed heuristically and appears to be slightly numerically unstable, we are not completely certain whether these are indeed the best keyrates that could be achieved. However, for the values that we managed to obtain at least, it appears that unfortunately our protocol with $\keyw_0=\keyw_1=1/2$ has worse detection-efficiency thresholds (for nonzero asymptotic keyrate) as compared to the results in~\cite{HST+20} based on noisy preprocessing alone, although at higher values of $\eta$ our protocol can yield slightly better keyrates.\footnote{Regarding the random-key-measurements technique alone, the results in~\cite{SGP+21} indicate that it does not significantly affect the detection-efficiency threshold as compared to the basic~\cite{PAB+09} protocol with optimized states and measurements, although it does somewhat improve the keyrates. We note that this comparison basically corresponds to the black curves in Fig.~\ref{fig:keyrateseta} here, although the photonic model used in~\cite{SGP+21} is a more detailed one~\cite{TWF+18} compared to the model here. Focusing on those curves in Fig.~\ref{fig:keyrateseta}, we see that our findings here are roughly similar as well (in our case the random-key-measurements technique yields a slightly worse threshold, but this may just be an artifact of having considered a coarser set of data points to evaluate).}

The comparatively poor detection-efficiency thresholds for our protocol in this model seems to be because of the error-correction value $h_\mathrm{hon}$ --- we find that for instance, the states and measurements that optimize the CHSH value yield a rather high value for at least one of the two terms $H(A_j | B_j;X_j = Y_j = 0)_\mathrm{hon}$ and $H(A_j | B_j;X_j = Y_j = 1)_\mathrm{hon}$ (see~\eqref{eq:ECrate}) describing the error-correction ``contributions'' from the two generation inputs. Therefore, there is a significant tradeoff between maximizing the CHSH value (and hence the entropy against Eve) versus minimizing the error-correction value $h_\mathrm{hon}$, which results in worse detection-efficiency thresholds as compared to the~\cite{HST+20} protocol which only used one input for key generation. Potential points for further keyrate improvements in the photonic model would hence be to optimize the ratio of the generation input choices or to introduce different amounts of noisy preprocessing for the two inputs, but such optimization is rather involved and we leave it for further work.

In principle, one could consider more detailed models of photonic implementations, for instance those developed in~\cite{TWF+18,HST+20} (such an analysis was performed in~\cite{SGP+21} for the random-key-measurements technique). However, since informally speaking these models account for a range of other imperfections not considered in the simple model here, they typically result in lower keyrates than the simple model. Given that the results shown here are not entirely promising, there may be limited prospects in pursuing that direction further unless we can find some improvements within the simpler model here.

\subsection{Optimality of results}
\label{sec:bestbnd}

For each of the bounds in Eq.~\eqref{eq:certbnds}, there is a feasible point of the optimization for $c_{\lambda}$ which yields a value within $0.005$ (or less, for higher values of $\p$) of the certified results shown in Eq.~\eqref{eq:certbnds}, so these bounds on entropy are very close to optimal in terms of absolute error.
In terms of the depolarizing-noise thresholds that they yield, taking the convex envelope of some of the feasible points shown in Fig.~\ref{fig:entbnds}
yields the result that the thresholds for $\p=0.2$ and $\p=0.3$ cannot be improved by more than about $0.1$ percentage points, so those thresholds are very close to optimal as well. 
However, larger values of $\p$ face the issue that the asymptotic keyrate becomes extremely low, which makes the horizontal intercept (i.e.~the depolarizing-noise threshold) very sensitive to changes in $\breve{\lin}_{\p}$ --- even a small absolute error in this bound results in a significant change in the threshold value. Therefore, the thresholds we obtained from the certified bounds with $\p=0.4$ and $\p=0.45$ in Eq.~\eqref{eq:certbnds} were only $9.32\pct$ and $9.10\pct$ respectively, worse than the results for $\p=0.3$. Heuristic computations suggest that the true thresholds for those cases might be approximately $9.46\pct$ and $9.50\pct$ respectively, but using our algorithm to certify these values would require it to converge to tolerances that currently appear impractical. Hence a different approach may be needed to find the true thresholds for these values of $\p$. From a practical perspective though, such improvements may be of limited use, because the very low asymptotic rates mean that the finite-size keyrate would likely be zero until extremely large sample sizes.

In any case, we note that for depolarizing noise at least, the threshold value cannot be improved much further by \emph{any} protocol choices within the framework we have presented in this section, e.g.~by using the full distribution as constraints (which also encompasses the use of modified CHSH inequalities~\cite{WAP21,SBV+21}), or adjusting the values of $\keyw_x$. This is essentially because our bounds are very close to the linear interpolation between the points $(2,\binh(\p))$ and $(2\sqrt{2},1)$ (as can be seen from Fig.~\ref{fig:entbnds}). Intuitively speaking, the bound on the entropy against Eve in the depolarizing-noise scenario cannot exceed this linear interpolation (because Eve can always perform classical mixtures of strategies in order to attain every point on this linear interpolation), which means that our bounds are close to the highest bounds that are even possible in principle.

Making this reasoning quantitative, we can obtain an explicit upper bound on the depolarizing-noise threshold. (We highlight that because we will do so by constructing an extremely generic attack, this upper bound holds for \emph{all} protocols of this form, regardless of choices of parameters such as the input distributions and noisy preprocessing. To some extent, it is surprising that there even exist any protocols that achieve thresholds close to the upper bounds implied by such a generic attack.) We first recall that (as defined in Sec.~\ref{sec:hon}) the measurement statistics in the depolarizing-noise model are given by real projective measurements on the Werner state $\varrho_\q \defvar (1-2\q)\pure{\Phi^+} + 2\q\, \id/4$. It is known~\cite{AGT06,Kri79} that these measurement statistics can be reproduced by a local-hidden-variable (LHV) model as soon as the state does not violate the CHSH inequality, i.e.~for $q\geq q_2 \defvar{(2-\sqrt{2})}/{4} \approx 14.6\pct$. (In fact, this is also true in a more general context where the honest parties have \emph{any} number of real projective measurements on the Werner state, via the value of the second Grothendieck constant $K_G(2)=\sqrt{2}$~\cite{AGT06,Kri79}. For general (i.e.~not necessarily real) projective measurements the analogous value is known to satisfy $\q_3 \lesssim 15.9\pct$, following from the best known bound on the third Grothendieck constant $K_G(3)$~\cite{HQV+17}.) 

In the device-independent setting, the existence of an LHV model yields an attack for Eve that gives her full knowledge of the measurement outcomes.
This means that for any depolarizing-noise value $\q$, we can construct the following attack for Eve: first, she generates a classical ancilla bit which is equal to $0$ with probability $\pBell \defvar (\q_2-\q)/\q_2$. If the bit is equal to $0$, Alice and Bob's devices simply perform the honest measurements on the maximally entangled state $\Phi^+$. Otherwise, the devices implement an LHV model that yields the same statistics as the honest measurements performed on the Werner state $\varrho_{\q_2}$, but which gives Eve full knowledge of the outcomes. This attack indeed reproduces the statistics corresponding to depolarizing noise~$\q$, since $\pBell\pure{\Phi^+} + (1-\pBell) \varrho_{\q_2} = \varrho_{\q}$ and the second strategy produces the same statistics as the honest measurements performed on $\varrho_{\q_2}$.

In the two cases, the entropies of Alice's outputs after noisy preprocessing are $H(\hat{A}_x|\tilde{E})_\mathrm{triv}=1$ (i.e.~Eve's side-information is trivial since the Alice-Bob state is pure) and $H(\hat{A}_x|\tilde{E})_\mathrm{LHV}=\binh(\p)$ (i.e.~the uncertainty arises purely from the noisy preprocessing) respectively, where $\tilde{E}$ denotes Eve's side-information excluding the ancilla bit. Incorporating the ancilla bit into Eve's side-information $E$, this attack achieves
\begin{align}
H(\hat{A}_x|E) = \pBell H(\hat{A}_x|\tilde{E})_\mathrm{triv} + (1-\pBell) H(\hat{A}_x|\tilde{E})_\mathrm{LHV} 
= \frac{\q_2-\q}{\q_2} + \frac{\q}{\q_2} \binh(p).
\end{align}
This expression hence serves as an upper bound on the best possible lower bound we could derive on the conditional entropy against Eve. Also, the conditional entropy against Bob's output (for Bob's optimal key-generation measurements on Werner states) is simply
\begin{align}
H(\hat{A}_x|B_x) = \binh(\p +(1-2\p) \q).
\label{eq:entBob}
\end{align}
Recalling that the asymptotic keyrate is given by the Devetak-Winter bound~\cite{DW05} (up to the sifting-related weights $\keyw_x$~\cite{SGP+21}), this yields a simple upper bound on the critical value of $\q$ that still allows a positive asymptotic keyrate (for protocols of this form, i.e.~applying noisy preprocessing, random key measurements, and considering the full output distribution, but only using one-way error correction):
\begin{align}
\q_\textrm{att}(\p) \defvar& \max \bigg\{ \q \, \bigg| \, \sum_x \keyw_x (H(\hat{A}_x|E) - H(\hat{A}_x|B_x))  \geq 0 \bigg\} \nonumber\\
=& \max \left\{ \q \, \middle| \, \frac{\q_2-\q}{\q_2} + \frac{\q}{\q_2} \binh(\p) -  \binh(\p +(1-2\p) \q) \geq 0 \right\},
\end{align}
where in the first line the entropies refer to those of the attack we have described.
We observe numerically that this upper bound is increasing with respect to $\p$, so we have $\q_\textrm{att}(\p)\leq \q_\textrm{att}(\p\to1/2)$. In the limit $\p\to 1/2$ we can write $\p=1/2-\delta$ and expand the expression in small $\delta$ to find
\begin{equation}
\q_\textrm{att}\!\left(\p\to\frac{1}{2}\right) = \frac{1+ 4 \q_2 -\sqrt{8 \q_2+1}}{8 \q_2} = 1 - \frac{\sqrt{7+4\sqrt{2}} - 1}{2\sqrt{2}} \approx 9.57\pct.
\end{equation}
Hence for protocols of the form described in this work, it is not possible for the depolarizing-noise threshold to exceed this value. For the parameter choice $\p=0.3$, we have $\q_\textrm{att}(0.3) \approx 9.51 \pct$, which is close to the threshold of $9.33 \pct$ for which we could certify a positive keyrate.\footnote{These two thresholds cannot match exactly, because the feasible points shown in Fig.~\ref{fig:entbnds} also imply that the tight bound $\breve{\lin}_{\p}({\constr})$ is not \emph{exactly} equal to the linear interpolation between $(2,\binh(\p))$ and $(2\sqrt{2},1)$, which essentially corresponds to the attack we describe here.}

Finally, it is worth noting that while the main focus of this work is depolarizing noise applied to the statistics corresponding to the ideal CHSH measurements on $\Phi^+$, the above analysis in fact generalizes to a substantially larger family of scenarios. Specifically, the same analysis applies for depolarizing noise applied to the statistics from any number of real projective measurements on $\Phi^+$, since the results of~\cite{AGT06,Kri79} yield the required LHV models.\footnote{A more general consideration would be the scenario of depolarizing noise applied to the statistics from an arbitrary two-qubit state, but it is not immediately obvious whether the thresholds in~\cite{AGT06,Kri79,HQV+17} apply to such states as well, so we leave this for future work.} (If Bob performs suboptimal generation measurements, then Eq.~\eqref{eq:entBob} still holds as a lower bound, so the argument carries through.) In addition, replacing $\q_2$ with $\q_3$ straightforwardly provides a threshold of $\q \lesssim 10.01 \pct$ for all possible projective measurements on that state, via the respective known bound in~\cite{HQV+17}. (Note that in all such cases where there are more than 2 possible measurements, there is no ``qubit reduction'' to make it easier to derive corresponding lower bounds, and hence the lower bounds have also not been very thoroughly explored.) An interesting further consideration is the case of general POVM measurements. Here, a rather loose bound $q_\textrm{POVM} \lesssim 27.3\pct$ is also known~\cite{HQV+17} for the threshold that allows an LHV description. However, pure POVMs on qubits can involve up to four outcomes, opening up the possibility of more general noisy preprocessing --- all doubly stochastic maps on probability vectors with four outcomes. We thus leave it for future work.

The above argument relies on the fact that depolarizing-noise statistics can be obtained by simple mixtures of ``extremal'' strategies. This is not necessarily the case for other noise models, such as limited detection efficiency in photonic experiments. Therefore, the same argument cannot be directly applied to obtain upper bounds on the thresholds for such forms of noise.

\section{Possible modifications}
\label{sec:mods}

\subsection{Coordinating input choices by public communication}
\label{sec:pubchann}

The random-key-measurement protocol has the drawback that the keyrate is effectively halved, since the generation rounds have ``mismatched'' inputs approximately half the time. It would be helpful to find a way to work around this issue. One possible approach could be to observe that in~\cite{ARV19}, it was assumed that the following operations can be performed in each round: the devices receive some shares of a quantum state, then\footnote{The assumption being made here is that since the quantum states are now entirely in the possession of Alice and Bob, the test/generation decision can no longer affect the distributed state --- if the distributed state could depend on whether it is a test or generation round, the protocol would be trivially insecure.} Alice and Bob publicly communicate to come to an agreement on \emph{both} of their input choices, and finally they supply these inputs to their devices. (This was necessary in~\cite{ARV19} because 
Alice and Bob's actions in that DIQKD protocol require both of them to know whether it is a test or generation round. In fact, our analysis can be viewed as the first EAT-based security proof for a ``genuinely sifting-based'' DIQKD protocol, in the sense that Alice and Bob do not coordinate which rounds are test rounds, and simply choose their inputs independently.)
If we assume that this is also possible in our scenario, then Alice and Bob could coordinate their inputs in the generation rounds instead of choosing them independently, thereby avoiding the sifting factor.

Unfortunately, it does not seem clear if such a proposal is entirely plausible in near-term experimental implementations. This is because it relies on the devices being able to store the quantum state for long enough for Alice and Bob to agree on their choice of inputs,
which is potentially challenging for current Bell-test implementations. As an alternative, we propose the following potential modification to the DIQKD protocol in~\cite{ARV19} --- instead of agreeing on the test rounds via public communication, Alice and Bob could use a small amount of pre-shared key to choose which rounds are test rounds, in the same way as in DIRE (for details on the amount of pre-shared key required, see the DIRE protocols in~\cite{ARV19,BRC20} or the discussion in Sec.~\ref{sec:preshared} below). This approach would essentially be a ``key {expansion}'' protocol that requires a small amount of pre-shared key to initialize. We remark that this is not a dramatic change in perspective, because a common method to authenticate channels (namely, message authentication codes) relies on having a small amount of pre-shared key, so the assumed existence of authenticated channels in the DIQKD protocol is likely to require some pre-shared key in any case. 

However, this basic notion cannot immediately be generalized to Protocol~\ref{prot:DIQKD} here, since requiring Alice and Bob to choose uniformly distributed ``matching'' inputs in the generation rounds would require a large amount of pre-shared key (roughly $(1-\gamma)n$ bits). Fortunately, in the following section, we propose a variation which overcomes this difficulty by ``recovering'' the entropy in the pre-shared key, thereby still achieving net key expansion.

\subsection{Protocol using pre-shared key}
\label{sec:preshared}

Here we describe a variant protocol that avoids the sifting factor \emph{without} requiring the brief quantum storage described above, through the use of a fairly long pre-shared key. The limitation of this variant is that the net increase in secret key is just a (constant) fraction of the amount of pre-shared key; however, the \emph{rate} of net key generation does not have the sifting factor of $1/2$.
Informally, the idea is to simply use the pre-shared key as Alice's input string $\str{X}$, which allows Bob to choose his generation inputs to match Alice's. Just as importantly, this also allows them to (almost) entirely omit the public announcement of their inputs --- hence $\str{X}$ remains private, and with some care it can be incorporated into the final key without losing the entropy it ``contains''. 

We now describe this idea in detail as Protocol~\ref{prot:preshared} below, followed by its security proof. The protocol supposes that Alice and Bob hold a pre-shared (uniform) key of $n$ bits, which we shall simply denote as $\str{X}$, since it will be exactly the string that Alice uses as her inputs. The appropriate value of $\lkey$ to choose will be described later in Theorem~\ref{th:preshared}. 
\begin{savenotes}
\begin{breakablealgorithm}
\caption{} 
\label{prot:preshared}
This protocol proceeds the same way as Protocol~\ref{prot:DIQKD}, except for the following changes:
\begin{itemize}
\item In each round, Alice's input $X_j$ is determined from the pre-shared key $\str{X}$, instead of being generated randomly in that round. Bob's input $Y_j$ is generated as follows: with probability $\gamma$ he chooses a uniformly random $Y_j \in \inYt$, otherwise Bob chooses $Y_j = X_j$. In addition, he generates another register $\Yt_j$ which equals $Y_j$ when $Y_j \in \inYt$ and equals $\perp$ otherwise.
\item Alice and Bob do not publicly announce the strings $\str{X}\str{Y}$. Instead, Bob only announces the string $\Ytstr$.\footnote{It might be possible to consider a slight variant which omits this step. However, knowing $\Ytstr$ allows Alice to compute $\str{Y}$, which may be relevant for error correction since it allows Alice to distinguish the test and generation rounds. In any case, it seems unclear whether the entropy of $\Ytstr$ can be usefully extracted even if it is kept secret, since Alice does not have access to it in that case.} Additionally, the sifting step is unnecessary, since there will be no rounds such that $Y_j \in \inYg$ and $X_j \neq Y_j$.
\item Privacy amplification is performed on the strings $\str{A}\str{X}$ and $\tilde{\str{A}}\str{X}$ instead of $\str{A}$ and $\tilde{\str{A}}$.
\end{itemize}
\end{breakablealgorithm}
\end{savenotes}
To prove the security of this protocol, we can simply follow almost exactly the same security proof as for Protocol~\ref{prot:DIQKD}, with some changes we shall now describe. Firstly, the value of $h_\mathrm{hon}$ (to be used when computing $\cmax$) is replaced by
\begin{align}
\tilde{h}_\mathrm{hon} 
= H(A_j|B_jX_jY_j)_\mathrm{hon} 
&= \frac{1-\gamma}{2} 
\sum_{z\in\inX} H(A_j | B_j;X_j = Y_j = z)_\mathrm{hon}  
\nonumber \\& \qquad 
+ \frac{\gamma}{4} 
\sum_{x\in\inX,y\in\inYt} 
H(A_j| B_j ;X_j = x, Y_j = y)_\mathrm{hon}, \label{eq:ECpreshared}
\end{align}
since the probabilities of $X_j = Y_j = z$ for $z\in\inX$ are now ${(1-\gamma)}/{2}$. (Note that no error-correction information needs to be sent from Alice to Bob regarding $\str{X}$, since both of them have a copy of that string.)

Also, since the strings used in the privacy-amplification step are now $\str{A}\str{X}$ and $\tilde{\str{A}}\str{X}$, this means that we need an equivalent of Eq.~\eqref{eq:EATbound}, with $\Hmin^{\es'}(\str{A}\Btstr\str{X}|\Ytstr E)$ in place of $\Hmin^{\es'}(\str{A}\Btstr | \str{X}\str{Y}E)$. To obtain this, we note that we can simply construct a virtual protocol in the analogous way to Protocol~\ref{prot:virtual}, then consider the same EAT channels $\map_j$ as before, but instead we shall identify $\Ct_j$ with $D_j$, $A_j\Bt_jX_j$ with $S_j$, and $\Yt_j$ with $T_j$ in Definition~\ref{def:EATchann}.
The Markov conditions are again fulfilled, since $\Yt_j$ is generated by trusted randomness in each round and independent of all previous data. To find an appropriate min-tradeoff function for these channels, we note that the output $(\map_j\otimes\idmap_{R})(\omega_{R_{j-1}R})$ of channel $\map_j$ always satisfies
\begin{align}
H(X_j | \Yt_j R) = H(X_j) = 1,
\end{align}
because $X_j$ is produced by trusted randomness independent of $\Yt_j R$. Therefore, we can use the chain rule to write
\begin{align}
H(A_j \Bt_j X_j | \Yt_j R) 
&= H(X_j | \Yt_j R) + H(A_j \Bt_j | X_j \Yt_j \Fj R) \nonumber \\ 
&\geq 1 + (1-\gamma) \lin_\p(w) + \gamma \lin_0(w) \nonumber\\
&\defvar 1 + \tilde{\g}(w), 
\label{eq:chaintrick}
\end{align}
where the function $\tilde{\g}$ (in contrast to $\g$) does {not} have the factor of $1/2$ introduced by sifting, since Alice does not ``erase'' the outputs of any rounds. We can thus construct a new min-tradeoff function $\tilde{f}_\mathrm{min}$ in the same way as in Sec.~\ref{sec:fmin}, but using $1 + \tilde{\g}(w)$ in place of $\g(w)$.\footnote{We remark that if we think of this replacement as happening in two steps, first replacing $\g$ by $\tilde{\g}$ and then adding a ``constant offset'' of $1$, then the latter has no effect on $\Var_{\mathcal{Q}_{f}}(\fmin)$ or the difference $\Max(\fmin)-\Min_{\mathcal{Q}_{f}} (\fmin)$, and hence does not change the finite-size correction to the keyrate except indirectly via changing the system dimensions and the range of $\cperp$. However, the first step of replacing $\g$ by $\tilde{\g}$ does slightly increase the finite-size correction (since $\tilde{\g}$ has a somewhat larger range).} The rest of the proof then proceeds as before, leading to the following security statement:
\begin{theorem}\label{th:preshared}
Protocol~\ref{prot:preshared} 
has the same security guarantees as those described in Theorem~\ref{th:DIQKD},
except with the following changes (with $\tilde{\g}$ being defined in~\eqref{eq:chaintrick}):
\begin{itemize}
\item $\cperp$ is chosen to be in 
$[1+\tilde{\g}(0),1+\tilde{\g}(1)]$ 
instead.
\item $h_\mathrm{hon}$ is replaced by $\tilde{h}_\mathrm{hon}$ as specified in Eq.~\eqref{eq:ECpreshared}.
\item In Eq.~\eqref{eq:keylength} for $\lkey$, $\g(\wexp-\dtol)$ is replaced by $1 + \tilde{\g}(\wexp-\dtol)$, and the values of $V$ and $K_\alpha$ are replaced by 
\begin{align}
\begin{aligned}
\tilde{V} &\defvar \sqrt{\Var_{\mathcal{Q}_{f}}(\tilde{f}_\mathrm{min})+2} + \log
129,
\\
\tilde{K}_\alpha &\defvar \frac{2^{(\alpha-1)(2\log8 + \Max(\tilde{f}_\mathrm{min})-\Min_{\mathcal{Q}_{f}} (\tilde{f}_\mathrm{min}))} }{6(2-\alpha)^3\ln2}
\ln^3\left(2^{2\log8 + \Max(\tilde{f}_\mathrm{min})-\Min_{\mathcal{Q}_{f}} (\tilde{f}_\mathrm{min})} + e^2\right),
\end{aligned}
\end{align}
where $\tilde{f}_\mathrm{min}$ is a function that satisfies
\end{itemize}
\vspace*{-.5cm}
\begin{align}
\begin{gathered} 
\Max(\tilde{f}_\mathrm{min})
= 1 + \frac{1}{\gamma}\tilde{\g}(1) + \left(1-\frac{1}{\gamma}\right)\cperp, \qquad
\Min_{\mathcal{Q}_{f}} (\tilde{f}_\mathrm{min}) 
= 1 + \tilde{\g}\!\left(\frac{2-\sqrt{2}}{4}\right), \\
\Var_{\mathcal{Q}_{f}}(\tilde{f}_\mathrm{min}) \leq 
\frac{2-\sqrt{2}}{4\gamma} 
\min \left\{\Delta_0^2, \Delta_1^2 \right\}
+ \frac{2+\sqrt{2}}{4\gamma} 
\max \left\{\Delta_0^2, \Delta_1^2 \right\}
, \text{ where } \Delta_w \defvar \cperp - 1 - \tilde{\g}(w).
\end{gathered}
\end{align}
\end{theorem}

Overall, recalling that Protocol~\ref{prot:DIQKD} required $n$ bits of pre-shared key, we see that the \emph{net} gain of secret key bits in Protocol~\ref{prot:preshared} 
is larger than that of Protocol~\ref{prot:DIQKD} by a factor of approximately (ignoring the changes to the finite-size corrections)
\begin{align}
\frac{n \left(\tilde{\g}(\wexp-\dtol) - \tilde{h}_\mathrm{hon}\right)}{n \Big(\g(\wexp-\dtol) - h_\mathrm{hon}\Big)} \approx 2
,
\end{align}
since it avoids the sifting factor. 
Informally, by keeping $\str{X}$ secret and incorporating it in the privacy amplification step, we have ``recovered'' the entropy that was present in the original pre-shared key.\footnote{Note that the underlying idea here critically relies on the fact that the bound $\lin_p$ is for the entropy of the output strings \emph{conditioned on $X_j$}. This allowed us to use the chain rule to obtain Eq.~\eqref{eq:chain}, which eventually led to the result that the seed entropy contained in $\str{X}$ simply ``adds on'' to the entropy of the output strings in the original Protocol~\ref{prot:DIQKD}. If $\lin_p$ had been a bound on, for instance, 
$H(A_j | \Yt_j R)$ instead of $H(A_j| X_j \Yt_j R)$, 
this argument would not have worked.} 

In practice, including the string $\str{X}$ in privacy amplification 
essentially doubles the input size for the hash function in that step, which raises its computational difficulty substantially (though not insurmountably).
One might wonder whether it would be possible to bypass this aspect --- for instance, by simply performing privacy amplification on $\str{A}$ and $\tilde{\str{A}}$ as before, then appending $\str{X}$ to the output. At first glance, this approach might appear plausible, since $\str{X}$ is not announced in Protocol~\ref{prot:preshared}. Unfortunately, it seems unclear how to certify that the publicly communicated error-correction string $\str{L}$ is independent of $\str{X}$ (in fact, it seems unlikely that this is true). Hence the idea of simply appending $\str{X}$ may not be secure. By instead incorporating it in privacy amplification in the specified manner, Protocol~\ref{prot:preshared} ensures that the entropy of $\str{X}$ is securely ``extracted'' into the final key.

As previously mentioned, the net increase in secret key given by one instance of Protocol~\ref{prot:preshared} is limited to a fraction of the amount of pre-shared key. 
However, it is possible in principle to recursively run Protocol~\ref{prot:preshared} in order to achieve unbounded key expansion --- one can use the key generated by one instance of Protocol~\ref{prot:preshared} to run it again with a longer pre-shared key and larger $n$ (since the security definition is composable, the soundness parameter 
will only increase 
additively in this process~\cite{PR14}). We stress that in doing so, one must always incorporate the seed into the privacy-amplification step exactly as specified in Protocol~\ref{prot:preshared} --- in particular, this means that the \emph{entire} key changes with every iteration, instead of simply having some new bits appended.
Some care is necessary regarding device memory across instances of this recursive procedure --- while it does not seem to be directly vulnerable to the memory attack of~\cite{BCK13}\footnote{This is because the only public communication in Protocol~\ref{prot:preshared} that can leak any information is the error-correction string 
(all other public communication is based on trusted randomness). In our security proof, we have bounded the min-entropy leakage at this step simply via the length of this string, without any assumptions about its structure, and hence we can still obtain a secure bound on the min-entropy of the input for privacy amplification in the final protocol instance. Note that this claim is strictly restricted to device reuse following the recursive process specified here --- once any key bits have been used for any other purpose, the attack again becomes a potential concern if the devices are reused.}, it is still important to ensure that the states measured in each instance of the protocol are independent of the key generated in the preceding instance, since this key is used to choose the device inputs (which must be independent of the state in order for our security arguments to hold). Again, this relies on the notion that the registers measured by the devices do not contain information about the key generated in the preceding instance.

There is another potential variant of this idea where a pre-shared key is instead used to generate \emph{both} input strings $\str{X}$ and $\str{Y}$, and the input-choice announcement is omitted entirely, with privacy amplification being performed on $\str{A}\str{X}\str{Y}$ and $\tilde{\str{A}}\str{X}\str{Y}$. This can be done by using $n$ bits to set the value of $X_j$ in all rounds as before, then using $\kappa \binh(\gamma) n$ bits to choose Bob's test rounds approximately according to the desired IID distribution, and finally using $\kappa' \gamma n$ bits to set the value of $Y_j$ in the test rounds (while the generation rounds simply have $Y_j$ set to $X_j$), where $\kappa,\kappa'>1$ are constants that can be chosen such that the approximations to the desired distributions are sufficiently accurate (see the randomness-expansion protocols in~\cite{ARV19,BRC20} for a more complete description of this process based on the \emph{interval algorithm}).\footnote{We break up the use of the seed into separate processes because it allows for better efficiency as compared to directly approximating the desired distribution of $\str{X}\str{Y}$ --- with the approach we describe, the ``inefficiency'' prefactors $\kappa,\kappa'$ of the interval algorithm only appear on the $\binh(\gamma)n,\gamma n$ terms instead of the full entropy of $\str{X}\str{Y}$.} This would hence require $(1 + \kappa \binh(\gamma) + \kappa' \gamma)n$ bits of seed randomness. A similar argument as above could then be performed by noting that (for $X_jY_j$ generated according to the ideal distribution) we have $H(X_j Y_j) 
= 1 + \binh(\gamma) + \gamma
$, so most of the seed entropy can be ``recovered'', up to the losses from the $\kappa,\kappa'$ factors. However, tracking the effects of using the interval algorithm to approximate the ideal distribution is cumbersome (albeit possible; see~\cite{BRC20}), and it is unclear if this variant offers any immediate advantage over Protocol~\ref{prot:preshared} for DIQKD --- though it may be useful for protocols that use other Bell inequalities or non-uniform input distributions, as mentioned in Sec.~\ref{sec:nonCHSH}.

On the other hand, it appears that this variant may have potential for the purposes of DIRE instead. (As mentioned in the introduction, this idea has also been independently proposed in~\cite{arx_BRC21}.) The main reason why the random-key-measurement approach in~\cite{SGP+21} could not be easily generalized to DIRE is that in order for Alice to select a uniformly random input in every round, she requires a (local) source of $n$ random bits, which is a free resource in DIQKD but not in DIRE --- if a proposed DIRE protocol consumes more random bits than it produces, then it has failed to achieve randomness \emph{expansion}\footnote{There is, however, the related but distinct task of device-independent randomness \emph{generation}~\cite{ZSB+20} (where the goal is to produce private randomness from an unbounded supply of public randomness that is independent from the devices), in which this would not be an issue.}. However, the protocol proposed in this section has the property that it ``recovers'' the entropy contained in the seed, which means that one can afford to use much larger seeds while still obtaining a net increase in secret key. Explicitly, the application of this idea to DIRE would hence be as follows: one begins with $2n$ uniformly random bits, which are then used as the input strings\footnote{For DIRE based on the CHSH inequality, Bob only requires two possible measurements instead of the four required for the DIQKD protocol here.} $\str{X}\str{Y}$ to the devices over $n$ rounds to obtain outputs $\str{A}\str{B}$. Modelling this process using EAT channels in a manner similar to above (see e.g.~\cite{LLR+21} for details), for each round we would have 
\begin{align}
H(A_j B_j X_j Y_j| R) 
&= H(X_j Y_j | R) + H(A_j B_j | X_j Y_j R) \nonumber \\ 
&= 2 + H(A_j B_j | X_j Y_j R),
\end{align}
which (given a bound on $H(A_j B_j | X_j Y_j R)$) allows one to bound the smoothed min-entropy of $\str{A}\str{B}\str{X}\str{Y}$ conditioned on $E$. By performing privacy amplification on $\str{A}\str{B}\str{X}\str{Y}$, one ``recovers'' all the entropy in the seed, due to the $H(X_j Y_j | R)$ term in the above equation. Overall, this proposed protocol allows one to use the improved entropy rate provided by the random-key-measurement approach~\cite{SGP+21}, in the context of DIRE instead of DIQKD.

\subsection{Collective attacks}

As a reference to compare our results against, we could consider whether a longer secure key could be obtained under the \emph{collective-attacks} assumption, which is the assumption that the device behaviour is IID in each round of the protocol (though Eve can still store quantum information for arbitrary periods).\footnote{To be precise, we mean that the part of the state held by Alice and Bob's devices is IID across the rounds, and in each round the devices have the same set of possible measurements. Since all purifications are isometrically equivalent, without loss of generality we can suppose that Eve also holds an IID purification of the Alice-Bob state.} To this end, we derive a security proof under this assumption. We defer the proof to Appendix~\ref{app:collective}, and just state the final key length formula here. 
As compared to Theorem~\ref{th:DIQKD}, the parameters involved in the formula are somewhat different: some of the previous parameters are no longer involved, though there are some new ones, which we qualitatively describe as follows (they are closely related to the notion of ``$\eps$-secure filtering'' described in~\cite{rennerthesis}).
{\center
\setlist[description]{leftmargin=2.4cm,rightmargin=1cm,labelindent=5mm,itemsep=0mm}
\begin{description}
\item \makebox[10mm]{$\eps_\mathrm{IID}$}: Informally, a bound on the probability that the virtual parameter estimation step (in the virtual protocol) accepts when given devices that produce insufficient min-entropy
\item \makebox[10mm]{$\dsou$}: ``Relaxation'' parameter that slightly enlarges the set of states for which we bound the entropy, in order to derive a nontrivial value of $\eps_\mathrm{IID}$
\end{description}
}
\noindent The formal theorem statement is:
\begin{theorem}\label{th:collective}
Take any 
$\ecEC,\ecPE,\ePA,\eh,\es \in (0,1]$, $\gamma\in(0,1)$, $\p\in[0,1/2]$, and $\dsou\in[0,\wexp - \dtol)$.
Define
\begin{align}
\eps_\mathrm{IID} \defvar 
\cdfBin{n}{1-(\wexp - \dtol - \dsou)\gamma}{\floor{(1-(\wexp-\dtol)\gamma)n}}.
\end{align}
Under the collective-attacks assumption, Protocol~\ref{prot:DIQKD} is $(\ecEC + \ecPE)$-complete and $(\max\{\eps_\mathrm{IID}, \ePA + 2\es\} + 2\eh)$-sound when performed with any choice of $\cmax$ and $\dtol$ such that Eq.~\eqref{eq:comEC} and Eq.~\eqref{eq:ecPE} hold, and $\lkey$ satisfying
\begin{align}
\lkey \leq
n\g(\wexp-\dtol-\dsou) - \sqrt{n} \,(2\log5)\sqrt{\log\frac{2}{\es^2}}
- \cmax - \ceil{\log\left(\frac{1}{\eh}\right)} - 2\log\frac{1}{\ePA} + 2.
\label{eq:lkeycoll}
\end{align}
\end{theorem}

However, the above theorem is simply a statement for Protocol~\ref{prot:DIQKD} under the assumption of collective attacks, and that protocol does not fully exploit some implications of that assumption. For instance, in Theorem~\ref{th:collective} there is implicitly an $O(\gamma)$ subtractive penalty to the keyrates (which was also present in Theorem~\ref{th:DIQKD}\footnote{In fact, Theorem~\ref{th:DIQKD} has another $O(\gamma)$ subtractive penalty from the use of the bound~\eqref{eq:chain}, but this was due to the technical limitations of the EAT and the fact that the bounds~\eqref{eq:fmax} and \eqref{eq:gproof} are slightly suboptimal.}) caused by having to include the test-round data in the $\cmax$ term. Yet under the collective-attacks assumption, the test rounds are completely independent of the generation rounds, which implies that the effect of $\gamma$ should instead be to reduce the keyrate by a \emph{multiplicative} factor of $(1-\gamma)$. Importantly, in the latter case it is possible to choose arbitrarily large test probabilities $\gamma$ without necessarily making the keyrates negative, which can dramatically improve the statistical bounds for parameter estimation. To formalize this idea, we consider Protocol~\ref{prot:optcoll} below, which attempts to minimize the finite-size correction as much as possible using the most optimistic assumptions that have been discussed thus far.
\begin{savenotes}
\begin{breakablealgorithm}
\caption{} 
\label{prot:optcoll}
This protocol proceeds the same way as Protocol~\ref{prot:DIQKD}, except for the following changes:
\begin{itemize}
\item Instead of independently choosing whether each round is a test or generation round, Alice chooses a uniformly random subset of size $m$ as test rounds before the protocol begins, 
and we define $\gamma$ as the value $m/n$. Alice also prepares the strings $\str{X}\str{Y}$ in advance, by choosing $X_j=Y_j\in\inYg$ uniformly at random in the generation rounds, and choosing $X_j\in\inX, Y_j\in\inYt$ uniformly at random in the test rounds.
\item In each round, Alice and Bob briefly store their received quantum states instead of immediately measuring them. Alice then publicly announces $X_jY_j$, which Alice and Bob then use as the inputs to their devices.\footnote{Here, in our attempt to minimize the finite-size effects, we are following the~\cite{ARV19} assumption mentioned previously: Alice and Bob can briefly store their received quantum states, in a manner such that the public communication cannot affect the stored states.}
\item In the error-correction step, Alice does not send error-correction data (and a corresponding hash) for the full string $\str{A}$, but rather only the subset of it consisting of the generation rounds, denoted as $\str{A}_g$. Bob's guess for this string will be denoted as $\tilde{\str{A}}_g$. The values of $\str{A}$ in the test rounds, denoted as $\str{A}_t$, are sent directly to Bob without compression or encryption, and Bob uses this string for parameter estimation.
\item Bob's accept condition is instead to check that $\hash(\str{A}_g) = \hash(\tilde{\str{A}}_g)$ and $\freq_{\str{c}_t}(1)\geq \wexp-\dtol$ 
hold, where $\str{C}_t$ denotes the substring of $\str{C}$ corresponding to the test rounds (in particular, this means the frequencies are computed with respect to a string of length $\gamma n$, not $n$).
\item Privacy amplification is performed only on the strings $\str{A}_g$ and $\tilde{\str{A}}_g$.
\end{itemize}
\end{breakablealgorithm}
\end{savenotes}
\newpage 

For this protocol, the value of $\cmax$ is to be computed based only on the number of generation rounds, 
since error correction is performed on the string $\str{A}_g$ rather than $\str{A}$. Focusing on the best possible theoretical bounds from Sec.~\ref{sec:EC}, this means we take $\cmax$ to be given by Eq.~\eqref{eq:optEC} with
\begin{align}
\Hmax^{\ez}(\str{A}|\str{B}\str{X}\str{Y})_\mathrm{hon} \leq (1-\gamma)n 
h_\mathrm{hon}
+ \sqrt{(1-\gamma)n} \, (2\log5)\sqrt{\log\frac{2}{\ez^2}},
\label{eq:optcollAEP}
\end{align}
where
\begin{align}
h_\mathrm{hon} 
=
\sum_{z\in\inX} \frac{1}{2} H(A_j | B_j;X_j = Y_j = z)_\mathrm{hon}  ,
\end{align}
since the test rounds are excluded. With this value of $\cmax$ in mind, we can state the security guarantees of this protocol, with the proof given in Appendix~\ref{app:optcoll} 
(we highlight that the dependence on several security parameters here is different as compared to the previous theorems --- {qualitatively, this is because for instance we no longer consider a virtual protocol when introducing parameters such as $\eps_\mathrm{IID}$, and also the various binomial distributions are different since the test-round subset now has a fixed size}):
\begin{theorem}\label{th:optcoll}
Take any 
$\ecEC,\ecPE,\ePA,\eh,\es \in (0,1]$, $\gamma\in(0,1)$, $\p\in[0,1/2]$, and $\dsou\in[0,\wexp - \dtol)$.
Define
\begin{align}
\eps_\mathrm{IID} \defvar 
\cdfBin{\gamma n}{1-\wexp+\dtol+\dsou}{\floor{(1-\wexp+\dtol)\gamma n}}. \label{eq:eIIDoptcoll}
\end{align}
Under the collective-attacks assumption, Protocol~\ref{prot:optcoll} is $(\ecEC + \ecPE)$-complete and $(\max\{\eps_\mathrm{IID}, \ePA + 2\es\} + \eh)$-sound 
when performed with $\cmax$ defined in terms of $\ecEC$ as described above, 
and $\dtol,\lkey$ satisfying
\begin{align}
\ecPE &\geq \cdfBin{\gamma n}{\wexp}{\floor{(\wexp-\dtol)\gamma n}}, \label{eq:ecPEoptcoll}\\
\lkey &\leq
(1-\gamma)n\lin_\p(\wexp-\dtol-\dsou) - \sqrt{(1-\gamma)n} \,(2\log5)\sqrt{\log\frac{2}{\es^2}}
\nonumber \\ &\qquad 
- \cmax - \ceil{\log\left(\frac{1}{\eh}\right)} - 2\log\frac{1}{\ePA} + 2. \label{eq:lkeyoptcoll}
\end{align}
\end{theorem}

\begin{figure}
\centering
\subfloat[\cite{RBG+17} parameters, $\p=0$]{
\includegraphics[width=0.48\textwidth]{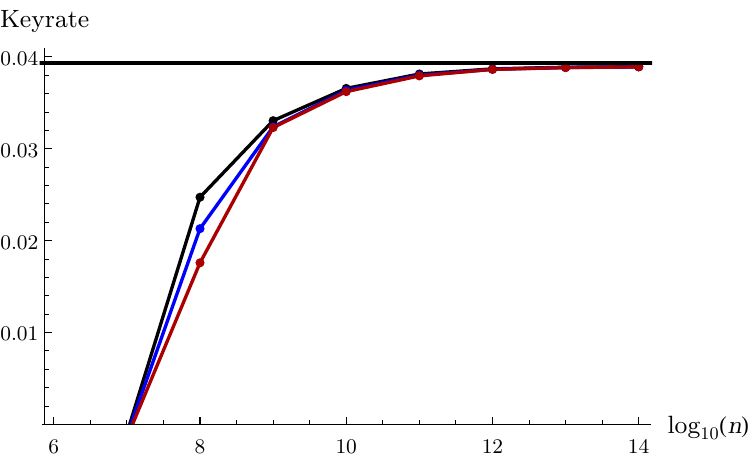}
} 
\subfloat[\cite{RBG+17} parameters, $\p=0.03$ (with heuristic $\lin_{\p}$)]{
\includegraphics[width=0.48\textwidth]{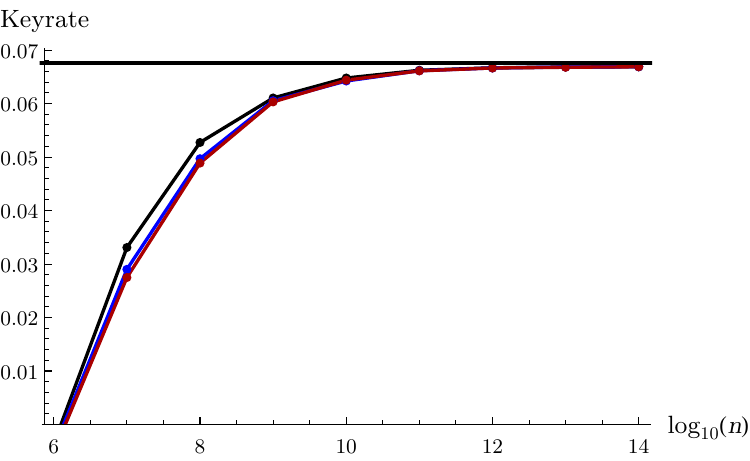}
}
\caption{Finite-size keyrates as a function of number of rounds in Protocol~\ref{prot:optcoll} (see Theorem~\ref{th:optcoll}),
using honest devices following the estimated parameters in~\cite{MvDR+19} for the loophole-free Bell test in~\cite{HBD+15}, for $\p=0$ and $\p=0.03$ (the latter is a rough estimate of the choice of $\p$ which yields the highest asymptotic keyrate for these experimental parameters). Note that the latter graph is computed using a heuristic estimate of $\lin_{\p}$ rather than a certified bound.
The error-correction protocol is taken to satisfy Eqs.~\eqref{eq:optEC}~and~\eqref{eq:optcollAEP}. The colours correspond to soundness parameters of $\esound=10^{-3}$, $10^{-6}$, and $10^{-9}$ for black, blue, and red respectively, and the completeness parameter is $\ecom=10^{-2}$ in all cases.
The horizontal line denotes the asymptotic keyrate. 
All other parameters in Theorem~\ref{th:optcoll} were numerically optimized. The required number of rounds to achieve positive keyrate is substantially lower than Protocol~\ref{prot:DIQKD} (see Fig.~\ref{fig:experiments}).
}
\label{fig:protIID}
\end{figure}

In Fig.~\ref{fig:protIID}, we plot the results of Theorem~\ref{th:optcoll}, focusing on the~\cite{RBG+17} experiment. This protocol has improved finite-size performance as compared to the original Protocol~\ref{prot:DIQKD} (under the collective-attacks assumption) due to at least two factors. Firstly, we can potentially use larger $\gamma$ values, as previously mentioned (some of the points shown in the figure correspond to values ranging up to $\gamma \approx 0.3$).
Secondly, we find that for fixed values of $\gamma,n,\wexp,\dtol,\dsou$, the binomial-distribution bounds for this protocol (Eqs.~\eqref{eq:eIIDoptcoll} and \eqref{eq:ecPEoptcoll}) are typically 
several orders of magnitude better
than their counterparts for Protocol~\ref{prot:DIQKD} (Eqs.~\eqref{eq:eIIDbound} and \eqref{eq:ecPEbound1}). Intuitively, this arises because in Protocol~\ref{prot:DIQKD}, the number of test rounds is itself a random variable, hence increasing the variance in e.g.~the number of rounds where $C_j=1$.
Practically speaking, this means Protocol~\ref{prot:DIQKD} requires noticeably larger values of $\dtol$ and $\dsou$ in order to achieve given completeness and soundness parameters, hence reducing the keyrate by a nontrivial amount.

However, we see that even with the optimistic assumptions that yield Theorem~\ref{th:optcoll}, the keyrate for the estimated experimental parameters we consider only becomes positive at fairly large $n$. This indicates that substantial further work is necessary in order to achieve a demonstration of positive finite-size keyrates.

\section{Conclusion and further work}
\label{sec:conclusion}

In this work, we have performed a finite-size analysis for a protocol that combines several of the most promising approaches towards improving keyrates for DIQKD. Furthermore, we develop an algorithm that computes arbitrarily tight lower bounds on the asymptotic keyrates for protocols of this form (i.e.~allowing for noisy preprocessing and random key measurements, but restricted to one-way error correction), which applies to all 2-input 2-output scenarios. This allows us to prove a new threshold of $9.33\pct$ for noise tolerance in the depolarizing-noise model, and we show (see Sec.~\ref{sec:bestbnd}) that for one-way protocols, any further improvements on this threshold can only be fairly small. Finally, we propose a modified protocol, based on a pre-shared key, that overcomes the disadvantage of the sifting factor in random-key-measurement protocols.

We remark that while the finite-size analysis shown here is for a protocol based on the CHSH inequality, our algorithm for the asymptotic keyrates applies more generally to 2-input 2-output scenarios. If some exploration with this algorithm suggests that an improvement can be obtained by using an inequality other than CHSH, then it would not be difficult to modify the finite-size analysis for such an inequality, as was done in e.g.~\cite{BRC20} for DIRE (essentially, it would just correspond to having a different bound $\lin_{\p}$). 

However, our results show that for the NV-centre experiment in~\cite{HBD+15} and the cold-atom experiment in~\cite{RBG+17}, an impractically large sample size (for those implementations) would still be needed in order to achieve a positive finite-size keyrate, even if one makes the optimistic assumption of collective attacks. 
A significant question that remains to be addressed is that of photonic experiments~\cite{SMC+15,GVW+15}, which achieve lower CHSH values but much larger sample sizes (for instance, the photonic DIRE demonstration in~\cite{LLR+21} implemented one run with $n=1.3824 \times 10^{11}$ over 19.2 hours, and one run with $n=3.168 \times 10^{12}$ over 220 hours). 
Unfortunately, for photonic models our heuristic results (Sec.~\ref{sec:1rndresults}) suggest that incorporating random key measurements is less helpful in improving the keyrate, because the experimental parameters that achieve maximal CHSH value also tend to result in higher error probability (with respect to Bob's outputs) for one of Alice's measurements than the other. Despite this challenge, we note that there is much freedom in parameter optimization for photonic experiments~\cite{MSS20}, and given the algorithm we developed here for bounding the asymptotic keyrates, it is now possible to analyze variants such as choosing different ratios for the two key-generating measurements, or different amounts of noisy preprocessing for each basis. We aim to continue studying this in future work.

We also note that for photonic implementations, a protocol modification termed \emph{random postselection} was recently proposed in~\cite{XZZ+22}, where it was shown to provide significant improvements in the detection-efficiency thresholds required for positive asymptotic keyrates, under the assumption of collective attacks. However, that modified protocol involves public announcements in each round that violate the Markov condition in the EAT, and hence the EAT cannot be directly applied to prove security of that protocol against general attacks. It remains to be seen whether this can be proven using some other approach, although one challenge to overcome would be the issue that for other protocols with a slightly similar public-announcement structure, it has been shown~\cite{TTB+16} that the achievable keyrates against general attacks are strictly lower than for collective attacks. Any approach for proving the security of the~\cite{XZZ+22} protocol against general attacks (with the same asymptotic keyrates as for collective attacks) would have to rely on some specific difference between that protocol and the ones analyzed in~\cite{TTB+16}.

As another extension of our work, our algorithm for bounding the asymptotic DIQKD keyrates also applies to DIRE. Since the results it returns are arbitrarily tight and easily incorporated into the EAT, this could be combined with the finite-size analysis of~\cite{BRC20,LLR+21} to improve the results of those works --- the proof methods used there yield slightly suboptimal keyrates, in that they either bound the min-entropy rather than the von Neumann entropy~\cite{BRC20}, or they only bound the entropy of one party's outputs and are restricted to the CHSH inequality~\cite{LLR+21}. Our algorithm would yield tight bounds on the two-party von Neumann entropy for 2-input 2-output Bell inequalities; furthermore, it allows the possibility of using the random-key-measurement approach (by applying the pre-shared key proposal) to improve the keyrates. 

\section*{Acknowledgements}

We are very grateful to Peter Brown for detailed information on the finite-size analysis in~\cite{BRC20,LLR+21}, as well as Omar Fawzi and Fr\'{e}d\'{e}ric Dupuis for discussions on the entropy accumulation theorem. 
We also thank Koon Tong Goh and Ignatius W.\ Primaatmaja for helpful discussion and feedback. 
Finally, we thank the reviewers and editor for many helpful suggestions in improving the manuscript.

E.~Y.-Z.~T.~and R.~R.~are supported by the Swiss National Science Foundation (SNSF) grant number 20QT21\_187724 via the National Center for Competence in Research for Quantum Science and Technology (QSIT), the Air Force Office of Scientific Research (AFOSR) via grant FA9550-19-1-0202, and the QuantERA project eDICT. 
P.~S.~is supported by by the Swiss National Science Foundation (SNSF).
C.~C.-W.~L is supported by the National Research Foundation (NRF) Singapore, under its NRF Fellowship programme (NRFF11-2019-0001) and Quantum Engineering Programme 1.0 (QEP-P2).

\section*{Computational platform and code} 

The min-tradeoff function computations were performed using the MATLAB package YALMIP \cite{yalmip} with the solver MOSEK~\cite{mosek}, while optimization of the finite-size keyrates was performed in Mathematica and MATLAB. Some of the calculations reported here were performed using the Euler cluster at ETH Z\"{u}rich. 
The code used to compute the certified lower bounds is available at the following URL:
\begin{center} 
\href{https://github.com/ernesttyz/qbitent}{https://github.com/ernesttyz/qbitent}
\end{center} 

\printbibliography

\newpage
\appendix

\section{Alternative map to describe noisy preprocessing}
\label{app:altNPP}

For Pauli measurements in the $X$-$Z$ plane, applying the Pauli $Y$ operator to the state before the measurement causes the measurement outcome to flip. Therefore, it can be seen that if one first subjects the state $\rho_{ABE}$ to the isometry $V_A \otimes \id_{BE}$, where
\begin{align}
V_A\ket{\psi}_{A} = \sqrt{1-\p} \ket{\psi}_{A}\ket{0}_T + \sqrt{\p} \, Y \ket{\psi}_{A}\ket{1}_T,
\label{eq:isoNPP}
\end{align}
then measures $A_x$ and directly stores the outcome in $\hat{A}_x$, 
this produces exactly the same reduced state on the registers $\hat{A}_x E$ as 
the noisy-preprocessing procedure.
Then since $\ket{\hat{\rho}}_{ABET}\defvar (V_A \otimes \id_{BE})\ket{\rho}_{ABE}$ is a pure state, by the same reasoning as before we have 
\begin{align}
H(\hat{A}_x|E) = D(\hat{\rho}_{ABT} \Vert \widetilde{\calZ}_x(\hat{\rho}_{ABT})) = D(\widetilde{\calG}({\rho}_{AB}) \Vert \widetilde{\calZ}_x(\widetilde{\calG}({\rho}_{AB}))),
\label{eq:altHduality}
\end{align}
where
\begin{align}
\widetilde{\calG}(\sigma_{AB}) &\defvar (V_A\otimes\id_{B}) \sigma_{AB} (V_A\otimes\id_{B})^\dagger, \\
\widetilde{\calZ}_x(\sigma_{ABT}) &\defvar \sum_a (\pvm_{a|x}\otimes\id_{BT}) \sigma_{ABT} (\pvm_{a|x}\otimes\id_{BT}). \label{eq:calZalt} 
\end{align}
This description of noisy preprocessing appears to more specifically exploit the reduction to Pauli measurements in the 2-input 2-output scenario, and hence may be less general.

\section{Alternative continuity bound}
\label{app:altcont}

In principle we could use the following approach: we first note that an intermediate step in the derivation of~\eqref{eq:altHduality} is the relation~\cite{Col12}
\begin{align}
H(\hat{A}_x|E) = H(\widetilde{\calZ}_x(\hat{\rho}_{ABT})) - H(\hat{\rho}_{ABT}),
\end{align}
and hence bounding how much $H(\hat{A}_1|E)$ changes in terms of the measurement angle is equivalent to bounding how much $H(\widetilde{\calZ}_1(\hat{\rho}_{ABT}))$ changes (since the $H(\hat{\rho}_{ABT})$ term is independent of the measurement).
We introduce the qubit channel
\begin{align}
\bar{\calZ}^\theta(\sigma) \defvar \sum_a \pvm_a(\theta) \sigma \pvm_a(\theta),
\end{align}
where the projectors 
$\pvm_a(\theta)$
describe a Pauli measurement in the $X$-$Z$ plane along angle $\theta$, in which case $\widetilde{\calZ}_1$ (for measurement angle $\thA$) is basically the channel $\bar{\calZ}^{\thA}$ applied only to the $A$ system. By the properties of the diamond norm, we have
\begin{align}
\norm{\bar{\calZ}^{\thA}_A(\hat{\rho}_{ABT}) - \bar{\calZ}^{\thA+\delta}_A(\hat{\rho}_{ABT})}_1 
\leq \norm{\bar{\calZ}^{\thA} - \bar{\calZ}^{\thA+\delta}}_\diamond 
\leq \dim(A)\norm{J(\bar{\calZ}^{\thA}) - J(\bar{\calZ}^{\thA+\delta})}_1,
\end{align}
where $J(\mathcal{E})$ denotes the (normalized) Choi state of the channel $\mathcal{E}$. 
In this case we have $\dim(A)=2$, and the trace distance between the Choi states can be computed in closed form (exploiting the fact that two of the Bell states are simultaneous eigenstates of the two Choi states). 
We could then apply the Fannes-Audenaert continuity bound to bound how much $H(\widetilde{\calZ}_1(\hat{\rho}_{ABT}))$ (and hence $H(\hat{A}_1|E)$) changes.\footnote{It was also possible to instead bound the change (in terms of trace distance) of the state on registers $\hat{A}_x E$ directly, using a slightly different channel, but it turns out to yield a slightly worse bound.} Unfortunately, the Fannes-Audenaert bound scales poorly at small values of trace distance, and hence this approach yields a much worse continuity bound than~\eqref{eq:contbnd} (in fact, the resulting bound has infinite derivative with respect to $\delta$ at $\delta=0$).

\section{Proof of Theorem~\ref{th:collective}}
\label{app:collective}

We observe that the completeness proof from Sec.~\ref{sec:com} directly applies, since we have not changed the honest behaviour. As for the soundness proof, we 
follow the argument in Sec.~\ref{sec:soundness} up until Eq.~\eqref{eq:secsplit}, upon which we again require a bound on the first term in that equation. To do so, we note that since the device behaviour is identical in every round, the true CHSH winning probability for the devices can be denoted as a constant value $\wtrue$ (though this value is not directly known to Alice and Bob). We can thus define the following exhaustive possibilities for the device behaviour (again, $\rho$ is the state at the end of Protocol~\ref{prot:virtual}):
{\setlist[enumerate]{leftmargin=2cm,labelindent=2cm}
\begin{enumerate}[label*=Case \arabic*:,ref=\arabic*]
\item \label{case:ECsmall2} For the state $\rho$, $\pr{\Og \land \Oh \land \OpPE} \leq \es^2$.
\item \label{case:wlow} $\wtrue < \wexp - \dtol - \dsou$.
\item \label{case:whigh} 
Neither of the above are true.
\end{enumerate}
}
\noindent In case~\ref{case:ECsmall2}, the first term of Eq.~\eqref{eq:secsplit} is immediately bounded by
\begin{align}
\pr{\Og \land \Oh \land \OpPE} \leq \es^2.
\end{align}
In case~\ref{case:wlow}, we observe that in each round, the probability of the CHSH game being played and the devices winning is $\wtrue\gamma$. Therefore, we have
\begin{align}
\pr{\freq_\ctstr(1) \geq (\wexp-\dtol)\gamma}
&=\pr{\freq_\ctstr(\neg 1) < 1-(\wexp-\dtol)\gamma} \nonumber\\
&\leq\pr{\freq_\ctstr(\neg 1)n \leq \floor{(1-(\wexp-\dtol)\gamma)n}} \nonumber\\
&= \cdfBin{n}{1-\wtrue\gamma}{\floor{(1-(\wexp-\dtol)\gamma)n}} \nonumber\\
&\leq \eps_\mathrm{IID}, 
\label{eq:eIIDbound}
\end{align}
where the last inequality follows from the fact that we have $1-\wtrue\gamma > 1-(\wexp - \dtol - \dsou)\gamma$, and $\cdfBin{n}{p}{k} \leq \cdfBin{n}{p'}{k}$ for $p \geq p'$.
(Again, one could obtain a simpler 
expression by noting $\pr{\freq_\ctstr(1) \geq (\wexp-\dtol)\gamma} 
\leq \pr{\freq_\ctstr(1) \geq (\wtrue+\dsou)\gamma}$ and then upper bounding the latter expression via the Chernoff bound (as long as $\dsou$ is chosen small enough to ensure $\dsou/\wtrue \leq 1$):
\begin{align}
\pr{\freq_\ctstr(1) \geq (\wtrue+\dsou)\gamma} 
\leq e^{-\frac{n \gamma \dsou^2}{3\wtrue}} 
\leq e^{-\frac{n \gamma \dsou^2}{3\left(\wexp-\dtol-\dsou\right)}},
\end{align}
but this gives significantly poorer results.) 
Overall, this allows us to bound the first term of Eq.~\eqref{eq:secsplit} in this case by
\begin{align}
\pr{\Og \land \Oh \land \OpPE} \leq \pr{\OpPE} \leq \pr{\freq_\ctstr(1) \geq (\wexp-\dtol)\gamma} \leq \eps_\mathrm{IID}.
\end{align}

As for case~\ref{case:whigh}, we can directly apply the AEP (Corollary~4.10 of~\cite{DFR20}) to obtain the following bound 
in place of Theorem~\ref{th:rawHmin}, with the slight difference that we have not conditioned on $\OpPE$ yet:
\begin{align}
\Hmin^{\es}(\str{A}|\str{X}\str{Y}E)_{\rho} >
n\g(\wexp-\dtol-\dsou) - \sqrt{n} \,(2\log5)\sqrt{\log\frac{2}{\es^2}}.
\label{eq:rawHminAEP}
\end{align}
The remainder of the analysis proceeds very similarly to 
Sec.~\ref{sec:soundness}. Specifically, since we have $\pr{\Og \land \Oh \land \OpPE} \geq \es^2$ in this case, we can conclude 
\begin{align}
\Hmin^{\es}(\str{A}|\str{X}\str{Y}\str{L}E)_{\rho_{\land \Og \land \Oh \land \OpPE}}
&\geq \Hmin^{\es}(\str{A}|\str{X}\str{Y}\str{L}E)_{\rho}
\nonumber \\
&\geq \Hmin^{\es}(\str{A}|\str{X}\str{Y}E)_{\rho} 
- \cmax - \ceil{\log\left(\frac{1}{\eh}\right)}.
\end{align}
Putting this together with Eq.~\eqref{eq:rawHminAEP}, we see that as long as we choose $\lkey$ satisfying Eq.~\eqref{eq:lkeycoll}, we will have
\begin{align}
\frac{1}{2}\left(\Hmin^{\es}(\str{A}|\str{X}\str{Y}\str{L}E)_{\rho_{\land \Og \land \Oh \land \OpPE}} - \lkey  + 2\right) \geq \log\frac{1}{\ePA}.
\end{align}
Again noting that $\pr{\Og \land \Oh \land \OpPE} \geq \es^2$ in this case, the Leftover Hashing Lemma then implies the first term of Eq.~\eqref{eq:secsplit} is bounded by
\begin{align}
&\frac{1}{2}\norm{\mPA(\rho_{\land \Og \land \Oh \land \OpPE})_{K_A E'} - \idk_{K_A} \otimes \mPA(\rho_{\land \Og \land \Oh \land \OpPE})_{E'}}_1 \nonumber\\
\leq{}& 2^{-\frac{1}{2} \left(\Hmin^{\es}(\str{A}|\str{X}\str{Y}\str{L}E)_{\rho_{\land \Og \land \Oh \land \OpPE}} - \lkey + 2\right)} + 2\es \nonumber\\
\leq{}& \ePA + 2\es.
\end{align}

Therefore, we finally conclude that under the collective-attacks assumption, the secrecy condition is satisfied by choosing
\begin{align}
\esecr = \max\{\es^2, \eps_\mathrm{IID}, \ePA + 2\es\} + \eh = \max\{\eps_\mathrm{IID}, \ePA + 2\es\} + \eh.
\end{align} 
As before, the protocol is $\eh$-correct, so we conclude that it is $(\max\{\eps_\mathrm{IID}, \ePA + 2\es\} + 2\eh)$-sound when performed with $\lkey$ satisfying Eq.~\eqref{eq:lkeycoll}.

\section{Proof of Theorem~\ref{th:optcoll}}
\label{app:optcoll}

We first show completeness by following a similar approach as in Sec.~\ref{sec:com}, 
except that here we do not need a conversion to a virtual parameter estimation procedure, because in this protocol Bob has access to the exact value of $\str{A}_t$.
Explicitly, observe that the abort condition for this protocol is the event $(\hash(\str{A}) \neq \hash(\tilde{\str{A}})) \lor \left(\freq_{\str{c}_t}(1) < \wexp-\dtol\right)$. Again, the probability of this event is upper bounded by the probability that $(\str{A}\neq\tilde{\str{A}}) \lor \left(\freq_{\str{c}_t}(1) < \wexp-\dtol\right)$.
For the honest behaviour, the probability that $\str{A}\neq\tilde{\str{A}}$ is again at most $\ecEC$ (by the guarantees of the error-correction procedure), while the probability that $\freq_{\str{c}_t}(1) < \wexp-\dtol$ is upper bounded by
\begin{align}
\pr{\freq_{\str{c}_t}(1) < \wexp-\dtol}_\mathrm{hon}
&\leq 
\pr{\freq_{\str{c}_t}(1)\gamma n \leq \floor{(\wexp-\dtol)\gamma n}}_\mathrm{hon} \nonumber\\
&= \cdfBin{\gamma n}{\wexp}{\floor{(\wexp-\dtol)\gamma n}}.
\end{align}
Hence by similar reasoning as in Sec.~\ref{sec:com} (although without needing the virtual parameter estimation, so one could also instead just use the union bound here), we see that the probability of the honest protocol aborting is at most $\ecEC + \ecPE$ with $\ecPE$ defined as in~\eqref{eq:ecPEoptcoll}.

As for soundness, we follow the argument in Sec.~\ref{sec:finite} up until Eq.~\eqref{eq:esecrnew}, though it is also no longer necessary to introduce the virtual parameter estimation. We also need to slightly modify the event definitions (and it is no longer necessary to consider the event $\OpPE$): 
{\setlist[description]{leftmargin=2cm,labelindent=2cm}
\begin{description}
\item $\Og$: $\str{A}_g = \tilde{\str{A}}_g$ 
\item $\Oh$: $\hash(\str{A}_g) = \hash(\tilde{\str{A}}_g)$
\item $\OPE$: $\freq_{\str{c}_t}(1)\geq \wexp-\dtol$ 
\end{description}
}
\noindent Again, the accept condition for this protocol can be stated as the event $\Oh \land \OPE$. 
For this protocol, we can prove a bound of the form~\eqref{eq:esecrnew} directly, without splitting it into the terms in Eq.~\eqref{eq:secsplit}. To do so, denote the state after the parameter-estimation step in Protocol~\ref{prot:optcoll} as $\rho$, and consider the following exhaustive possibilities:
{\setlist[enumerate]{leftmargin=2cm,labelindent=2cm}
\begin{enumerate}[label*=Case \arabic*:,ref=\arabic*]
\item \label{case:ECsmall3} 
For the state $\rho$, $\pr{\Oh \land \OPE} \leq \es^2$.
\item \label{case:wlow3} $\wtrue < \wexp - \dtol - \dsou$.
\item \label{case:whigh3} Neither of the above are true.
\end{enumerate}
}
In case~\ref{case:ECsmall3}, the left-hand-side of Eq.~\eqref{eq:esecrnew} is immediately bounded by $
\es^2$.
In case~\ref{case:wlow3}, we observe that in each of the $\gamma n$ test rounds (note that now we focus only on the test rounds, instead of the full output strings as we did in the previous proofs), the probability of winning the CHSH game is $\wtrue
$. Therefore, we have
\begin{align}
\pr{\freq_{\str{c}_t}(1) \geq \wexp-\dtol}
&=\pr{\freq_{\str{c}_t}(
0
) < 1-\wexp+\dtol} \nonumber\\
&=\pr{\freq_{\str{c}_t}(0)\gamma n < (1-\wexp+\dtol)\gamma n} \nonumber\\
&\leq\pr{\freq_{\str{c}_t}(0)\gamma n \leq \floor{(1-\wexp+\dtol)\gamma n}} \nonumber\\
&= \cdfBin{\gamma n}{1-\wtrue}{\floor{(1-\wexp+\dtol)\gamma n}} \nonumber\\
&\leq \eps_\mathrm{IID},
\end{align}
which allows us to bound the left-hand-side of Eq.~\eqref{eq:esecrnew} in this case by
\begin{align}
\pr{\Oh \land \OPE} \leq \pr{\OPE} = \pr{\freq_{\str{c}_t}(1) \geq \wexp-\dtol} \leq \eps_\mathrm{IID}.
\end{align}

Finally, for case~\ref{case:whigh3}, we note that since Eve's quantum side-information $E$ is genuinely in (IID) tensor-product form under the collective-attacks assumption, we can denote the subsets corresponding to test and generation rounds as $\str{E}_t$ and $\str{E}_g$ respectively. The AEP in~\cite{DFR20} then gives
\begin{align}
\Hmin^{\es}(\str{A}_g|\str{X}_g\str{Y}_g\str{E}_g)_{\rho} >
(1-\gamma)n\lin_\p(\wexp-\dtol-\dsou) - \sqrt{(1-\gamma)n} \,(2\log5)\sqrt{\log\frac{2}{\es^2}},
\end{align}
and 
we proceed similarly: since we have $\pr{\Oh \land \OPE} \geq \es^2$ in case~\ref{case:whigh3}, we can conclude
\begin{align}
\Hmin^{\es}(\str{A}_g|\str{A}_t\str{X}\str{Y}\str{L}{E})_{\rho_{\land \Oh \land \OPE}}
&\geq \Hmin^{\es}(\str{A}_g|\str{A}_t\str{X}\str{Y}\str{L}{E})_{\rho}
\nonumber \\
&\geq \Hmin^{\es}(\str{A}_g|\str{A}_t\str{X}\str{Y}{E})_{\rho} 
- \cmax - \ceil{\log\left(\frac{1}{\eh}\right)} \nonumber \\
&= \Hmin^{\es}(\str{A}_g|\str{X}_g\str{Y}_g\str{E}_g)_{\rho}
- \cmax - \ceil{\log\left(\frac{1}{\eh}\right)},
\end{align}
where the last line holds because the test rounds are independent of the generation rounds.
Hence as long as we choose $\lkey$ satisfying Eq.~\eqref{eq:lkeyoptcoll}, we will have
\begin{align}
\frac{1}{2}\left(\Hmin^{\es}(\str{A}_g|\str{A}_t\str{X}\str{Y}\str{L}{E})_{\rho_{\land \Oh \land \OPE}} - \lkey  + 2\right) \geq \log\frac{1}{\ePA}.
\end{align}
Again noting that $\pr{\Oh \land \OPE} \geq \es^2$ in this case, the Leftover Hashing Lemma then implies 
\begin{align}
&\frac{1}{2}\norm{\mPA(\rho_{\land \Oh \land \OPE})_{K_A E'} - \idk_{K_A} \otimes \mPA(\rho_{\land \Oh \land \OPE})_{E'}}_1 \nonumber\\
\leq{}& 2^{-\frac{1}{2} \left(\Hmin^{\es}(\str{A}_g|\str{A}_t\str{X}\str{Y}\str{L}{E})_{\rho_{\land \Oh \land \OPE}} - \lkey + 2\right)} + 2\es \nonumber\\
\leq{}& \ePA + 2\es.
\end{align}

Therefore, we finally conclude that under the collective-attacks assumption, the secrecy condition is satisfied by choosing
\begin{align}
\esecr = \max\{\es^2, \eps_\mathrm{IID}, \ePA + 2\es\} = \max\{\eps_\mathrm{IID}, \ePA + 2\es\}.
\end{align} 
As before, the protocol is $\eh$-correct, so we conclude that it is $(\max\{\eps_\mathrm{IID}, \ePA + 2\es\} + \eh)$-sound when performed with $\lkey$ satisfying Eq.~\eqref{eq:lkeyoptcoll}.

\end{document}